# Interpretable Machine Learning for High-Strength High-Entropy Alloy Design

Anurag Bajpai[1,2*], Ziyuan Rao[1,3], Abhinav Dixit[2], Krishanu Biswas[2*] and Dierk Raabe[1*]

[1]*Max-Planck-Institut for Sustainable Materials, 40237 Düsseldorf, Germany*

[2]*Department of Materials Science and Engineering, Indian Institute of Technology, Kanpur, Uttar Pradesh 208016, India*

[3]*National Engineering Research Center of Light Alloy Net Forming, Shanghai Jiao Tong University, Shanghai 200030, China*

**Corresponding Authors:** *Anurag Bajpai, a.bajpai@mpie.de; Krishanu Biswas, kbiswas@iitk.ac.in; Dierk Raabe, d.raabe@mpie.de*

## Abstract

High-entropy alloys (HEAs) are metallic materials with solid solutions stabilized by high mixing entropy. Some exhibit excellent strength, often accompanied by additional properties such as magnetic, invar, corrosion, or cryogenic response. This has spurred efforts to discover new HEAs, but the vast compositional search space has made these efforts challenging. Here we present a framework to predict and optimize the yield strength of face-centered cubic (FCC) HEAs, using CoCrFeMnNi-based alloys as a case study due to abundant available data. Our novel Residual Hybrid Learning Model (RELM) integrates Random Forest and Gradient Boosting, enhanced by material attribute data, to handle sparse, skewed datasets for real-world alloys. A hybrid Generative Adversarial Network-Variational Autoencoder model explores new alloy compositions beyond existing datasets. By incorporating processing parameters, which determine the microstructure and thus strength, RELM achieves an $R^2$ score of 0.915, surpassing traditional models. SHapley Additive Explanations (SHAP) and Partial Dependencies enhance interpretability, revealing composition-processing-property relationships, as validated by experiments, including X-ray diffraction, SEM analysis, and tensile testing. The model discovered two novel $Co_{20}Cr_{16}Fe_{20}Mn_{16}Ni_{24}Al_4$ and $Co_{24}Cr_{12}Fe_{12}Mn_{16}Ni_{28}Al_4Si_4$ HEAs with a maximum possible yield strength of 842 and 937 MPa, significantly exceeding previously reported values for these alloy systems. This study pioneers interpretable machine learning in alloy design, providing a rigorous, data-driven approach to discovering, processing, and optimizing real-world materials. The findings highlight the critical role of both compositional and post-fabrication processing parameters in advancing the understanding of composition-processing-property relationships.

## 1. Introduction

Alloy design is a foundational pillar of materials science that has driven advancements from the Bronze Age to the Space Age in various industrial and consumer sectors, including energy, health, transportation, information, safety, and infrastructure. Traditionally, material development has required the gradual addition of alloying elements to a primary base metal to attain desired properties. This approach has successfully produced a huge variety of metallic alloys suited to a broad range of engineering applications (*1*, *2*). However, the discovery of high-entropy alloys (HEAs) has fundamentally disrupted this paradigm, introducing a radically new approach. While conventional alloys typically feature one or two principal elements, HEAs consist of multiple principal elements in equiatomic or near-equiatomic proportions (*3*). This unique composition rule not only results in a huge and unexplored chemical design space, but the vital underlying processing steps that lend metallic alloys their complex microstructure, characterized by lattice defects and determining most of their properties, remain unknown for this new materials class. The high configurational entropy in HEAs serves to stabilize solid solution phases while suppressing brittle intermetallic compounds (*4*-*6*). However, the mechanical properties of metallic materials are primarily determined by their non-equilibrium microstructure, shaped by processing methods such as casting, rolling, and heat treatments. The complex interplay between equilibrium features like entropy and enthalpy with phase formation kinetics, lattice defects, and processing in HEAs poses significant challenges, rendering traditional experimental and computational methods inefficient or even impractical for their design and optimization (*7*-*9*).

Established computational techniques such as density functional theory (DFT) (*10*-*12*), molecular dynamics (MD) (*13*-*15*), and the calculation of phase diagrams (CALPHAD) (*16*-*18*) have provided valuable insights into the phase stability and certain properties of HEAs. However, these methods are typically limited by their reliance on simplified compositions and microstructures, as well as by their lack of predictive capabilities regarding the synthesis-processing-microstructure relationships that transform a lattice structure into a real-world material (*19, 20*). We therefore propose a novel approach for the discovery and design of HEAs that leverages machine learning (ML) techniques with an emphasis on scientific interpretability in terms of composition-processing-property linkages. ML has demonstrated substantial potential in accurately predicting material properties by leveraging known materials datasets. This significantly reduces the necessity for extensive experimental trials and computationally intensive simulations (*21*, *22*). This capability is particularly crucial for HEAs, considering their sheer limitless range of potential compositions. ML approaches have already demonstrated their efficacy in accurately predicting various characteristics of HEAs, including phase formation, hardness, and yield strength (*23-26*). Prior research has utilized artificial neural networks (ANNs) to predict the hardness of AlCoCrFeMnNi alloys (*27*), gradient boosting trees (GB-Trees) to estimate the elastic properties of $Al_{0.3}CoCrFeNi$ (*28*), and random forest regressors to forecast the yield strength of HEAs at various temperatures (*29*). While these ML techniques have enhanced the efficiency of alloy design, they encounter significant challenges, including avoiding overfitting, preserving interpretability, and capturing the intricate, non-linear structure-property relationships of HEAs. Obtaining meaningful scientific knowledge from data-driven models remains a significant obstacle, particularly in materials science, where comprehending the relationships between composition, processing, and properties is crucial for rational alloy design, particularly concerning mechanical properties. Previous research has mainly used methods such as SHapley Additive exPlanations (SHAP) (*30*) and feature importance rankings (*31*) to interpret predictions by ML models. While these tools can offer causal-like insights, they frequently fail to provide a thorough, actionable, mechanism-based foundation for the exploration of new alloys.

Ensemble learning has made recent progress in addressing these challenges by merging the results of different models to improve the accuracy and reliability of predictions (*32*). This strategy has proven effective in reducing variability and systematic errors linked to individual models. However, in the context of HEAs, where understanding the underlying physical mechanisms is as crucial as achieving accurate predictions, traditional ensemble methods still encounter some limitations. Due to the complex properties of HEAs, it is important to have both accurate predictions and models that are easy to comprehend. However, classic ensemble learning methods sometimes sacrifice interpretability. To address these challenges, we introduce a novel ensemble learning framework, called the Residual Ensemble Learning Model (RELM), specifically tailored for the discovery and optimization of HEAs. This framework is resilient to sparse materials datasets while maintaining a focus on scientific interpretability. Our approach employs residual hybrid learning, wherein a primary model predicts the target property and a secondary model learns from the residual errors of the first. This strategy integrates the principles of bagging and boosting (*33*), effectively balancing variance and bias, and enhancing the model's ability to capture subtle, intricate patterns within the data. Our framework is tailored to predict the yield strength of face-centered cubic (FCC) HEAs as a demonstrator case, chosen for their superior ductility compared to body-centered cubic (BCC) and hexagonal close-packed (HCP) HEAs, thereby addressing the widely recognized strength-ductility trade-off. In addition, the stacking fault energy map in FCC structures is rather simple, making it easier to predict mechanical behavior and aiding in the rational design and optimization of alloys (*34*).

Our research aims to achieve several key alloy design objectives: (*i*) integrate critical processing parameters, such as the degree of deformation during cold rolling, annealing temperature, and annealing duration, to enhance the scientific rigor of our predictive framework and extend the methodology from mere structure prediction to real-world material prediction, wherein processing and microstructure are crucial; (*ii*) address the challenge of sparse materials data using methods for enhancing the robustness and reliability of our ML models without the loss of vital information; (*iii*) demonstrating the superiority of residual hybrid learning over conventional ensemble approaches by capturing the complex, multi-dimensional relationships within the data. It not only enhances predictive accuracy but also facilitates gaining insights into the underlying mechanisms, making it well-suited for FCC HEAs development; and (*iv*) advancing ML models interpretability beyond conventional SHAP analysis by employing partial pair dependencies to construct multi-dimensional compositional and processing maps. Ultimately, such maps can form the basis for the design of future HEAs by providing a visual and quantitative means of exploring and interpreting the complex interdependencies of variables and their interactions that affect material properties.

## 2. Results and Discussion

Figure 1 illustrates the overview of our RELM workflow to discover high-strength FCC HEAs and analyze their composition-processing-property correlations. This is an integrated, three-module workflow for a) residual ensemble learning model development; b) RELM interpretation to obtain actionable composition-processing-property relationships; and c) generative alloy design of new FCC HEA compositions, with RELM predictions are validated through experiments. The following subsections will discuss these outcomes in detail.

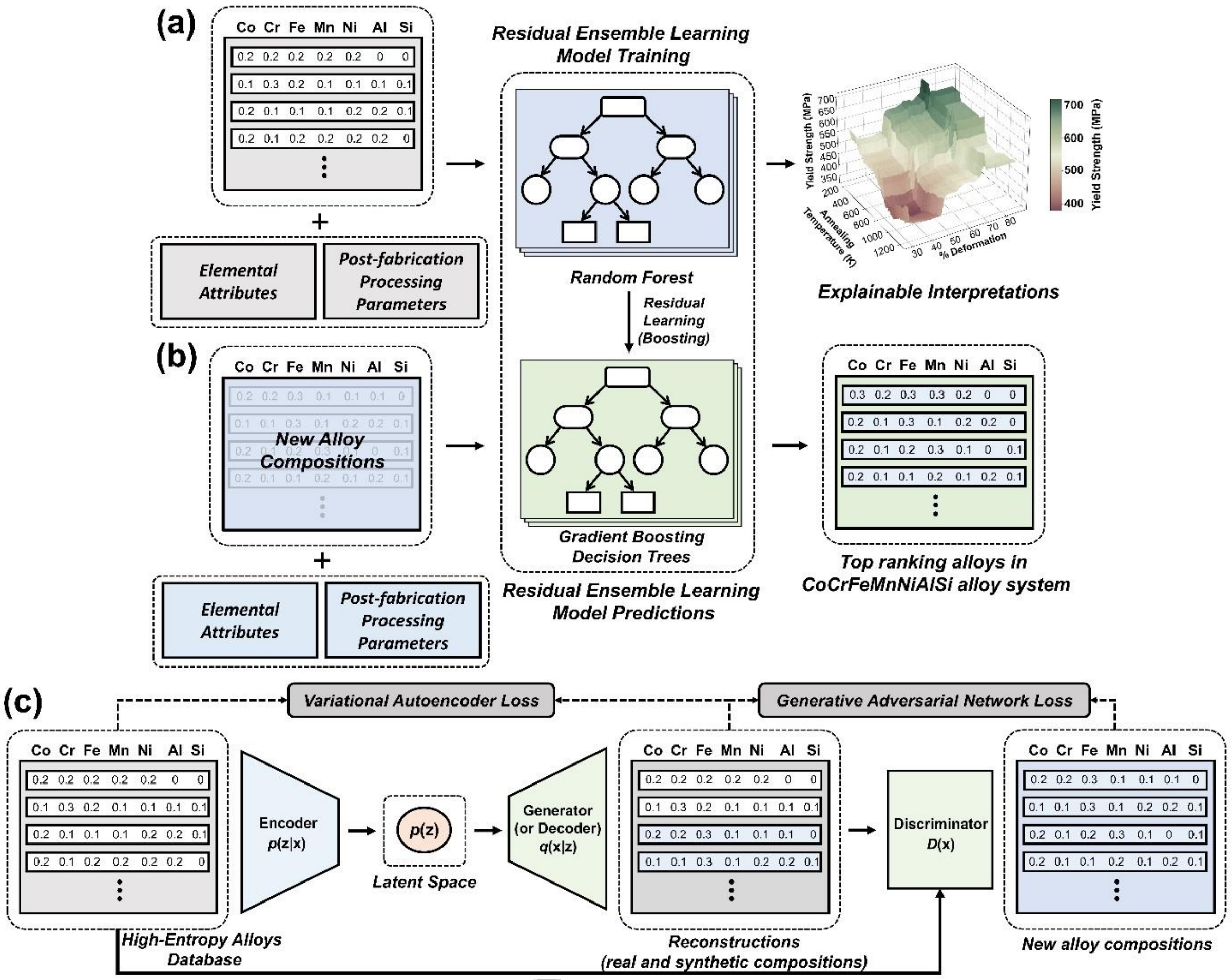

**Figure 1** – Workflow of the developed residual ensemble learning model (RELM) framework for targeted composition design of high-strength FCC HEAs. (a) Establishment of the RELM framework, model training with various feature spaces (over 10 seeds) and establishment of composition-processing-property interpretation from RELM; (b) RELM model predictions utilizing the mean yield strength prediction value from 10 models (for 10 seeds) as the final prediction, while the corresponding variance as the uncertainty in predictions; (c) FCC HEA composition generation scheme using generative adversarial network-variational autoencoder (GAN-VAE) hybrid model to discover unknown FCC HEA compositions using RELM. The encoder $p(z|x)$ takes compositions of FCC HEAs as input and develops a lower-dimensional latent space distribution $p(z)$ and the generator (or decoder) $p(x|z)$ with parameters $\theta$ act as a generator for suggesting new alloy compositions based on the learned latent $z$ representation. Finally, the discriminator $D(z)$ evaluates the distinction between real and generated compositions.

## 2.1 Residual Ensemble Learning

Identifying alloy composition alone is usually insufficient to predict real-world material properties, including yield strength. In general, the composition can be used to predict thermodynamic equilibrium phase states and certain intrinsic features such as lattice structure or Young's modulus, parameters that are needed for a predictive model for properties like yield strength. However, post-fabrication processing parameters play a significant role in modulating the yield strength of FCC HEAs. For example, the degree of deformation during cold or hot rolling determines the character, density and topological distribution of dislocations and other defects, all metrics that highly affect yield strength, which is essentially a dislocation-based measure that states how strongly an alloy withstands inelastic yielding (*35, 36*). The annealing temperature and duration are other critical factors, as they dictate the extent of recovery, recrystallization and phase transformation, with similar implications for the inherited defect density and distribution which in turn determine the yield strength (*35*). Thus, data comprising chemical compositional information and post-fabrication processing parameters are both equally considered in establishing the initial descriptor space (DS1) (see Methods in Supplementary Information for more details).

Several ML algorithms were trained on DS1 across 10 random seeds and 5-fold cross-validation and the hyperparameters were optimized through Bayesian optimization (see Methods in Supplementary Information for more details). Figures SF4 (Supplementary Information) present parity plots that compare the predicted and experimental yield strength of FCC HEAs for different base regressors. The performance metrics indicate that the Random Forest (RF) and Gradient Boosting (GB) regressors have improved performance in correlating alloy composition and processing parameters to yield strength. The RF regressor achieved an $R^2$ score of 0.755, while the GB regressor achieved an $R^2$ score of 0.796 on the test dataset. Despite the relatively strong performance of these models, the predictions remain somewhat moderate. Notably, the Neural Network (NN) model showed sub-optimal performance. While NNs are potent ML models capable of capturing the intricate relationships among input features, they require a substantial volume of data to perform optimally. In the presence of limited data points, such models struggle to establish these complex associations, resulting in a lower $R^2$ score. Figures SF5 and SF6 (Supplementary Information) provide the residual distribution for the training and test datasets. Residual analysis is essential for assessing the overall performance of an ML model. While metrics like the coefficient of determination ($R^2$) and root mean square error (RMSE) (see Methods in Supplementary Information for more details) provide a general measure of the model's performance, the residual analysis offers deeper insights into its local performance. An accurate ML model should exhibit a residual distribution with a mean as close to zero as possible. The residual analysis plots reveal that both the accuracy and the error distribution are more favorable for the RF and GB models compared to other models. This trend is also observed in the test dataset, where the GB model exhibits a much more contracted residual distribution even compared to the RF model. Conversely, the NN model shows greater variance in residual distribution, indicating less reliable performance.

Building on these observations, we developed the Residual Hybrid Learning Model (RELM) based on the residual learning strategy (see Methods in Supplementary Information for more details), using the RF and GB models as base learners. A hybrid approach like RELM leverages the strengths of different modeling techniques. Random Forest (RF) offers robustness and generalization through bagging, while Gradient Boosting (GB) focuses on minimizing residual errors through boosting. By combining these techniques, the hybrid model gains the stability of RF and the precision of GB. This synergy allows the model to effectively capture the complex interactions between alloy composition and processing parameters, utilizing RF's capacity to handle diverse features and GB's ability to fine-tune predictions (*37*). This synergy can result in a more accurate and interpretable model for predicting yield strength, providing deeper insights into how different factors contribute to material properties. The performance of RELM on the training and test datasets, summarized over 10 random seeds, is illustrated in Figure SF2(a). Compared to the individual RF and GB models, the RELM predictions are more tightly clustered around the diagonal line, indicating more accurate predictions, consistent with a high $R^2$ value of 0.848 on the test dataset. Additionally, the residual analysis shows a more contracted distribution compared to the base RF and GB models.

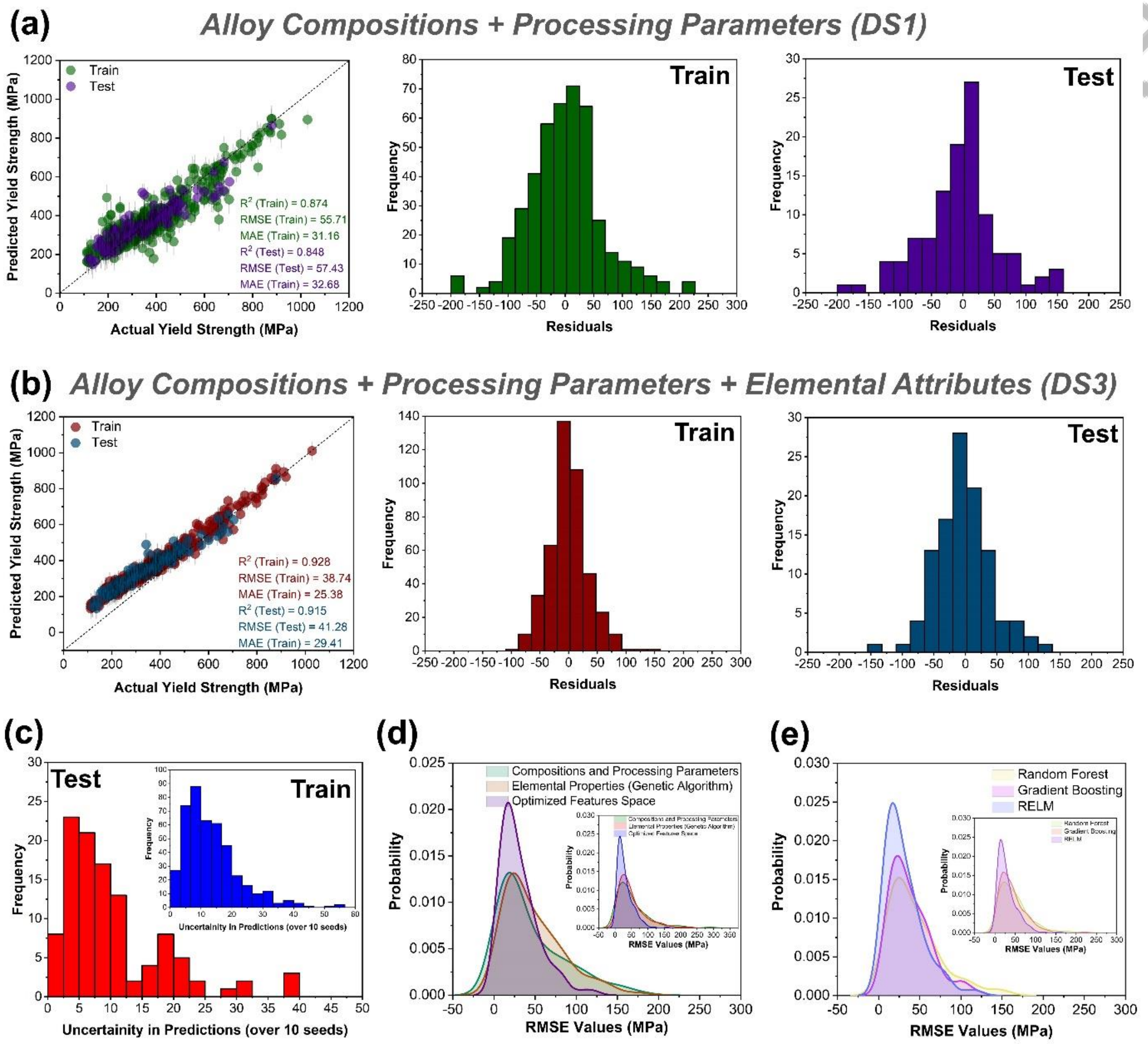

**Figure 2** – Performance of RELM framework. (a) parity plot for RELM with compositions and processing parameters as features; (b) parity plot for physics descriptor enhanced RELM. The model is trained on 10 seeds and the metrics are averaged (10 different training and test datasets). The error bars indicate the uncertainty (standard deviation in prediction across 10 seeds) in the prediction values. The adjacent figures show the distribution of the residuals (prediction errors) for the training and test dataset; (c) distribution of uncertainties associated with the predictions by physics descriptor enhanced RELM framework for test and training (inset) datasets; (d) kernel density estimate (KDE) plots for the probability distribution of root mean square error (RMSE) values for different descriptor spaces based on the physics descriptor enhanced RELM framework (inset provides the KDE distribution for the training dataset); (f) KDE plots for the probability distribution of RMSE values for individual random forest and gradient boosting models as well as the physics descriptor enhanced RELM framework (inset provides the KDE distribution for the training dataset).

### 2.2 Physics descriptors enhanced RELM

Although compositional chemistry and processing parameters appear effective in mapping the yield strength of FCC HEAs with reasonable accuracy, the predictive performance can be further enhanced by incorporating material attributes that directly influence mechanical properties, particularly yield strength. This additional information helps to refine the predictive model, leading to greater accuracy in mapping the yield strength. Accordingly, 16 relevant material attributes were selected (as listed in Table ST3, Supplementary Information) for this purpose. Scatter plots (Figure SF8, Supplementary Information) and hexagonal binning plots (Figure SF9, Supplementary Information) were used to visualize the relationship between yield strength and the various material attributes. These visualizations show that most data points are unique and widely distributed, though some features are significantly correlated and therefore redundant. Additionally, the hexagonal binning plots indicate that no single feature strongly correlates with yield strength, highlighting that in the physics descriptor subspace (DS2), multiple features must be used to map HEA yield strength with reasonable accuracy.

Further, outliers, *i.e.*, the points that do not constitute the overall dataset population, skew statistical measures like mean and variance, with possible loss in the performance of ML models. Elimination or appropriately handling outliers ensures that models are trained on representative data, thereby maintaining the integrity and reliability of predictions (*38*). This is particularly important in the context of HEAs, where the compositions and processing parameters can vary widely. However, removing outliers from an already sparse dataset can lead to a loss of valuable scientific information for the model to interpret as well as chances of overfitting. Therefore, the Yeo-Johnson transformation (*39*), an extension of the Box-Cox transformation (*40*), was used to stabilize variance and normalize the data, making it more suitable for modeling. Unlike other transformations that may change the sign or order of data values, the Yeo-Johnson transformation preserves the data's interpretability. It stabilizes the variance and makes data more normally distributed. The value of the transformation parameter ($\lambda$) for the transformation was estimated using maximum likelihood estimation (MLE) to best normalize the data (see Methods in Supplementary Information for more details). Figure SF11 (Supplementary Information) presents box plots for the transformed dataset, showing a reduction in the number of outliers across several features. The remaining outliers, even after transformation, were removed, reducing the dataset size to 532 data points.

Effective feature selection is another prerequisite to applying these attributes in ML models to cut down computational time and minimize the risk of overfitting (*41*). Given the varied sample distributions across different feature datasets, multiple feature selection methods were tested upon to identify the optimal feature subset for modeling. Details of these feature selection strategies and the results are provided in Methods (Supplementary Information). To identify the best feature subset, all nine feature subsets, derived from various feature selection methods, were used to train the base regressors, with the results summarized in Figure SF17(a-c) and Table ST5 (Supplementary Information). The figures and metrics clearly show that the feature subset selected using the Genetic Algorithm (GA) yielded the best performance across all models. The superior performance of GA-based feature selection deserves further attention. GA excels by systematically exploring a wide range of feature combinations, and identifying optimal or near-optimal subsets. Unlike methods that rely on heuristics or assumptions, GA iteratively refines candidate feature subsets through crossover and mutation, adapting to complex, non-linear relationships between features and the target variable. Moreover, GA naturally accounts for feature interactions and dependencies, a crucial advantage when dealing with real-world datasets, where methods like Correlation Analysis or Feature Importance might miss these intricate relationships (*42*).

After optimizing the material attributes feature space (DS2), the final feature subset DS2, including mixing entropy ($\Delta S_{mix}$), molar volume ($V_m$), elastic modulus ($E$), number of itinerant electrons ($e/a$), Pauling's electronegativity ($\chi_P$), modulus mismatch ($\eta$) and cohesive energy ($E_C$), was concatenated with DS1. This resulted in the final feature space (DS3), encompassing composition, processing parameters, and material attributes. The RELM model was then trained using this extended feature space, with performance results shown in Figure 2(b). The physics descriptor enhanced RELM trained on DS3 outperformed the prior RELM, with an overall improvement of 13.8% in the $R^2$ score. The residual distribution for the physics descriptor enhanced RELM model indicates that residuals are highly concentrated near zero, signifying enhanced performance. The uncertainty was quantified by calculating RMSE (or standard deviations) for RELM predictions across 10 random seeds. Figure 2(c) quantifies the uncertainty arising from model predictions (epistemic uncertainty), revealing that these uncertainties are predominantly skewed towards smaller values. Figure 2(d) illustrates the distribution of RMSE for RELM predictions across the three feature spaces (DS1, DS2, and DS3). The performance of RELM shows significant improvement across the distribution with the integration of physical descriptors into the compositional and processing datasets. This enhancement is attributed to the more accurate data representation afforded by the inclusion of physical descriptors, enabling the RELM to better interpret and model the underlying relationships. Figure 2(e) compares the RMSE distribution for the physics descriptor-enhanced RELM with those of individual RF and GB base regressors. The RELM framework demonstrates superior performance, as evidenced by the relatively narrower spread of RMSE values. This observation aligns with the broader understanding that ensemble models often outperform individual models while also providing more reliable estimates of prediction-specific uncertainty (*43*, *44*). These observations suggest that physics descriptor enhanced RELM is not only accurate but also exhibits high generalization capability across different datasets.

## 2.3 Scientifically Interpretable RELM and composition-processing-property design spaces

The statistical performance of ML models critically depends on selecting the right features, as demonstrated in the preceding discussion. However, it is equally important that these features not only optimize model accuracy but also provide meaningful scientific insights into the underlying phenomena and develop composition-processing-property correlations. To elucidate the relationships between input features and yield strength, we first employed the Shapley Additive Explanations (SHAP) algorithm to interpret the predictions made by the RELM. The SHAP algorithm, rooted in game theory, quantifies the contribution of each feature to the model's output by calculating SHAP values, which traditionally measure the impact of each player in a cooperative game (*30*).

Figure 3(a) shows the mean SHAP values for the selected features on the test dataset, while Figure 3(b) provides a violin plot illustrating the impact of each feature on yield strength predictions. The abscissa represents the SHAP values, with positive or negative values indicating an increase or decrease in yield strength, respectively. The features are listed along the vertical axis, and each marker corresponds to a data point. The color of each point reflects the magnitude of the feature value, with red indicating higher values and blue indicating lower values. The horizontal spread of SHAP values for each feature indicates its influence on the prediction outcome; the wider the spread, the greater the feature's importance. From this analysis, it is clear that five features — annealing temperature (T), annealing time (t), manganese concentration (% Mn), elastic modulus mismatch ($\eta$), and percentage of plastic bulk deformation (% deformation) — exhibit the widest relative coverage of SHAP values, underscoring their paramount importance in predicting yield strength. This highlights a significant shift from earlier studies, which predominantly depended on thermodynamic and structural parameters alone, without adequately considering post-fabrication processing conditions. Our analysis clearly demonstrates the critical role of these parameters in the accurate estimation of the yield strength of HEAs using ML.

While SHAP values provide a global view of feature importance and the directional impact of each feature on yield strength (Figure 3(c)), they do not fully reveal the qualitative relationships between specific features and yield strength. Therefore, partial dependence plots were applied here for further clarification of these interrelations. While partial dependencies typically analyze one feature at a time, we extended our analysis to 3-dimensional partial pair dependencies (PPDs) (Figure 3(d)) to gain a more comprehensive understanding of the underlying composition-processing-property correlations in FCC HEAs.

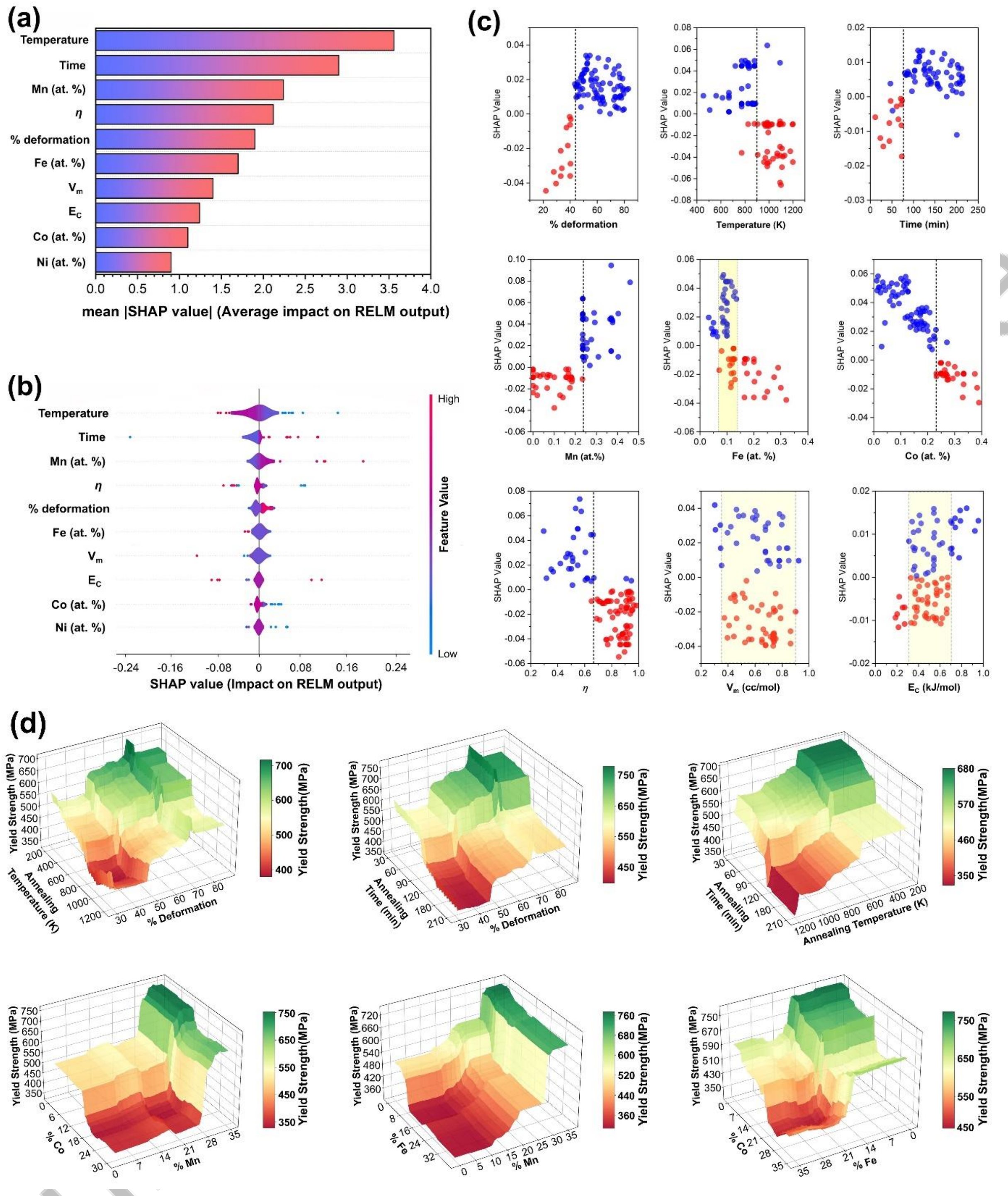

**Figure 3 –** Analysis of influencing factors on the FCC HEA yield strength using Shapley Additive Explanations (SHAP). (a) Ranked mean absolute value of SHAP values and (b) the SHAP summary plot of the 10 most important features for yield strength; (c) the SHAP values distribution (negative in red, positive in blue) of the sets of three most important processing parameters, alloy constituents and elemental properties for each datapoint the test dataset; (d) Partial pair dependence plots for the combined effects of the three processing parameters as well as the three most influential alloy constituents on the yield strength of FCC HEAs.

The composition and selection of alloy constituents play a crucial role not only in phase evolution and microstructural design but also in price and sustainability. The results obtained from the PPD analysis Figure 3(d) indicate that manganese (Mn), iron (Fe), and cobalt (Co) are particularly influential in determining the yield strength of FCC HEAs. Mn and Co have a direct positive correlation with yield strength, while an increase in Fe content correlates with a decrease in yield strength. In the context of Cantor alloys and FCC HEAs containing Mn, Co, and Fe, Mn and Co contribute to solid solution strengthening by introducing alloying elements within the crystal lattice. This leads to increased lattice distortion, thereby elevating the frictional stress that impedes dislocation motion, ultimately enhancing yield strength. Moreover, Mn and Co can facilitate the formation of secondary phases, such as Laves phases, which further reinforce the alloy by obstructing dislocation and grain boundary movement (*45*, *46*). Conversely, the addition of Fe tends to reduce yield strength, primarily due to its tendency to limit the incorporation of elements like Co, which are essential for solid solution strengthening. Fe is less effective than Co or Mn in promoting solid solution strengthening in CoCrFeMnNi based HEAs because it induces less lattice distortion and exerts a smaller lattice friction against dislocation motion (*47*).

Processing parameters also play a pivotal role in determining the mechanical properties of FCC HEAs. Yield strength is conventionally related to grain size through the Hall-Petch equation, with smaller grain sizes typically enhancing yield strength with an inverse square root relationship in conventional constitutive models (*48*). The annealing temperature ($T$) profoundly influences yield strength by driving recovery and recrystallization processes that reduce the initial dislocation density, a parameter that affects yield strength with a square root relation in mean field models. As annealing temperature increases, the reduction in dislocation density leads to lower yield strength, a relationship accurately captured by the RELM. However, if the temperature is insufficiently high, full recrystallization may be incomplete, resulting in higher yield strength due to the prevalence of some of the deformed regions carrying a high inherited dislocation density. Annealing time (*t*) complements annealing temperature, as prolonged annealing promotes diffusion and phase transformations with a square root relation in mean field approximations. Extended annealing can enlarge grains and reduce dislocation density, ultimately lowering yield strength. However, excessive annealing can cause over-annealing, leading to significant grain growth and further reduction in yield strength. Conversely, shorter annealing times and lower temperatures can enhance yield strength by fostering the formation of short-range ordered (SRO) atomic configurations, which provide additional resistance to dislocation motion and act as a strengthening mechanism in multicomponent alloys (*49*, *50*). The percentage of deformation (or dimensional reduction) during rolling (% deformation) is another critical factor influencing yield strength. Deformation introduces dislocations and defects into the crystal lattice, thus enhancing yield strength via dislocation-dislocation interaction. However, excessive deformation may lead to dislocation recovery, potentially reducing yield strength. Dislocation motion in HEAs is inherently more difficult than in pure metals or conventional alloys due to the severe lattice distortions caused by interactions among the multiple solutes, causing much higher friction stress than in conventional materials. Increasing the number of alloying elements exacerbates lattice distortion, due to differences in atomic sizes and shear moduli among the constituent atoms (*51*). This severe lattice distortion imparts a unique structure to HEAs, where each atom behaves as a solute, distinguishing HEAs from traditional alloys.

These insights gained from SHAP and PPD analyses underscore the importance of considering key mechanical and processing parameters in the design of new HEAs tailored for specific mechanical properties. The RELM's ability to interpret materials physics through these analyses demonstrates its potential to drive physics-informed alloy development, advancing our understanding of composition-processing-property relationships and enabling the rational design of next-generation HEAs.

### 2.4 Generative Alloy Design

To thoroughly evaluate the performance and scientific implications of the RELM framework, it is necessary to test the model on previously unknown alloy systems. Four alloy systems were chosen as prospective candidates for this evaluation. The primary alloy system, CoCrFeMnNi, was chosen first to identify a new alloy composition with superior yield strength compared to existing alloys within this material family. Subsequently, two additional alloying elements, Al and Si, were incorporated to form the other three alloy sub-systems under investigation: CoCrFeMnNiAl, CoCrFeMnNiSi, and CoCrFeMnNiAlSi. The selection of the elements Al and Si was deliberate, aimed at evaluating both the most representative and the most skewed distributions among the alloying elements in the input dataset. Si, being one of the alloy constituents with the most skewed frequency distribution (as shown in Figure SF1, Supplementary Information), was specifically chosen to assess the true potential of the RELM model.

To generate new alloy compositions for these systems, we employed a Generative Adversarial Network-Variational Autoencoder (GAN-VAE) hybrid model (see Methods in Supplementary Information for more details). The GAN-VAE approach integrates the benefits of both GAN and VAE inside a unified framework. GANs excel in producing samples of high fidelity, while VAEs offer a meticulously organized and easily understandable latent space for data manipulation (*52*, *53*). This hybrid approach is more robust than either model alone, with the VAE stabilizing the often challenging training of GANs. Within the GAN-VAE framework, the VAE component maps the compositions to a latent space and then reconstructs them, whilst the GAN component generates novel compositions that are both realistic and diverse. This is especially advantageous when studying HEAs to explore their complex composition space and understand the correlation between composition and yield strength. The structured latent space guarantees credibility, while the GAN enhances these compositions to produce high-fidelity samples. The GAN-VAE model enhances the resilience of AI models and the interpretability of their predictions by generating artificial data that closely resembles real compositions while exploring unexplored regions of the composition space. This capability is especially valuable in HEAs, providing insights into how different elements influence material properties.

Figure 4 shows the two-dimensional GAN-VAE latent space representation of the input dataset, which contains 342 FCC HEAs. Given that the GAN-VAE framework was trained using only the compositional information of the alloys, it is probable that it has successfully captured certain aspects linked to composition that contribute to the significance and physical relevance of the latent space. This study has specifically focused on CoCrFeMnNi-based FCC HEAs. As illustrated in Figure 4, the GAN-VAE effectively differentiates between the various alloy systems, represented by separate islands corresponding to the CoCrFeMnNi, CoCrFeMnNiAl, CoCrFeMnNiSi, and CoCrFeMnNiAlSi alloy systems. Based on this distribution, 5,000 new FCC alloy compositions were sampled for each of the four systems. The trained RELM model was then used to predict the yield strength of these alloys, with the highest predicted yield strength compositions marked by stars in each sub-figure (Figure SF20, Supplementary Information).

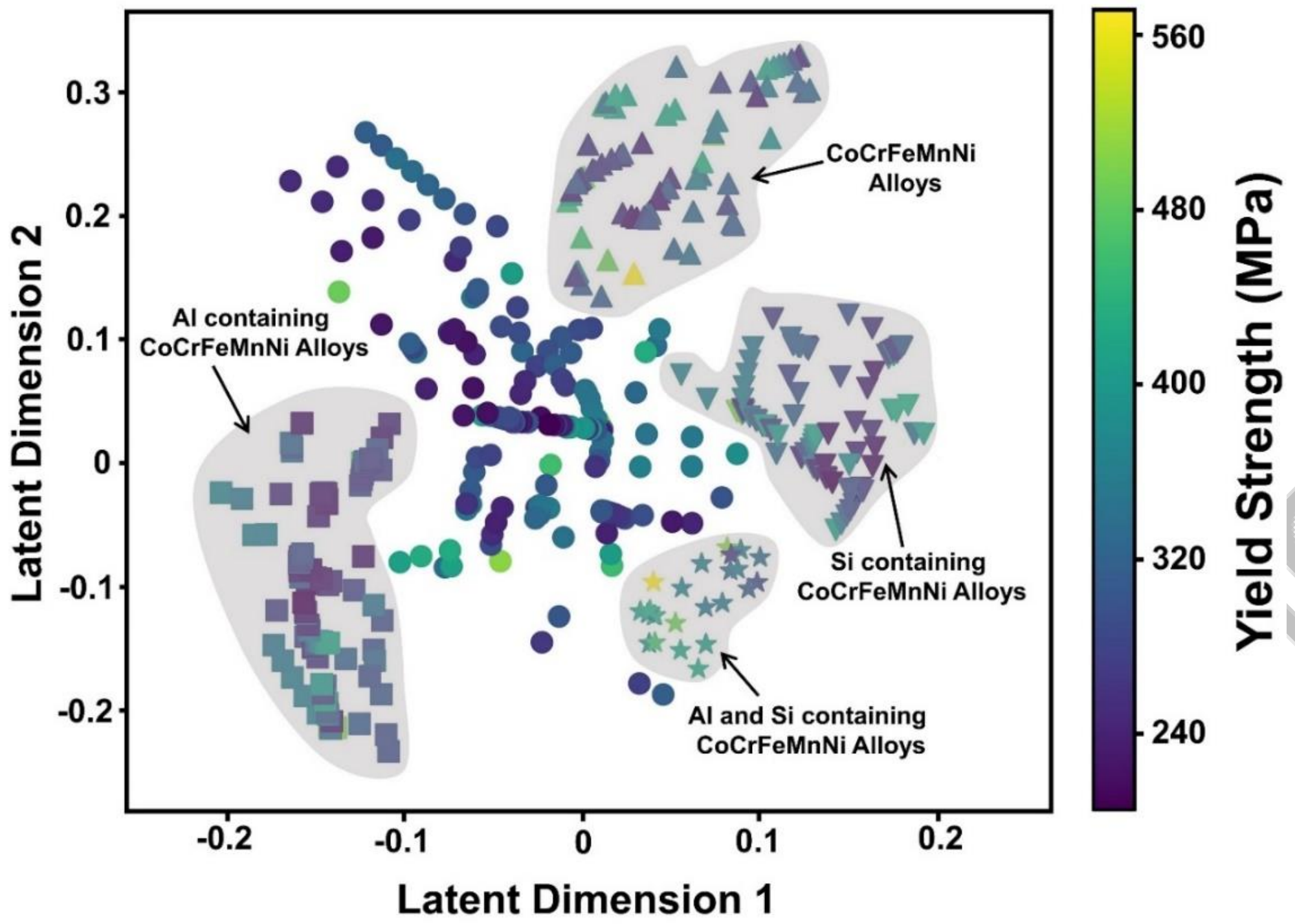

**Figure 4 –** Two-dimensional Generative Adversarial Network-Variational Autoencoder (GAN-VAE) latent space description of the input dataset containing 543 FCC HEAs. Different markers represent the latent distribution of CoCrFeMnNi, CoCrFeMnNiAl, CoCrFeMnNiSi and CoCrFeMnNiAlSi alloys. The circular markers indicate the distribution of the remaining FCC HEAs in the input dataset.

### 2.5 Experimental Validation of the RELM and GAN-VAE Model Predictions

Following the identification of the four alloy compositions discussed earlier, experimental investigations were conducted to validate the predictions and insights derived from the RELM framework. Figure SF22 (Supplementary Information) presents the X-ray diffraction (XRD) patterns for the as-cast and homogenized alloys, confirming that all alloys exhibit a single-phase FCC structure. Notably, a leftward shift of the highest intensity peak was observed with the addition of Al and/or Si, indicating an expansion of the FCC lattice. This shift is more pronounced with the addition of Si compared to Al, which can be attributed to the smaller atomic radius of Si (~111 pm) compared to Al (~143 pm). The resulting greater lattice expansion caused by Si might be further enhanced by its more covalent bonding nature, both effects that induce significant strain within the lattice, resulting in a larger overall increase in the lattice parameter. Further microstructural and compositional analyses were performed using Scanning Electron Microscopy (SEM). Figure SF23 (Supplementary Information) shows Backscattered Electron (BSE) micrographs of the four alloys, revealing a single-phase microstructure. The corresponding Energy Dispersive Spectroscopy (EDS) maps indicate that the alloys are chemically homogeneous following thermal homogenization, further validating the GAN-VAE predictions.

The RELM framework delivers far more than merely the prediction of potential alloys with high yield strength; it also provides detailed guidance on the post-fabrication thermo-mechanical processing conditions required to optimize yield strength. Uniaxial tensile tests were conducted to further validate these predictions on the chosen alloys. Table 1 presents the processing conditions alongside both the predicted and experimentally measured yield strengths for the four alloys. The conditions that produced the maximum yield strength are highlighted in red, demonstrating a strong correlation between the model's predictions and experimental outcomes. The conditions producing the best mix of yield strength and ductility are highlighted in green. To further validate the processing maps generated using the SHAP and PPD analyses, the RELM framework was employed to predict the yield strength of $Co_{20}Cr_{16}Fe_{20}Mn_{16}Ni_{24}Al_4$ and $Co_{24}Cr_{12}Fe_{12}Mn_{16}Ni_{28}Al_4Si_4$ alloys under varying post-fabrication processing parameters. Figure 5(a) illustrates heatmaps showing the RELM-predicted yield strength as a function of changing deformation percentage during cold rolling (PD), annealing temperature (T), and annealing time (t). The contour heatmaps reveal good agreement with the conclusion reached from PPD analysis regarding the combined effects of post-fabrication processing parameters on yield strength. Further experiments have been carried out to confirm these findings through systematic variation of only one parameter at a time, shown in Table 1 and also highlighted by white markers in Figure 5(a).

**Table 1** – Values of predicted yield strength and experimental yield strength as well as % ductility. The values in red indicate the best combinations of annealing temperature, % deformation and annealing time for $Co_{24}Cr_{12}Fe_{16}Mn_{16}Ni_{32}$, $Co_{20}Cr_{16}Fe_{12}Mn_{20}Ni_{28}Si_4$, $Co_{20}Cr_{16}Fe_{20}Mn_{16}Ni_{24}Al_4$ and $Co_{24}Cr_{12}Fe_{12}Mn_{16}Ni_{28}Al_4Si_4$ alloys to obtain highest yield strength. Detailed analysis of the effect of annealing temperature, % deformation and annealing time on the yield strength of $Co_{20}Cr_{16}Fe_{20}Mn_{16}Ni_{24}Al_4$ and $Co_{24}Cr_{12}Fe_{12}Mn_{16}Ni_{28}Al_4Si_4$ alloys is shown and the values in green indicate the best combinations of annealing temperature, % deformation and annealing time for best combination of yield strength and % ductility for $Co_{20}Cr_{16}Fe_{20}Mn_{16}Ni_{24}Al_4$ and $Co_{24}Cr_{12}Fe_{12}Mn_{16}Ni_{28}Al_4Si_4$ alloys.

| **$Co_{24}Cr_{12}Fe_{16}Mn_{16}Ni_{32}$** | | | | | | |
|---|---|---|---|---|---|---|
| **% Deformation** | **Annealing Temperature (K)** | **Annealing Time (min)** | **Predicted Yield Strength (MPa)** | **Experimental Yield Strength (MPa)** | **% Error** | **Experimental % Ductility** |
| 60 | 973 | 60 | 690 | 776 | +11.12 | 14 |
| **$Co_{20}Cr_{16}Fe_{12}Mn_{20}Ni_{28}Si_4$** | | | | | | |
| **% Deformation** | **Annealing Temperature (K)** | **Annealing Time (min)** | **Predicted Yield Strength (MPa)** | **Experimental Yield Strength (MPa)** | **% Error** | **Experimental % Ductility** |
| 50 | 1073 | 90 | 883 | 821 | -7.83 | 12 |
| **$Co_{20}Cr_{16}Fe_{20}Mn_{16}Ni_{24}Al_4$** | | | | | | |
| **% Deformation** | **Annealing Temperature (K)** | **Annealing Time (min)** | **Predicted Yield Strength (MPa)** | **Experimental Yield Strength (MPa)** | **% Error** | **Experimental % Ductility** |
| 60 | 673 | 90 | 778 | 842 | +7.79 | 16 |
| 60 | 873 | 90 | 938 | 838 | -12.04 | 21 |
| 60 | 1073 | 90 | 645 | 669 | +3.32 | 29 |
| 60 | 1273 | 90 | 295 | 274 | -7.63 | 58 |
| 30 | 873 | 90 | 341 | 384 | +11.1 | 31 |
| 50 | 873 | 90 | 515 | 469 | -9.75 | 22 |
| 80 | 873 | 90 | 510 | 575 | +11.3 | 26 |
| 60 | 873 | 30 | 624 | 565 | -10.5 | 18 |
| 60 | 873 | 60 | 543 | 582 | +6.65 | 16 |
| 60 | 873 | 120 | 438 | 472 | +7.06 | 26 |
| **$Co_{24}Cr_{12}Fe_{12}Mn_{16}Ni_{28}Al_4Si_4$** | | | | | | |
| **% Deformation** | **Annealing Temperature (K)** | **Annealing Time (min)** | **Predicted Yield Strength (MPa)** | **Experimental Yield Strength (MPa)** | **% Error** | **Experimental % Ductility** |
| 50 | 773 | 90 | 854 | 937 | +8.83 | 12 |
| 50 | 973 | 90 | 808 | 766 | -5.48 | 18 |
| 50 | 1073 | 90 | 528 | 593 | +10.97 | 30 |
| 50 | 1273 | 90 | 308 | 341 | +9.72 | 53 |
| 30 | 973 | 90 | 439 | 492 | +10.67 | 31 |
| 50 | 973 | 90 | 701 | 632 | -5.56 | 28 |
| 70 | 973 | 90 | 627 | 574 | -9.27 | 34 |
| 90 | 973 | 90 | 568 | 538 | -10.9 | 29 |
| 50 | 973 | 30 | 586 | 617 | +4.92 | 22 |
| 50 | 973 | 60 | 608 | 642 | +6.86 | 19 |
| 50 | 973 | 90 | 545 | 588 | +10.7 | 21 |
| 50 | 973 | 120 | 498 | 549 | +9.77 | 29 |

Figure SF24 (Supplementary Information) shows the engineering stress-strain curves for the two alloys, $Co_{20}Cr_{16}Fe_{20}Mn_{16}Ni_{24}Al_4$ and $Co_{24}Cr_{12}Fe_{12}Mn_{16}Ni_{28}Al_4Si_4$, under different post-fabrication processing conditions. The resulting yield strength values, evaluated at 0.2% proof stress, are summarized in Table 1 along with the relative errors in the predictions made by the RELM framework. The results show that the RELM model performs well, even for the Si-containing alloy ($Co_{24}Cr_{12}Fe_{12}Mn_{16}Ni_{28}Al_4Si_4$) (with an average error of 8.63%), despite Si being a less frequent alloying element in the input dataset. This highlights the robustness of the RELM framework in capturing composition-processing-property correlations, even in cases where data is sparse. Additionally, the maximum achievable yield strength of $Co_{24}Cr_{12}Fe_{12}Mn_{16}Ni_{28}Al_4Si_4$ surpasses that of $Co_{20}Cr_{16}Fe_{20}Mn_{16}Ni_{24}Al_4$, although $Co_{20}Cr_{16}Fe_{20}Mn_{16}Ni_{24}Al_4$ offers a better balance of strength and ductility (Figure SF25, Supplementary Information).

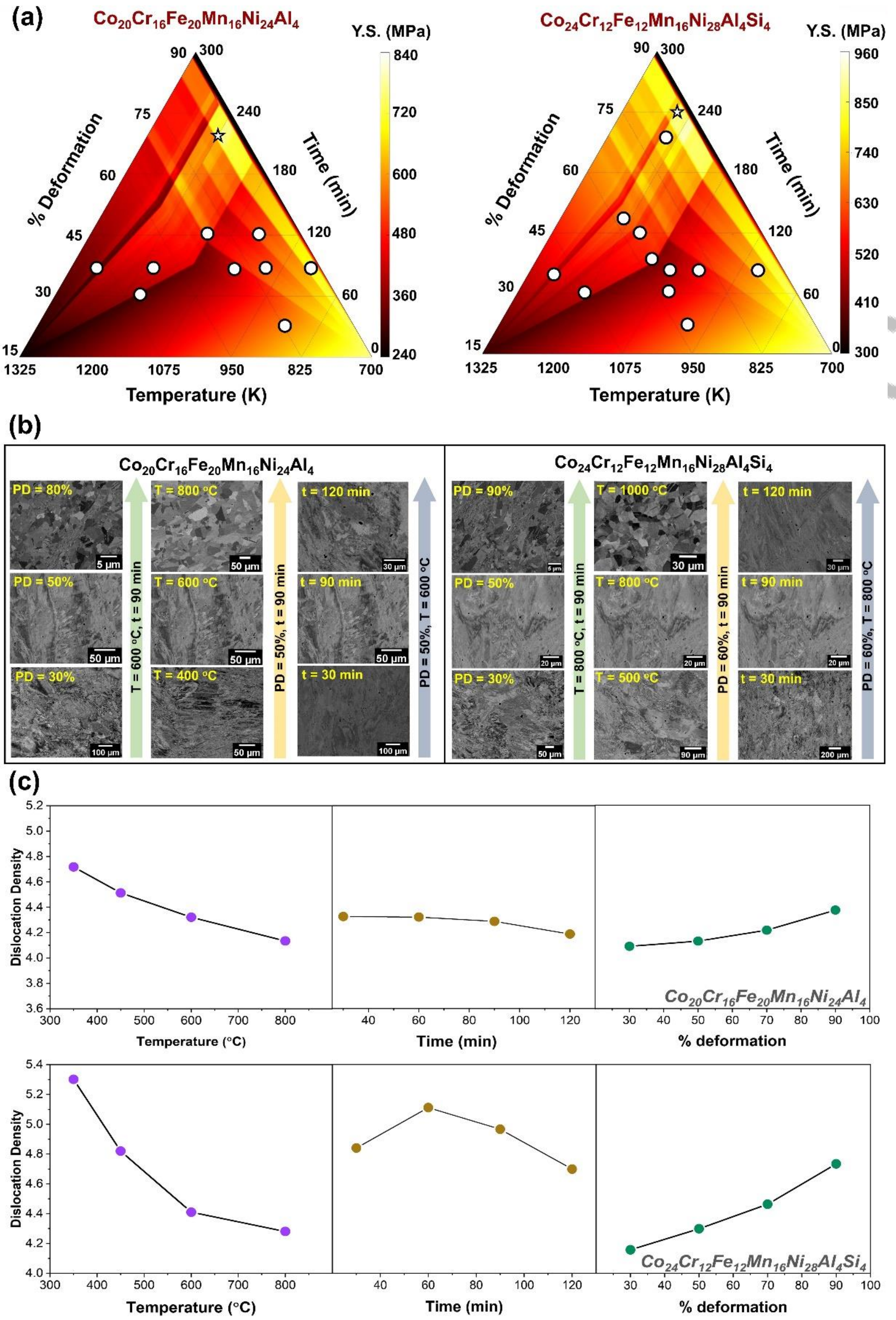

**Figure 5** – (a) Distribution of yield strength values for $Co_{20}Cr_{16}Fe_{20}Mn_{16}Ni_{24}Al_4$ and $Co_{24}Cr_{12}Fe_{12}Mn_{16}Ni_{28}Al_4Si_4$ alloys, as a function of annealing temperature, % deformation and annealing time, predicted by the RELM framework. The white markers indicate the combination of points chosen for experimental validation. The star marker indicates the combinations of processing parameters that exhibit the highest yield strength for both alloys; (b) Variation of the microstructure evolution and; (c) Variation of dislocation density as a function of annealing temperature, % deformation and annealing time for $Co_{20}Cr_{16}Fe_{20}Mn_{16}Ni_{24}Al_4$ and $Co_{24}Cr_{12}Fe_{12}Mn_{16}Ni_{28}Al_4Si_4$ alloys.

Detailed SEM investigations were performed to understand the effects of Al and Si, as well as the processing parameters, on the microstructural evolution in $Co_{20}Cr_{16}Fe_{20}Mn_{16}Ni_{24}Al_4$ and $Co_{24}Cr_{12}Fe_{12}Mn_{16}Ni_{28}Al_4Si_4$ HEAs. Figure 5(b) presents BSE micrographs of the two alloys under various thermo-mechanical processing conditions. In both alloys, increasing annealing temperature leads to more pronounced recrystallization, resulting in decreased yield strength. It may be clearly seen from the micrographs that an increase in % of deformation results in partial recrystallization, even at sub-recrystallization temperatures. High deformation rates introduce significant stored energy in the form of dislocations and other crystalline defects, which drive recrystallization. This is indeed observed during deformation where new grains are nucleated and grow as the material continues to get deformed. The extent of partial recrystallization is less in $Co_{24}Cr_{12}Fe_{12}Mn_{16}Ni_{28}Al_4Si_4$ compared to $Co_{20}Cr_{16}Fe_{20}Mn_{16}Ni_{24}Al_4$, likely due to the increased atomic size mismatch and lattice distortions introduced by Si. These factors lead to higher internal stress fields, making the alloy more resistant to recovery and recrystallization, thus explaining the higher yield strength observed under most processing conditions. However, the effects of % deformation and annealing time, particularly at temperatures below the recrystallization threshold, are more subtle and challenging to resolve through SEM alone.

To gain further insights into the effects of post-fabrication thermo-mechanical processing parameters on yield strength, X-ray diffraction analysis was performed on the processed alloys. Figure SF26 (Supplementary Information) illustrates the phase evolution in $Co_{20}Cr_{16}Fe_{20}Mn_{16}Ni_{24}Al_4$ and $Co_{24}Cr_{12}Fe_{12}Mn_{16}Ni_{28}Al_4Si_4$ alloys under varying annealing temperatures, % deformation, and annealing times. Additionally, the dislocation density was evaluated using XRD analysis. The relationship between dislocation density and yield strength is based on the mechanisms of dislocation movement and dislocation-dislocation interactions *via* core reactions and elastic forces within the material. Dislocations interact with each other and with other microstructural features, such as grain boundaries, precipitates, and solute atoms, impeding dislocation motion and thus requiring higher applied stresses to continue deformation. This relationship is often described by the Taylor hardening model (*54*), expressed for HEAs as:

$$\sigma_y = \sigma_o + M\alpha Gb\sqrt{\rho} \qquad (1)$$

where, $\sigma_o$ is the friction stress, which represents the intrinsic strength of the material without the contribution of dislocations, $M$ is the Taylor factor, which accounts for the crystallographic orientation of the grains, $\alpha$ is a constant (typically between 0.2 and 0.5) that describes the strength of dislocation interactions, $G$ is the shear modulus of the material and $b$ is the Burgers vector, representing the magnitude and direction of the lattice distortion caused by dislocation. Figure 5(c) presents the dislocation density values (evaluated from XRD) for different processing conditions for $Co_{20}Cr_{16}Fe_{20}Mn_{16}Ni_{24}Al_4$ and $Co_{24}Cr_{12}Fe_{12}Mn_{16}Ni_{28}Al_4Si_4$. The results indicate that the overall dislocation density is higher for $Co_{24}Cr_{12}Fe_{12}Mn_{16}Ni_{28}Al_4Si_4$ than for $Co_{20}Cr_{16}Fe_{20}Mn_{16}Ni_{24}Al_4$, correlating with the higher yield strength values observed for most processing conditions. In particular, the trend in dislocation density closely emulates the variations in yield strength observed through SHAP and PPD analyses, further confirming the material physics-aware nature of the RELM framework. While the microstructure showed no apparent changes with varying annealing time (below recrystallization temperatures), the dislocation density initially increased and then decreased as annealing time progressed. During the early stages of annealing, dislocations accumulate, resulting in increased dislocation density. As annealing continues, recovery mechanisms, such as dislocation climb of edge dislocations and double cross-slip of screw dislocations, come into play, allowing them to mutually cancel out, thus reducing the overall dislocation density. Since annealing was conducted in the recovery region below the recrystallization temperature, recrystallization did not influence dislocation density changes (*49, 50*). In a broader temperature range, as illustrated in the deformation maps generated by PPD analysis, prolonged annealing at higher temperatures leads to a more significant decrease in yield strength (and dislocation density) due to recrystallization, where new grains form free of dislocations, resulting in a softer material. The experimental results strongly validate the effectiveness of the RELM framework, demonstrating not only its accuracy in predicting yield strength but also its capability to provide deep scientific insights into the deformation dynamics of FCC HEAs.

## 3. Conclusions

We present a comprehensive framework for predicting and optimizing the yield strength of FCC HEAs using advanced ML techniques, specifically a novel Residual Hybrid Learning Model (RELM). The RELM framework, bolstered by the integration of an innovative GAN-VAE hybrid generative alloy design approach, demonstrates a robust capability to predict the mechanical properties of HEAs with high accuracy, even when dealing with limited or skewed data distributions. More specifically, the study leads to the discovery of several new alloy variants in the CoCrFeMnNi alloy system, namely $Co_{24}Cr_{12}Fe_{16}Mn_{16}Ni_{32}$, $Co_{20}Cr_{16}Fe_{12}Mn_{20}Ni_{28}Si_{4}$, $Co_{20}Cr_{16}Fe_{20}Mn_{16}Ni_{24}Al_{4}$ and $Co_{24}Cr_{12}Fe_{12}Mn_{16}Ni_{28}Al_{4}Si_{4}$, showing higher yield strength than those alloys reported for these alloy sub-systems. The combination of compositional and post-fabrication processing parameters is shown to be crucial in accurately mapping the yield strength of FCC HEAs. This emphasizes that predicting new metallic alloys based on lattice structure data alone is not sensible, and consideration of processing that equips alloys with characteristic microstructures that lend the materials their real-world mechanical properties is essential. Further, the RELM framework not only identifies potential alloy compositions with superior yield strength but also provides valuable insights into the underlying mechanisms influencing these properties. This was further validated through detailed experimental investigations, including X-ray diffraction, SEM analysis, and tensile testing, confirming the model's predictions and making them interpretable. The experiments demonstrate that the RELM framework effectively captures the complex composition-processing-property relationships in HEAs, particularly in alloys containing elements with less frequent distributions in the dataset, such as Si. A significant contribution of this work is the enhanced interpretability achieved through the use of SHAP and partial pair dependencies. These provide a clear understanding of how alloying elements and processing parameters, influence yield strength. The findings highlight the importance of including mechanical and post-fabrication parameters in machine learning models for materials design, shifting the focus from using empirical thermodynamic and structural parameters alone to a more holistic approach that considers the entire processing history of the alloy. The interpretable RELM framework, in conclusion, bridges the gap between data-driven models and fundamental materials science, paving the way for the rational design of next-generation HEAs.

## ACKNOWLEDGMENTS

We thank M. Adamek, D. Klapproth, J. Wichert and F. Schlüter from Max-Planck-Institute for Sustainable Materials for their assistance in the experimental work.

**Funding:** A.B. was supported by Max Planck Gesellschaft (Max Planck Society) fellowship. Z.R. was supported by Deutsche Forschungsgemeinschaft (DFG, German Research Foundation project ID 405553726-TRR 270).

**Author contributions:** A.B. conceived the study, designed the residual ensemble learning framework, performed the experiments, analyzed the results, produced the final figures and wrote the manuscript. A.B. and A.D. collected the initial data set. Z.R., K.B. and D.R. supervised the study and reviewed the manuscript. All authors discussed the results and commented on the manuscript.

**Competing interests:** The authors declare no competing interests.

**Data and materials availability:** The training dataset is curated from the previous publications (2–63 in Supplementary Information) and can be found in (*55*). The necessary data produced by our model in this work can be found in the supplementary materials. The original codes used to perform this work is available on GitHub (*55*).

## Supplementary Information

# Interpretable Machine Learning for High-Strength High Entropy Alloy Design

Anurag Bajpai*, Ziyuan Rao, Abhinav Dixit, Krishanu Biswas* and Dierk Raabe*

Corresponding Authors: *Anurag Bajpai*, a.bajpai@mpie.de; *Krishanu Biswas*, kbiswas@iitk.ac.in; *Dierk Raabe*, d.raabe@mpie.de

## This PDF file contains:

1. Materials and Methods
2. Supplementary Figures: SF1 to SF27
3. Supplementary Tables: ST1 to ST6
4. References

## Materials and Methods

The succeeding subsections will explain the various methods used to develop and analyse the RELM modules in detail.

### *Initial data preparation and pre-processing for RELM*

The performance of the ML model is largely dependent on the quantity and quality of the dataset. Among FCC HEAs, the most popular family, CoCrFeNiMn, has attracted a large amount of attention since its discovery by Cantor in 2004 (*1*). This has led to a substantially large number of investigations on this alloy system. Therefore, a dataset of FCC HEAs derived from the CoCrFeNiMn alloy system, comprising of a total of 14 elements, has been curated from the available literature (*2-63*) for training the RELM. The dataset includes 342 alloy compositions, out of which 44 were ternary, 51 were quaternary, 162 were quinary, 75 were senary and 10 were septenary compositions. All the data has been considered for alloys prepared only through vacuum arc melting to eradicate the influence of fabrication technique on yield strength. For alloys whose yield strength was not available in the reported literature, the value of the yield strength has been estimated using the empirical relation $H_V \approx 3\sigma_y$ (*64*), where $H_V$ and $\sigma_y$ indicates Vicker's hardness and yield strength in MPa, respectively. The frequency distributions of the constituent elements and yield strength are provided in Figure SF1 and Figure SF3, respectively. Table ST1 provides the statistical description of the constituent elements and yield strength in the collected dataset.

In addition to chemical compositions, the post-fabrication processing conditions of HEAs play an essential role in microstructure-property correlations in HEAs. Therefore, several processing parameters that are believed to influence the yield strength and are industrially applied; namely deformation percentage during cold rolling ($PD$), annealing temperature ($T$) and annealing time ($t$) were also included in the candidate feature pool. The statistical description of the processing parameters is provided in Figure SF2 and Table ST1. The incorporation of the processing parameters in the dataset resulted in the expansion of the dataset size to 543 datapoints. Since composition and processing parameters comprise the primary feature set (henceforth named DS1), no feature selection was performed at this stage.

### *Residual Ensemble Learning Model (RELM) Construction and Evaluation*

The selection of an ML algorithm to train the model is an equally critical aspect of any statistical model-building exercise. This selection also hinges on the underlying transparency-accuracy trade-offs. While simple models, such as linear regression, are transparent and, therefore, also completely explainable and interpretable, more complex models, such as neural networks (NNs), can have a large set of model parameters, leading to improved fitting performance but little transparency and are often treated as black boxes. On the other hand, tree-based ensemble models, such as random forest (RF) models, provide a balance between the model interpretability and predictive accuracy. Based on the "no free lunch theorem" for ML model selection optimization, which states that no single model can interpret all materials properties, eight different ML algorithms were used, including lasso regression (Lasso), ridge regression (Ridge), support vector regression (SVR), k-nearest neighbor (KNN), random forest (RF), gradient boosting regression (GB), and artificial neural network (Neural Net). The neural network models were implemented in *TensorFlow* using the *Keras* subroutine, while the remaining models were implemented using the *scikit learn* subroutine. All the models were trained using 80% of the data and tested by 20% of the data samples. To take the randomness effect in the dataset into account, the model performance was validated by testing for 10 different seeds (random states of train-test dataset split) and comparing the mean and standard deviation of evaluation metrics on the 10 unseen test sets, leading to more robust learning and predictions.

Extensive hyperparameter tuning was done for each model using Bayesian optimization. Bayesian optimization is a computational method that aims to optimize the performance of an ML model by iteratively selecting the most promising set of parameters. The fundamental concept underlying Bayesian optimization involves constructing a probabilistic model to represent the objective function.

This model captures the association between hyperparameters and the evaluation measure, such as mean squared error. This approach aims to strike a balance between exploration, which involves testing new hyperparameters, and exploitation, which focuses on regions that show promise. The iterative procedure persists until a specified termination condition is satisfied, such as a certain number of iterations or the attainment of a particular level of performance. Bayesian optimization has been found to show more efficiency compared to grid search and random search methods (*65*). The algorithm exhibits intelligent behavior by autonomously selecting hyperparameters for evaluation, minimizing the number of trials required to identify optimal values. In this investigation, Bayesian optimization is performed using the Gaussian process as the surrogate model and 5-fold cross-validation. Cross-validation helps in estimating how well a model will generalize to unseen data. The *k*-fold cross-validation technique involves partitioning the dataset into *k* segments, where *k*-1 segments are utilized for training, and one segment is reserved for prediction. The procedure is iterated *k* times, and the final prediction result is determined by calculating the average value of the evaluation error. This technique guarantees the independence of each training iteration, enhancing the model's generalization performance during hyperparameter tuning and model assessment. The optimization was run for 500 iterations to obtain the most promising set of hyperparameters with the random state set to 0 to ensure the reproducibility of hyperparameter optimization results. The chosen hyperparameter space and the optimized hyperparameters for the best-performing seed across all the chosen base regressors are provided in Table ST2.

After training the 8 chosen ML regressors, the two best-performing models were chosen to create an ensemble framework. However, unlike the conventional ensemble learning approach, which relies on average predictions of two or more models, the present investigation uses the concept of residual ensemble learning where the first model (the second best performing model) is trained on the target variable and the second model (the best-performing model) is trained on the errors (residuals) in the predictions of the first model. This strategy allows for the design of a hybrid bagging-boosting framework. This two-step regression model is referred to as Residual Ensemble Learning Model (RELM).

To assess the predictive accuracy of the model during the stages of model development, testing, and prediction, we constantly employ three distinct metrics: coefficient of determination ($R^2$), root mean squared error (RMSE), and mean absolute error (MAE). The coefficient of determination ($R^2$) measures the proportion of the variance in the dependent variable that is predictable from the independent variable(s). $R^2$ values range from 0 to 1, where 1 indicates that the model perfectly explains the variance, and 0 indicates that the model explains none of the variance. Root Mean Squared Error (RMSE) measures the average magnitude of the prediction errors, giving higher weight to larger errors. RMSE is the square root of the average of squared differences between predicted and actual values. Lower values indicate better model performance. Mean Absolute Error (MAE) measures the average magnitude of errors in a set of predictions, without considering their direction (i.e., positive or negative errors). Like RMSE, lower MAE values indicate better model performance but without penalizing larger errors more heavily. These metrics have been evaluated using the formulae provided through Equations SE1-SE3.

$$R^2 = \frac{\sum_{i=1}^{N}(\hat{x}_i - \bar{x})^2}{\sum_{i=1}^{N}(x_i - \bar{x})^2} \quad \text{(SE1)}$$

$$RMSE = \sqrt{\frac{\sum_{i=1}^{N}(x_i - \hat{x}_i)^2}{N}} \quad \text{(SE2)}$$

$$MAE = \frac{\sum_{i=1}^{N}|\hat{x}_i - x_i|}{N} \quad \text{(SE3)}$$

where, $x_i$ and $\hat{x}_i$ are the variable *x's* true and predicted values, respectively. *N* is the total number of input data points and $\bar{x}$ is the mean of the true values of *x*.

Prior to model training, it is imperative to establish a commonly accepted tenet, namely, that the heterogeneity in the order of magnitudes and range of input vectors, potentially leading to algorithmic bias. To circumvent this issue, it is crucial to ensure that the processing parameters are normalized to a comparable scale using Equation SE4.

$$X_{norm,i} = \frac{X_i - X_{min}}{X_{max} - X_{min}} \quad \text{(SE4)}$$

where $X_{norm,i}$ is the normalized value of $i^{th}$ feature, $X_{min}$ and $X_{max}$ are the minimum and maximum values of $i^{th}$ feature. All data on processing parameters was hence normalized to $[0, 1]$.

### *Feature Space expansion using physics descriptor*

While composition and processing parameters can potentially provide an accurate mapping to the yield strength of HEAs, a simple input dataset like chemical composition can seldom prove to be insufficient. The ML models based on the physical or chemical features that contribute to material mechanistic research and material design are extensively reported in the literature. This, therefore, requires the expansion of the feature space with elemental descriptors relating to the mechanical properties, phase formation, and chemical properties that influence the yield strength of FCC HEAs. A second dataset was constructed consisting of thermodynamical, structural and mechanical parameters that can influence the yield strength of HEAs (henceforth referred to as DS2). These have been summarised in Table ST3, along with their formulae. These elemental attributes are not only efficiently computable and accessible but also unique for each alloy in the composition space. Due to a difference in fundamental atomic mechanics, the kind of phase significantly impacts the mechanical behavior of HEAs. The phase can be quantified using thermodynamic and structural parameters such as the formation enthalpy ($\Delta H_{mix}$), mixing entropy ($\Delta S_{mix}$), atomic size mismatch ($\delta$) and valence electron concentration (VEC) (*66*). Thus, their inclusion in the feature space for a mechanical property estimation problem is logical.

### *Outliers Processing for DS2*

Figure SF10 shows the box plots for the 16 elemental attributes chosen in DS2. The visual representation of the data distribution is depicted by a box that encompasses the interquartile range (IQR), while the whiskers extend to a length of 1.5 times the IQR. Any data point that falls outside of this specified range is classified as an outlier. It is evident that there are several outliers in this dataset. In machine learning, the presence of outliers has the potential to induce overfitting, hence exerting a detrimental impact on the predictive accuracy of the model and leading to poor generalization performance when applied to unseen data. The most common approach to dealing with these outliers is to remove them from the dataset. However, since our dataset is limited to only 357 alloy compositions, it is unwise to remove outliers. This will reduce the dataset size and limit the prediction capability as well as the generalizability of the ML model.

A pragmatic approach to dealing with outliers without reducing the dataset size is to use transformations that reduce the impact of outliers in the dataset. We use the Yeo-Johnson transformation that can indirectly address the influence of outliers. The Yeo-Johnson transformation stabilizes the variance and makes data more normally distributed. By transforming the data to approximate normality, the impact of outliers is diminished, which improves the robustness of statistical models. It involves applying a power function to the dataset, as shown by Equation SE5 (*67*).

$$Y(x) = \begin{cases} ((x+1)^{\lambda} - 1)/\lambda\,, & if\ \lambda \neq 0\ and\ x \geq 0 \\ -\ln(-x+1)\,, & if\ \lambda = 2\ and\ x < 0 \\ -\left((-x+1)^{2-\lambda} - 1\right)/(2-\lambda)\,, & if\ \lambda \neq 2\ and\ x < 0 \\ \ln(x+1)\,, & if\ \lambda = 0\ and\ x \geq 0 \end{cases} \qquad \text{(SE5)}$$

Here, $\lambda$ is the transformation parameter, and $Y(x)$ is the transformed value of $x$. The degree of transformation can be modified by manipulating the parameter $\lambda$. When the value of $\lambda$ is smaller than 1, the transformation assigns a lower weight to outliers, hence diminishing their influence on the overall distribution. The λ parameter has been estimated using maximum likelihood estimation (MLE) to best normalize the data. This adaptive nature allows the transformation to be more effective in handling outliers.

## *Feature Selection from material attributes descriptor space (DS2)*

Although all the attributes provided in Table ST3 hold significant relationships to yield strength from the materials science perspective, they may not be beneficial from the ML model training. Several features may be rendered redundant during model training. Redundancy in the feature space often leads to overfitting or underfitting and, therefore, inaccurate predictions. Simultaneously, reducing the number of features in modeling by feature selection methods can make ML models more physically interpretable and can improve the efficiency of model operation.

Different feature selection techniques were utilized in the present investigation, as no one feature selection method holds itself above others. The selected feature selection techniques can be divided into three categories: (a) feature space-based selection, (b) model and threshold-based selection and (c) threshold-free model-based selection.

***Feature space based selection***. Pearson correlation analysis is a technique that measures the degree of correlation between two features. Pearson's correlation coefficient ($PCC$), determined using Equation SE6, provides the quantitative interrelation between any two features (*68*).

$$PCC_{xy} = \frac{1}{N-1}\frac{\sum_{i=1}^{N}(x_i-\bar{x})-(y_i-\bar{y})}{\sigma_x\sigma_y} \qquad \text{(SE6)}$$

where $N$ are the total number of data, $\bar{x}$ and $\bar{y}$ are the mean values of two input features $x$ and $y$, and $\sigma_x$ and $\sigma_y$ refer to the standard deviation of the two features. The $PCC$ ranges between -1 and 1, with 0 indicating no correlation between the two features. The $PCC$ value of 1 indicates a strong positive correlation, and -1 indicates a strong negative correlation (*69*). $|PCC_{xy}| \geq 0.8$ is taken as the threshold for feature elimination in this study, as highly correlated features indicate mutual redundancy. Among the highly correlated pair of features, the feature with a high correlation to the yield strength was considered, and the other was removed from the feature space (Figure SF13). Although the PCC method can help to evaluate the correlation between individual features, it does not establish a logical link between features and target values.

***Model and threshold-based selection***. This includes techniques that rely on feature importance to ascertain the impact of features on the model performance. These include feature importance, permutation importance and mutual information based feature selection. All these strategies involve training an ML model and assigning scores to individual features, which reflect their respective contributions to the predicted performance of the model. Algorithms such as decision trees or ensemble approaches like Random Forests are frequently employed for the computation of this metric. Features that result in substantial reductions in the error are seen as being of greater significance (Figures SF14(a, b)). The concept of permutation importance, on the other hand, is a statistical measure used to assess the importance of variables in a predictive model. It assesses the importance of features by quantifying the extent to which the performance of the model is diminished when the values of a certain feature are randomly permuted. A notable decrease in performance is indicative of the significance of certain features (Figures SF14(c)). Finally, the mutual information based feature selection involves the utilization of mutual information to quantify the interdependence between two stochastic variables, specifically a feature and the target variable (yield strength here), thereafter opting for the features that exhibit the highest mutual information scores. A threshold has to be subsequently chosen based on the results to reduce the size of the feature space (Figures SF14(d)).

***Threshold free model-based selection***. These include backward elimination, recursive feature elimination, best subset selection and genetic algorithm-based feature selection. The backward elimination strategy is a systematic approach employed in machine learning to iteratively enhance a model by removing the features that have the least significance. The process commences by incorporating all features present in the model. Then, throughout each iteration, the feature with the lowest significance score is identified and then eliminated from the feature set. Subsequently, the model is subjected to retraining using the remaining features, followed by an evaluation of its performance (Figures SF14(e)). The aforementioned procedure is iterated until a predetermined termination criterion is satisfied, such as the attainment of a desired degree of enhancement in performance or the completion of a specified number of iterations.

Recursive Feature Elimination (RFE) (Figures SF14(f)) is another feature selection technique that aims to systematically identify the most important features of a machine learning model. In RFE, the process starts with all features included, and it iteratively removes the least significant feature(s) based on a specified criterion until the desired number of features or a stopping condition is met. During each iteration, the model is trained and evaluated, and feature importance scores or a similar metric are used to determine which feature(s) should be eliminated. RFE is a backward elimination method, but unlike traditional backward elimination, it considers multiple features simultaneously, making it well-suited for identifying feature interactions and complex relationships within the data. Best subset feature selection (Figures SF15), on the other hand, is a method that exhaustively explores all possible combinations of features to find the subset that results in the best model performance. It systematically trains and evaluates models using varying feature combinations and selects the one that optimizes a chosen performance metric.

Genetic algorithms (GAs) are used for feature selection by representing potential feature subsets as binary chromosomes, with each gene indicating whether a feature is included (1) or excluded (0). The process begins with initializing a population of such chromosomes and evaluating their quality using a fitness function, typically based on classification or regression performance metrics. GAs then employ selection mechanisms to choose chromosomes for reproduction, crossover operations to recombine genes from selected chromosomes, and mutation to introduce diversity. These operations create new generations of chromosomes, and the process repeats over multiple iterations or generations. The algorithm converges towards an optimal or near-optimal feature subset, determined by the best-performing chromosome in the final population. This selected feature subset is used to build a machine learning model for prediction or classification, offering the advantage of handling high-dimensional datasets and capturing complex feature interactions. In this investigation, Genetic selection with a Gradient boosting regression estimator is considered. The number of population and number of generations was taken 10 and 1000, respectively with a mutation rate of 0.3 which gives the probability of mutation for each gene (Figures SF16). Table ST4 summarizes the outcome of the various feature selection methods indicating the filtered features for various methods.

To identify the best feature subset for DS2, all nine feature subsets, derived from various feature selection methods used, were used to train eight ML models, with the results summarized in Figure SF18(a-c). Performance metrics are provided in Table ST5. Interestingly, the RF and GB models continued to be the best-performing base regressors, likely due to their versatility in mapping complex non-linear relationships between input features and the target variable without overfitting, without compromising on interpretability. Given the continued strong performance of the RF and GB models, the original architecture of the RELM was retained and used to predict the yield strength of FCC HEAs using the GA-derived feature subset. Figure SF18(d) shows the parity plot between predicted and experimental yield strength using the RELM and GA-derived feature subset, with blue markers for the training dataset and red markers for the test dataset. The performance metrics for both datasets indicate that the elemental attributes show comparable performance to RELM trained on DS1. Figures SF18(e, f) show the residual distribution for the training and test datasets, respectively, showing that the errors are concentrated near zero for the training dataset, and a similar distribution is interpolated for the test dataset.

## *RELM Interpretation and Explanation*

It is often difficult to directly observe the implicit relationships between the inputs and outputs of the "black-box" ML models. In this regard, the significance of model interpretation and explanation is multifaceted. First and foremost, incorporating transparency and trust in ML models provides stakeholders with a comprehensive understanding of the underlying rationale behind certain predictions. This is especially crucial in domains of materials discovery. Furthermore, this process aids in the detection of biases within models and the evaluation of their fairness, guaranteeing that predictions do not exhibit discriminating behavior, particularly concerning the variables that are considered sensitive.

In this study, we use a combination of SHAP (SHapley Additive exPlanations) and partial pair dependencies to develop key scientific interpretations from RELM. This allowed us to get a comprehensive understanding of the primary alloying elements, processing conditions and material attributes influencing the yield strength of FCC HEAs. Lundberg *et al*. introduced SHAP which utilizes game theory principles to provide explanations for the predictions made by ML models. SHAP measures a feature's importance by quantifying the prediction error while perturbing a given feature value. If the prediction error is large, the feature is important otherwise the feature is less important (*70*). Classically, shapely values are calculated using Equation SE7.

$$\varphi_i = \sum_{S \in F\{i\}} \frac{|S|!(|F|-|S|-1)!}{|F|!} \left[ f_{SU\{i\}}\left(x_{SU\{i\}}\right) - f_s(x_s) \right] \qquad \text{(SE7)}$$

where $F$ is the set of all features, $S$ is the subset of features, $f_{SU\{i\}}$ is model trained with the feature present, $f_{S\{i\}}$ is a model trained with feature withheld, $x_s$ is input features in subset $S$. The SHAP values are visualized using violin plots. The violin plot represents the contribution of a given feature towards the different output values as a function of the feature value. Thus, the violin plot is colored according to the feature value.

Apart from SHAP, partial pair dependencies were used to understand the local representation of input features on the yield strength. The SHAP metric assesses the collective impact of all input features on the target variable. In contrast, the partial dependencies are capable of illustrating the nature of the connection between target values and input characteristics, whether it is linear, monotonic, or exhibits a more intricate pattern. Furthermore, it provides an account of the impact that each feature has on the target values (*71*).

For a model $f(X)$, where $X = (X_1, X_2, \dots, X_p)$ are features, the partial dependence of feature $X_j$ on the prediction $y$ is given by:

$$PD(X_j) = E_{X_{\backslash j}}[f(X_j, X_{\backslash j})] \quad \text{(SE8)}$$

Where, $X_{\backslash j}$ denotes all features except $X_j$; $E_{X_{\backslash j}}$ represents the expectation over the joint distribution of $X_{\backslash j}$. This formulation allows to isolate the effect of $X_j$ by averaging over the variability of other features. Subsequently, the connection between one or two feature variables and the corresponding predicted values can be investigated.

## *Generative Alloy Design*

Considering the large number of possible composition combinations for FCC HEA systems and the small experimental datasets available (357 alloy compositions), the challenge is to directly sample new compositions with the desired properties. We developed a combined generative adversarial network and variational autoencoder (GAN-VAE) strategy to sample new alloy compositions. The hybrid GAN-VAE model is designed to learn a continuous latent space representation of material compositions in an unsupervised manner. This model combines the strengths of VAE and GAN to not only generate new compositions but also provide a meaningful and interpretable latent space.

GANs consist of two neural networks, a generator and a discriminator, which are trained in a competitive setting. The generator creates synthetic samples that aim to resemble the real data, while the discriminator evaluates these samples to determine whether they are real (from the training set) or fake (generated). This adversarial training process continues until the generator produces highly realistic data that the discriminator cannot easily distinguish generated data from real data.

VAEs are a type of generative model that, unlike GANs, explicitly model the distribution of the data. A VAE consists of an encoder, a decoder (or generator), and a latent space. The encoder maps input data to a latent space distribution, typically a Gaussian distribution. The decoder then reconstructs the input data from samples drawn from this latent space. The VAE is trained to minimize the reconstruction loss (how closely the output resembles the input) and a regularization term (which ensures that the latent space distribution remains close to a standard normal distribution). This dual-objective approach helps in creating a smooth and continuous latent space that can be sampled to generate new, coherent data points.

***The GAN-VAE Architecture***

The GAN-VAE framework consists of 3 nodes:

- **Encoder:** Maps input compositions to a latent space, providing both the mean and log variance of the latent distribution. It regularizes the latent space to follow a standard normal distribution *via* KL divergence.
- **Generator/Decoder**: Decodes latent vectors to generate alloy compositions. It plays a dual role: reconstructing input compositions in the VAE part and generating new compositions in the GAN part.
- **Discriminator**: Evaluates the authenticity of compositions, distinguishing between real (from the dataset) and fake (generated by the generator).

The generator is a feedforward neural network with 3 dense layers with 64, 86, and 52 neurons, each followed by *LeakyReLU* activation with $\alpha = 0.02$ and Batch Normalization. The final layer uses a *tanh* activation function to output values within the range [-1, 1], which is compatible with our normalized data. The discriminator is also a feedforward neural network with 3 dense layers having 52, 86, and 64 neurons and *LeakyReLU* activations, culminating in a *sigmoid* activation function that outputs a probability indicating whether the input is real or generated. We combine the generator and discriminator to form the GAN. The discriminator is compiled first with binary cross-entropy loss as the loss function and accuracy metrics. We then create a combined model where the discriminator's weights are frozen (i.e., not trainable) when training the generator. This setup ensures that during the combined model training, only the generator's weights are updated to improve its ability to produce realistic samples.

To ensure that compositions can be accurately mapped to and reconstructed from the latent space, we also trained an autoencoder. The autoencoder comprises the encoder (with 2 dense layers having 64 and 52 neurons) and the generator developed earlier (acting as a decoder). It is trained to minimize the mean squared error (MSE) between the input compositions and the reconstructed compositions. The Adam optimizer with a learning rate of 0.0002 is used. The training loss of the autoencoder is tracked and plotted to ensure that the reconstruction loss decreases over time (Figure SF17). After training the GAN and autoencoder, we use the encoder to project the known compositions into the latent space. This step allows us to visualize the latent space and understand how the compositions are distributed. We then generate 5,000 new compositions by sampling from a standard normal distribution (the latent space) and passing these samples through the trained generator. The generated compositions are then denormalized to bring them back to the original scale of the alloy compositions.

The reconstruction loss function used for the autoencoder is the Mean Squared Error (MSE) loss. MSE loss calculates the average of the squared differences between the predicted values (reconstructed compositions) and the actual values (original compositions).

$$MSE = \frac{1}{n}\sum_{i=1}^{n}(y_i - \hat{y}_i)^2 \qquad \text{(SE9)}$$

where:

$n$ is the number of samples,

$y_i$ is the original composition of the $i$-th alloy,

$y_i$ is the predicted (reconstructed) composition of the $i$-th alloy.

Besides, in Variational Autoencoders (VAEs), the loss function is a combination of two terms: the reconstruction loss and the Kullback-Leibler (KL) divergence. While the reconstruction loss ensures that the autoencoder can accurately reconstruct the input data, the KL divergence regularizes the distribution of the latent space to match a prior distribution, typically a standard normal distribution. The KL divergence is a measure of how one probability distribution diverges from a second, expected probability distribution. In the context of VAEs, it measures how much the encoded latent space distribution deviates from the desired prior distribution, usually a standard normal distribution.

The formula for KL divergence between two distributions $P$ and $Q$ is:

$$D_{KL}(P||Q) = \sum_i P(i) log \frac{P(i)}{Q(i)} \quad \text{(SE10)}$$

In the VAE, the KL divergence term can be expressed as:

$$D_{KL}(q(z|x)||p(z)) = \frac{1}{2}\sum_{j=1}^{d}(\mu_j^2 + \sigma_j^2 - log\, \sigma_j^2 - 1) \quad \text{(SE11)}$$

where $\mu_j$ and $\sigma_j$are the mean and standard deviation of the latent variable $z$ for each dimension $j$. $q(z|x)$ is the encoder's output distribution, and $p(z)$ is the prior distribution (usually $N(0,1)$). Therefore, the overall loss function in a VAE combines the reconstruction loss (MSE, here) and the KL divergence:

$$VAE\ Loss = Reconstruction\ Loss + D_{KL}(q(z|x)||p(z)) \quad \text{(SE12)}$$

## *Experimental Methods*

The alloys chosen for experimental validation of the RELM predictions were produced by arc melting in a pure Ar atmosphere using raw materials of at least 99.9% purity. All the alloys were re-melted and turned at least five times to ensure homogeneity. Thereafter, alloys were cast into 30x30x50 $mm^3$ billets using suction casting. The billets were then homogenized at appropriate temperatures (obtained through CALPHAD calculations, Figure SF22) by sealing them in quartz tubes under a vacuum, followed by quenching them in water. This was followed by thermo-mechanical treatment wherein the samples were cold rolled for the required percentage deformation, followed by annealing at the required temperature and the required time.

Uniaxial tensile tests were performed at room temperature at an initial strain rate of $1 \times 10^{-3}$ $s^{-1}$. The flat dog-bone specimens for each thermo-mechanically treated alloy were cut by electron discharging machining along the rolling direction. The tensile specimens had a total length of 50 mm (gauge length of 30 mm, gauge width of 5 mm) and a thickness of 2 mm to probe the bulk tensile properties. At least three samples for each condition were tested to confirm reproducibility.

X-ray diffraction (XRD) measurements were carried out on the homogenized as well as thermo-mechanically treated samples with a Bruker D8 Advance instrument in Bragg-Brentano geometry using Co as the X-ray source and dimensions of the specimens of 20×20×2 $mm^3$. Phase identification was carried out using the HighScore software from PANalytical. The microstructure of the alloys was investigated by scanning electron microscopy (SEM). Samples for SEM investigation were hot-mounted, ground, and polished. A Zeiss Sigma SE microscope was employed for backscattered electron (BSE) imaging and energy-dispersive X-ray spectroscopy (EDS) mapping at 15 kV.

## Supplementary Figures

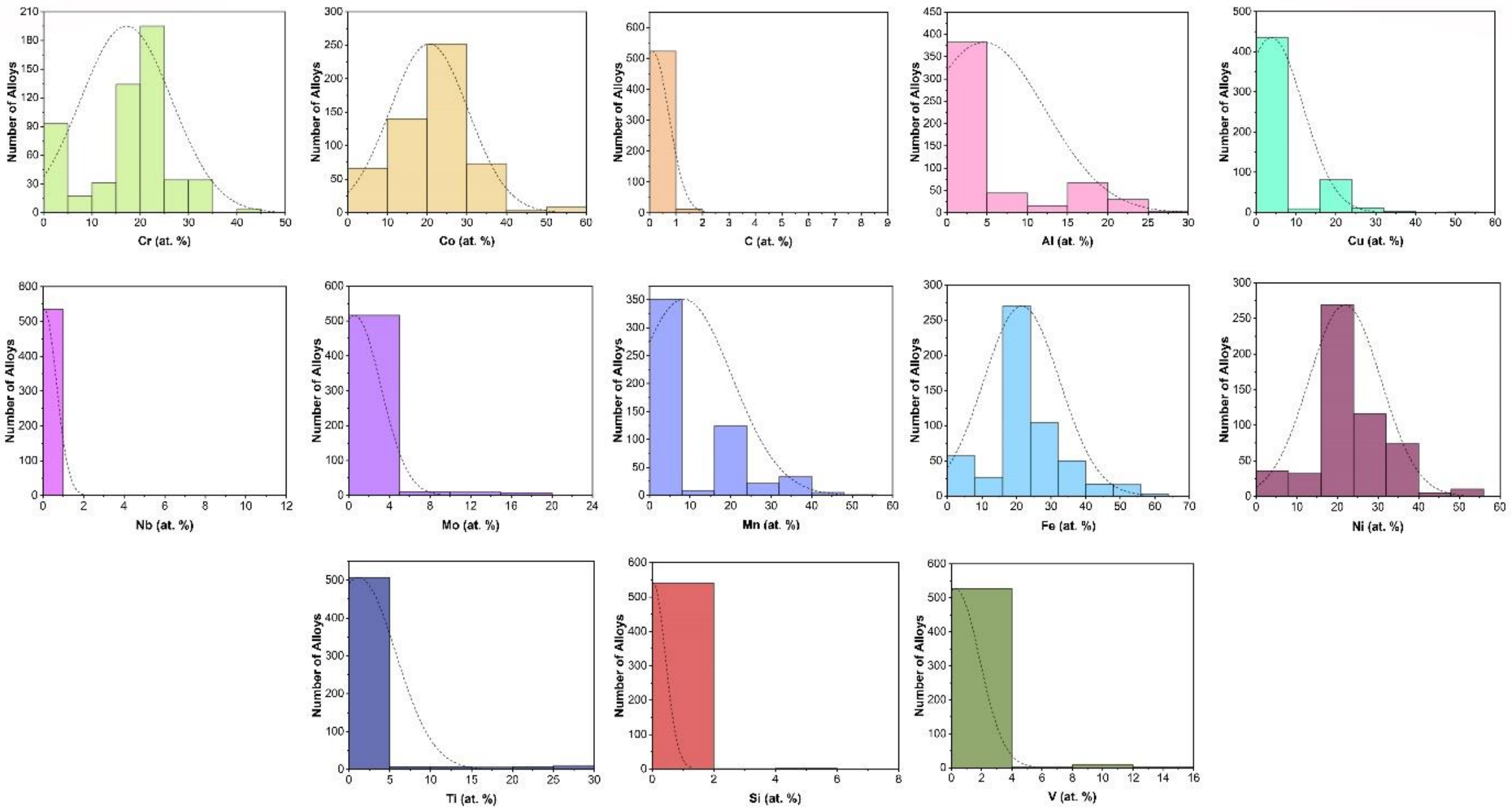

**Figure SF1 –** Distribution of alloy constituents as a function of their composition (atomic %) for 543 FCC HEAs. The red dashed curve shows the average normal distribution of the alloy constituent.

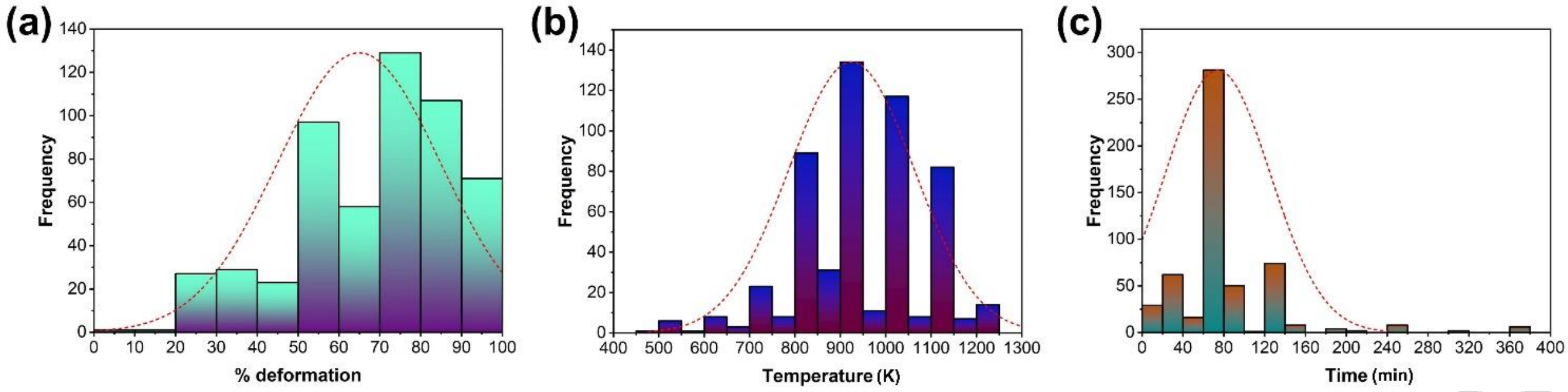

**Figure SF2 –** Frequency distribution of processing parameters for 543 FCC HEAs. The red dashed curve shows the average normal distribution of the processing parameter.

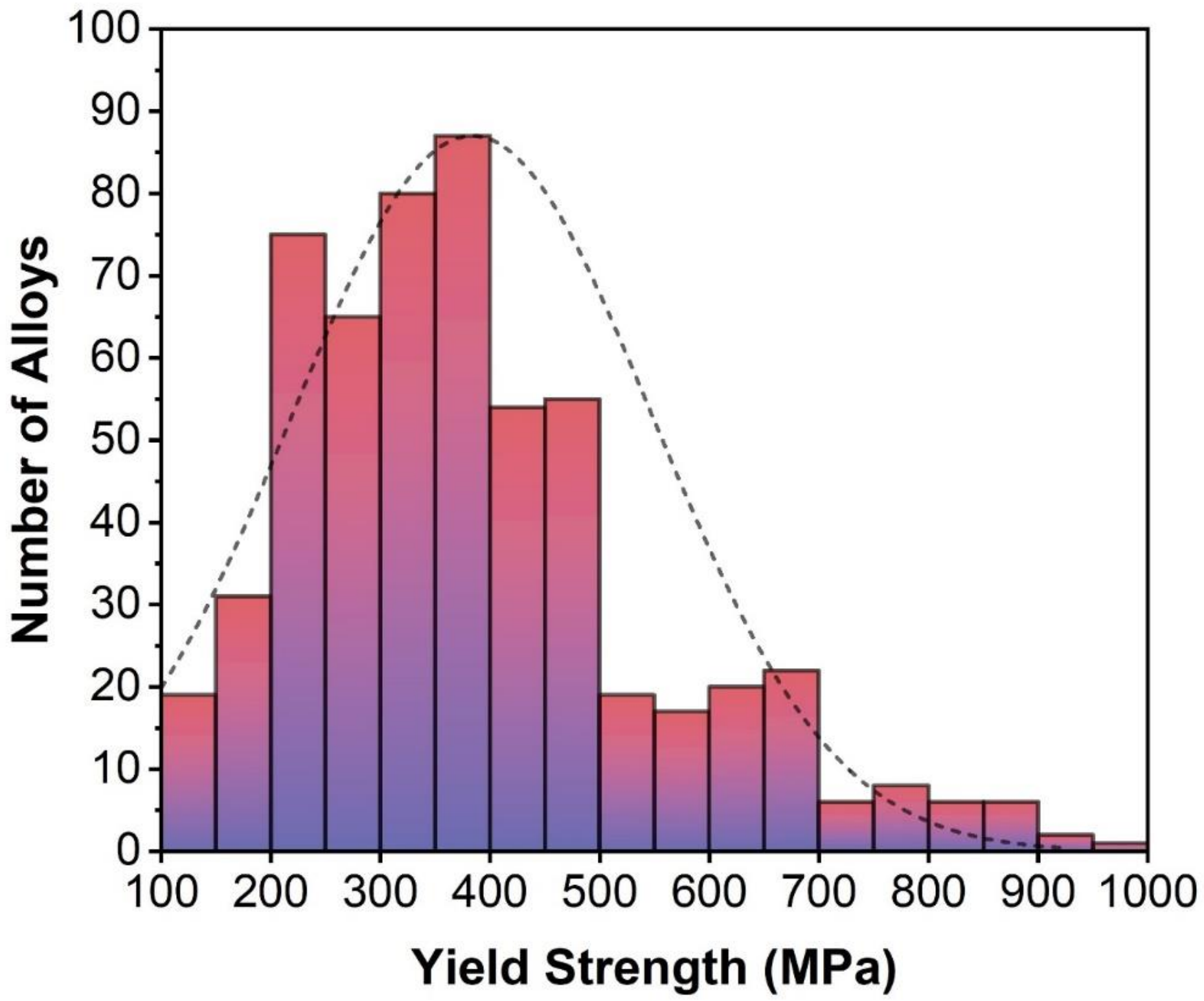

**Figure SF3 –** Frequency distribution of yield strength values for 543 FCC HEAs. The red dashed curve shows the average normal distribution of the yield strength values.

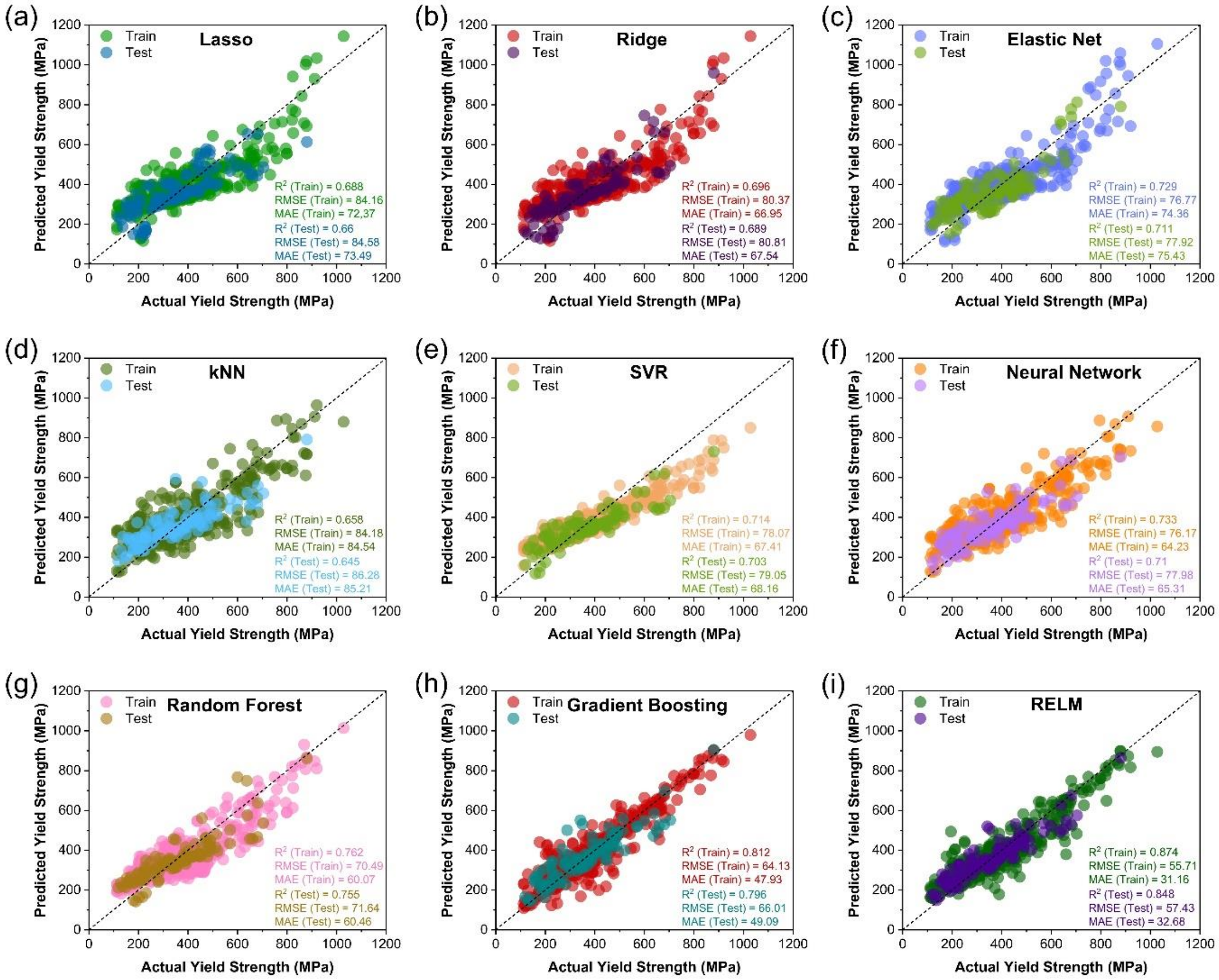

**Figure SF4 –** Parity plots showing the performance of various ML models on the training and the test datasets using DS-1 feature space. The models are trained on 10 seeds and the metrics are averaged (10 different training and test datasets). The parity plots for all the models are shown for the seed value, which provides the best performance of the RELM.

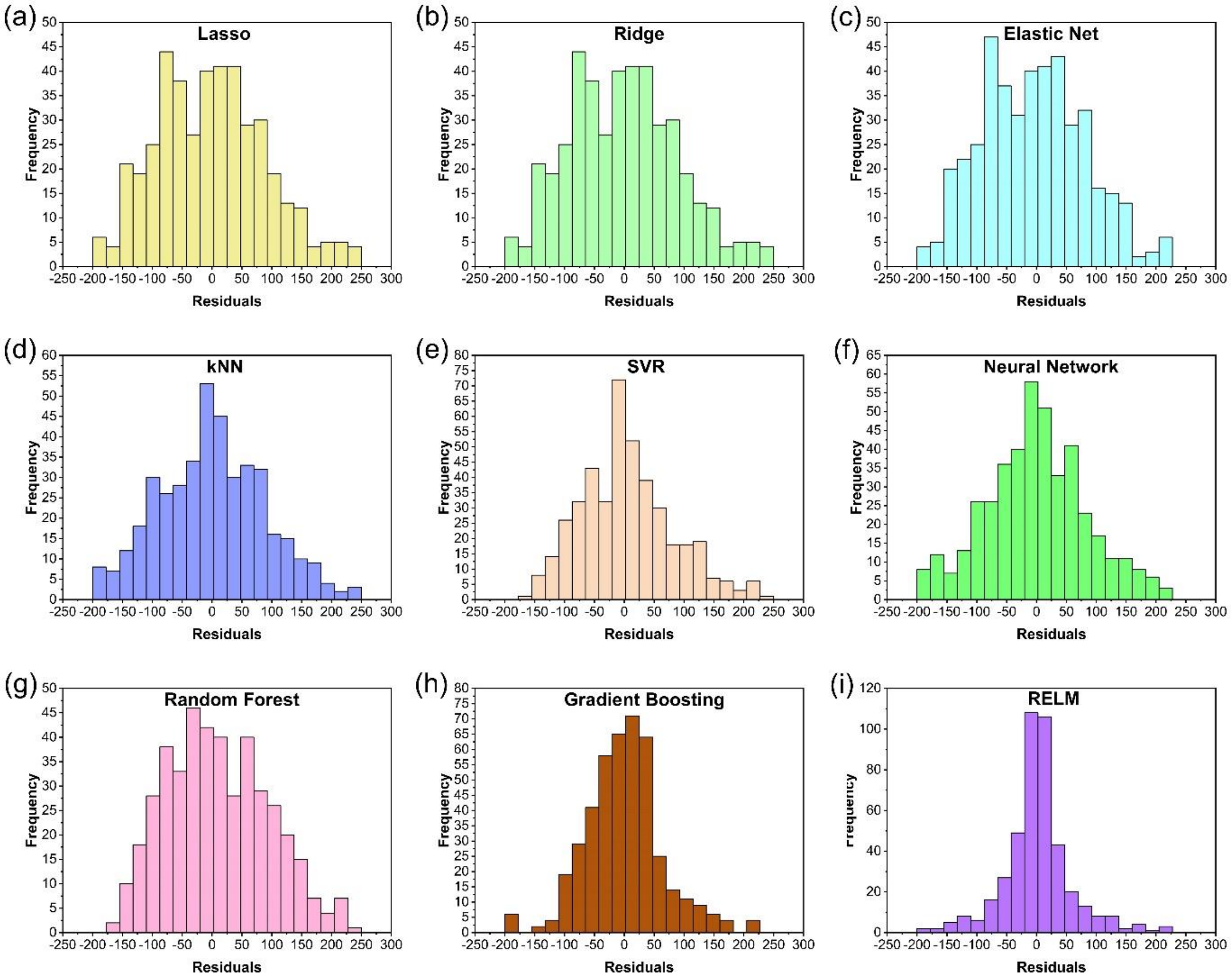

**Figure SF5 –** Residual analysis plots showing the distribution of errors for the training dataset with various ML models using DS-1 feature space.

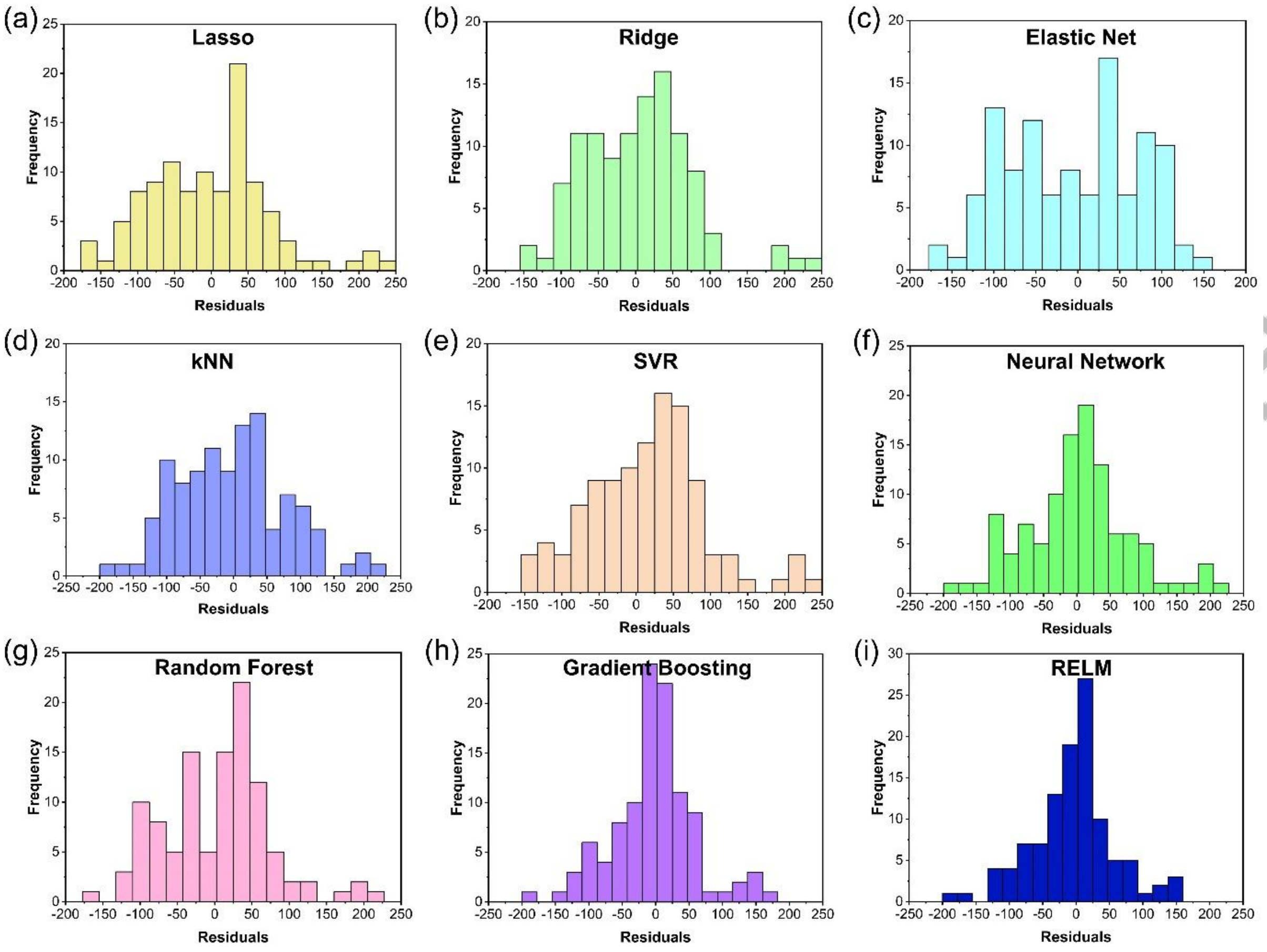

**Figure SF6 –** Residual analysis plots showing the distribution of errors for the test dataset with various ML models using DS-1 feature space.

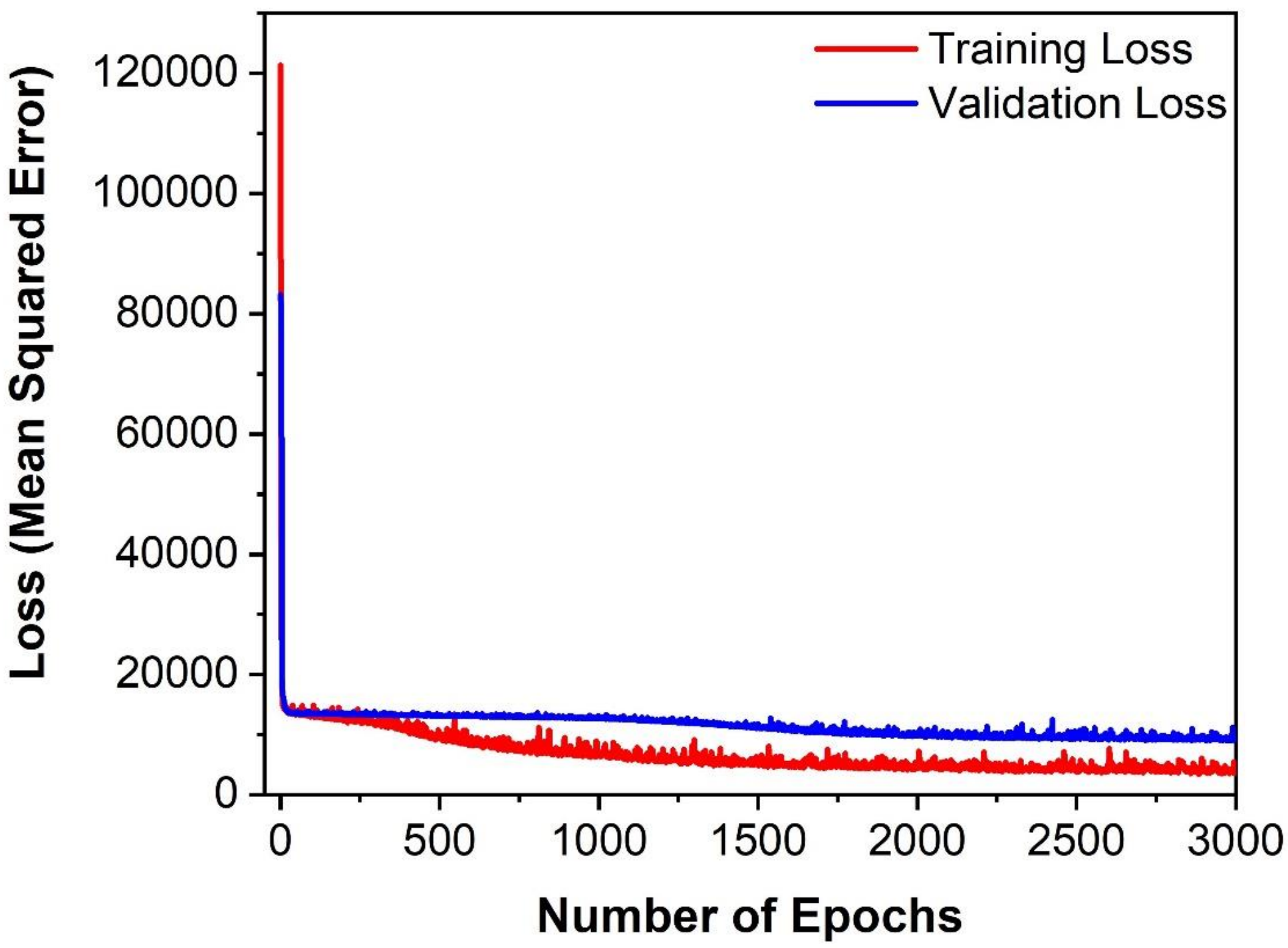

**Figure SF7 –** Variation of loss with the number of epochs for the training and validation datasets for the Neural Network (NN) model trained on DS1. The NN was trained through hyper-parameter optimization using Bayesian optimization and 5-fold cross-validation.

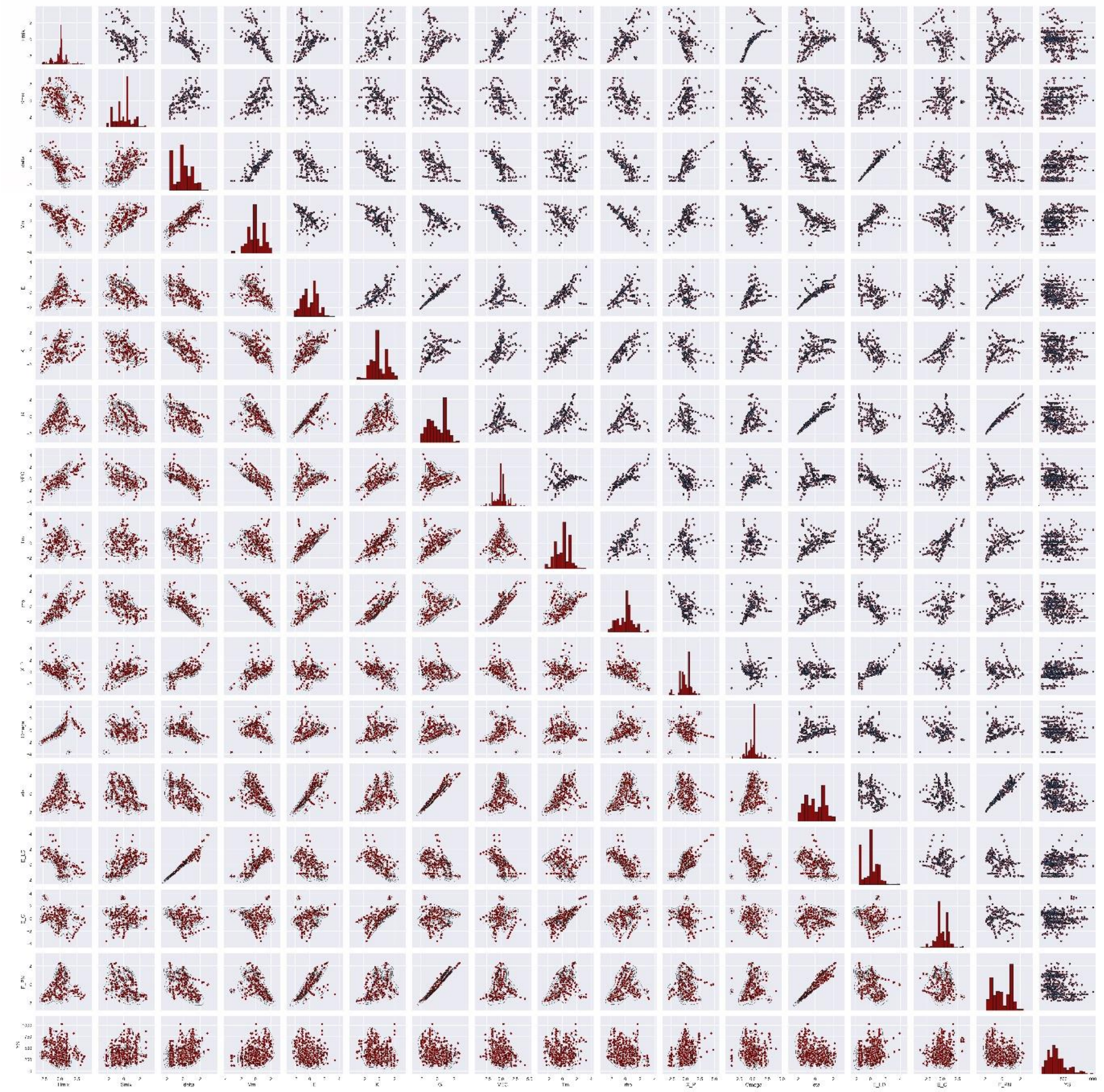

**Figure SF8** – Scatter matrix plotting the distribution of various material attributes.

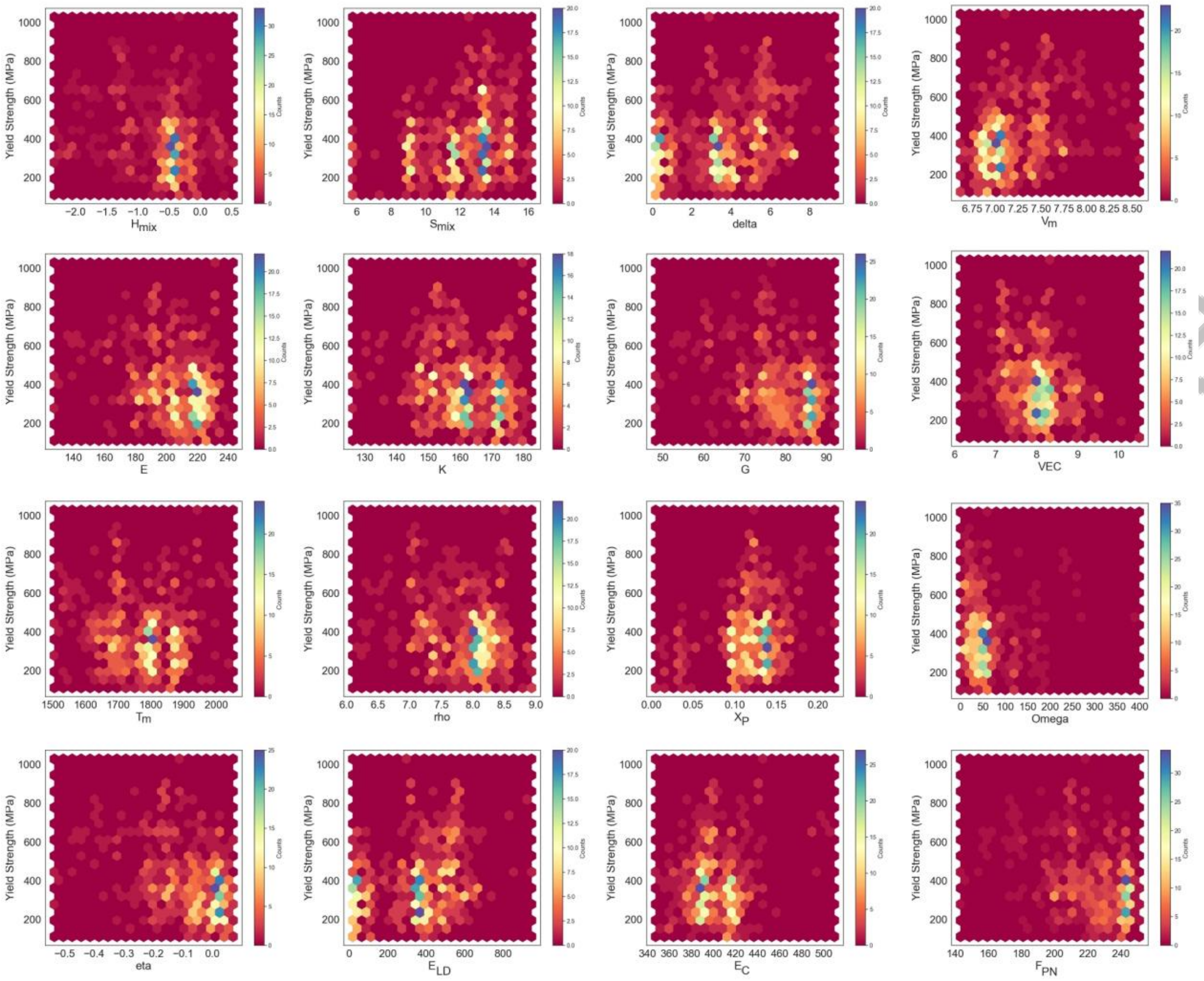

**Figure SF9 –** Visualization of DS2 through hexagonal binning plots for variation of yield strength values with different material properties

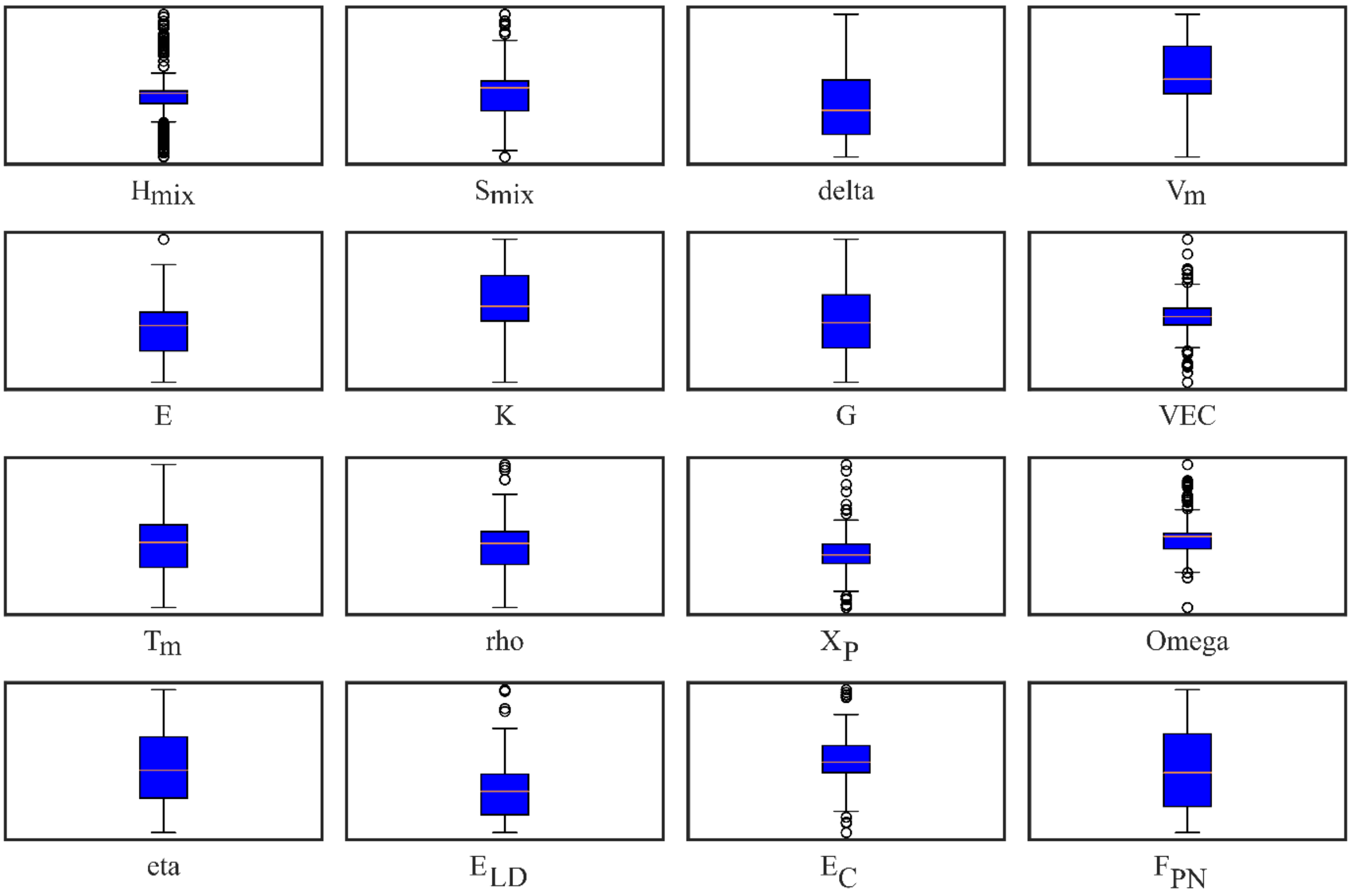

**Figure SF10** – Box plots showing the distribution of various material attributes. The values away from the whiskers indicate the presence of outliers.

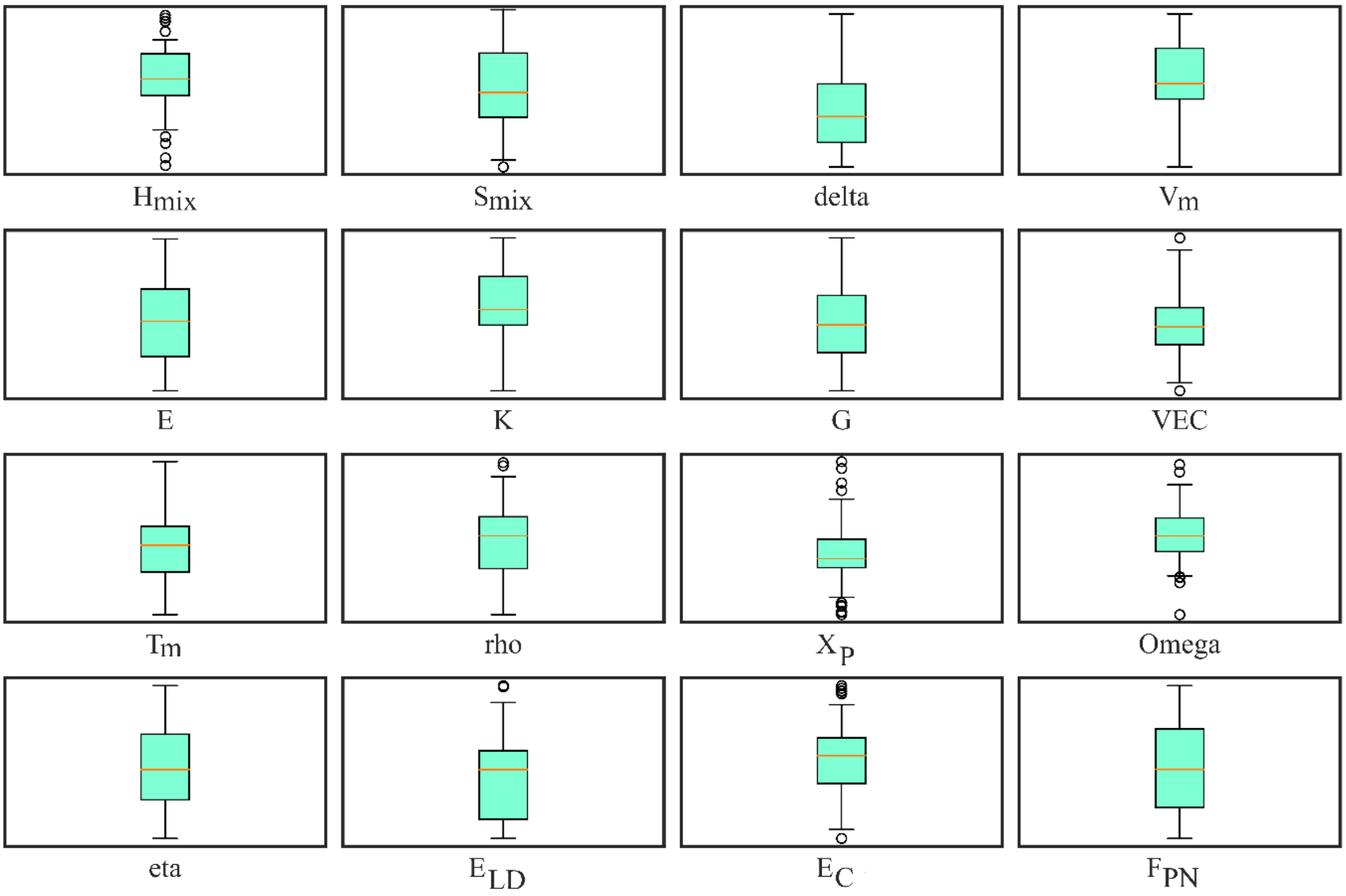

**Figure SF11** – Box plots showing the distribution of various material attributes post Yeo-Johnson transformation. The values away from the whiskers indicate the presence of outliers.

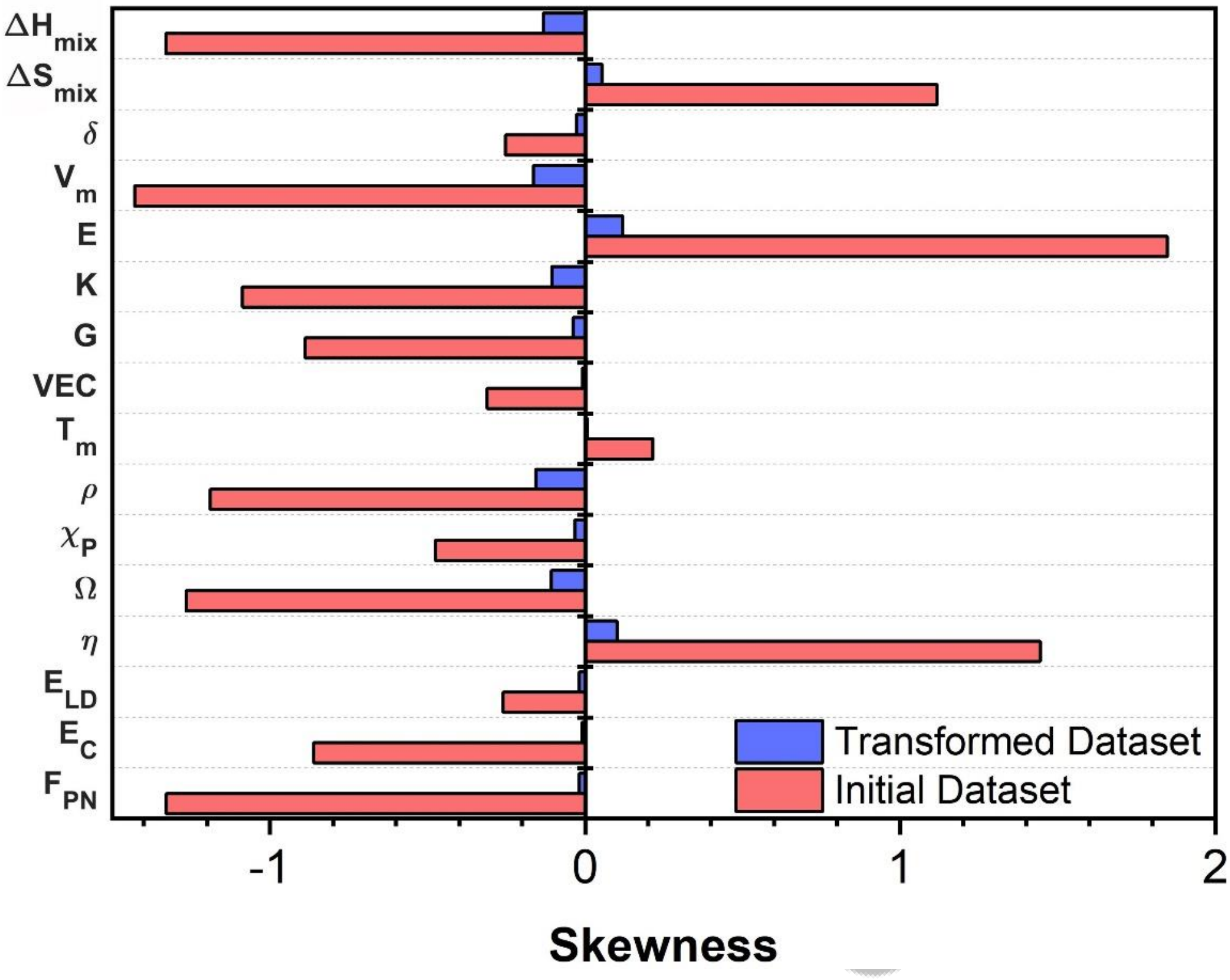

**Figure SF12** – Skewness (from the standard normal distribution) in the data distribution of elemental attributed in the original dataset and the Yeo-Johnson transformed dataset.

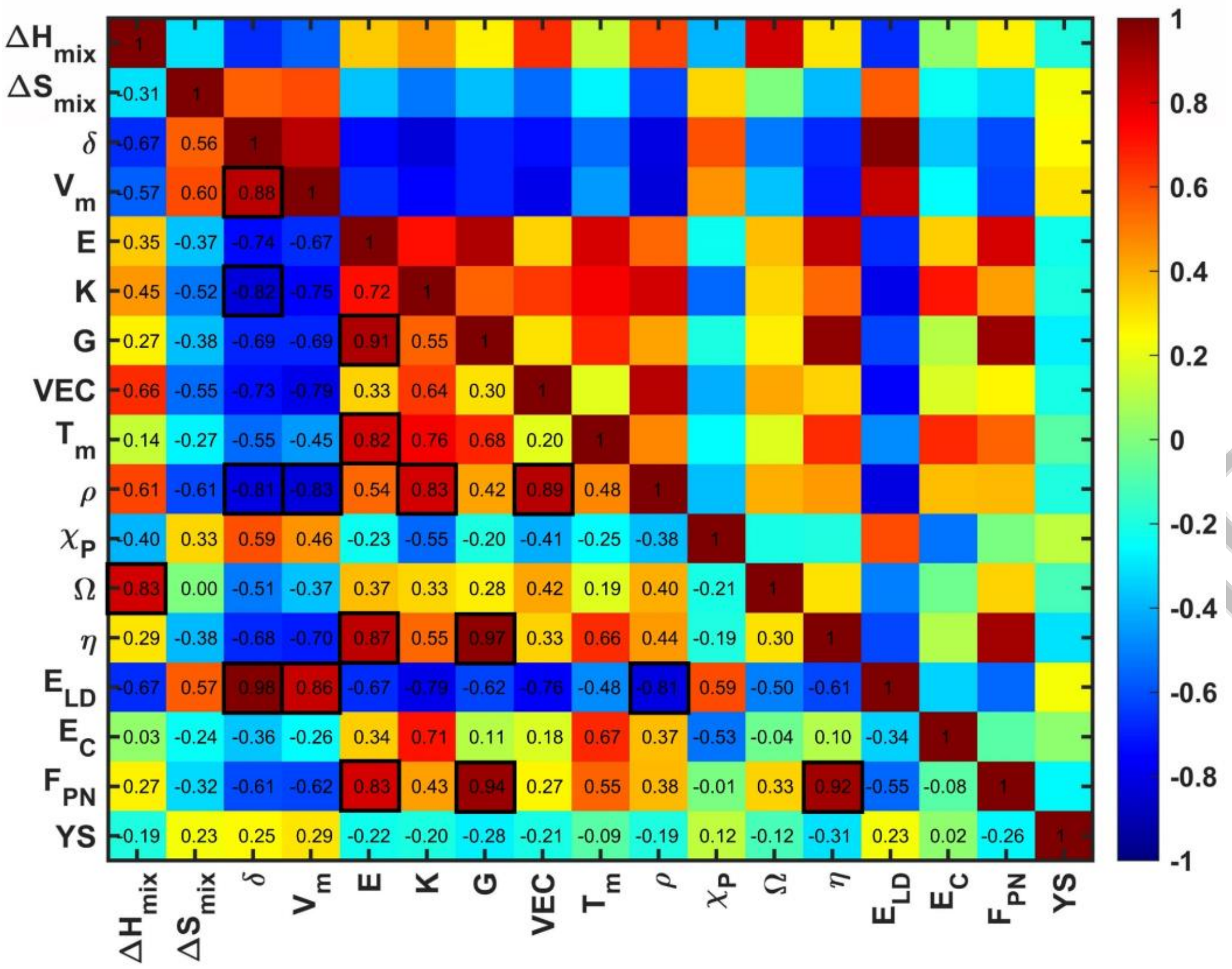

**Figure SF13 –** Heatmap showing the Pearson correlation coefficient values between different material properties.

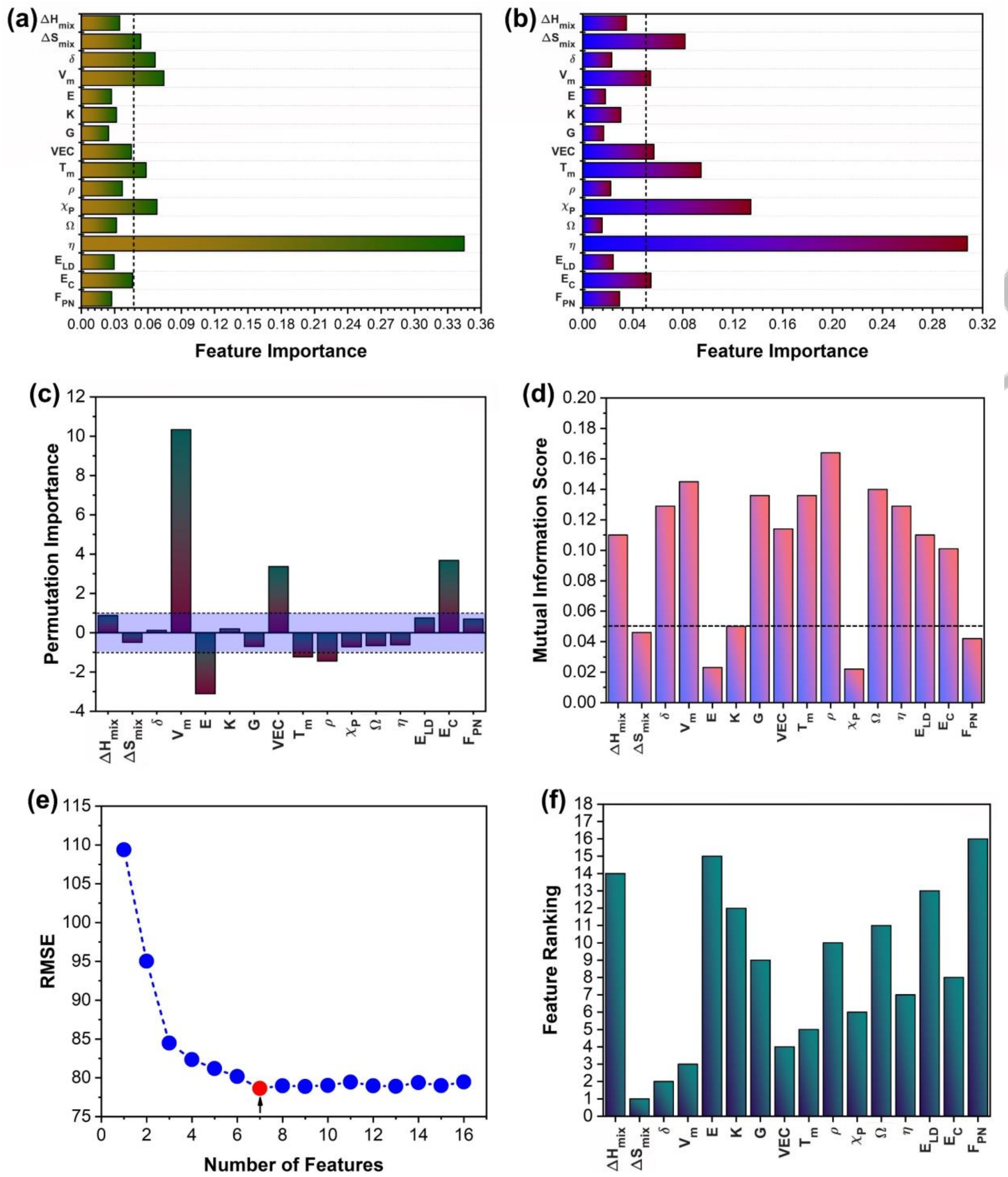

**Figure SF14 –** (a) Random Forest derived feature selection; (b) Gradient boosting derived feature selection based on the importance of features on the corresponding dry-run model for predicting yield strength; (c) Permutation Importance; (d) Mutual Importance; (e) Backward Selection and (f) Recursive Feature Selection using the gradient boosting algorithm.

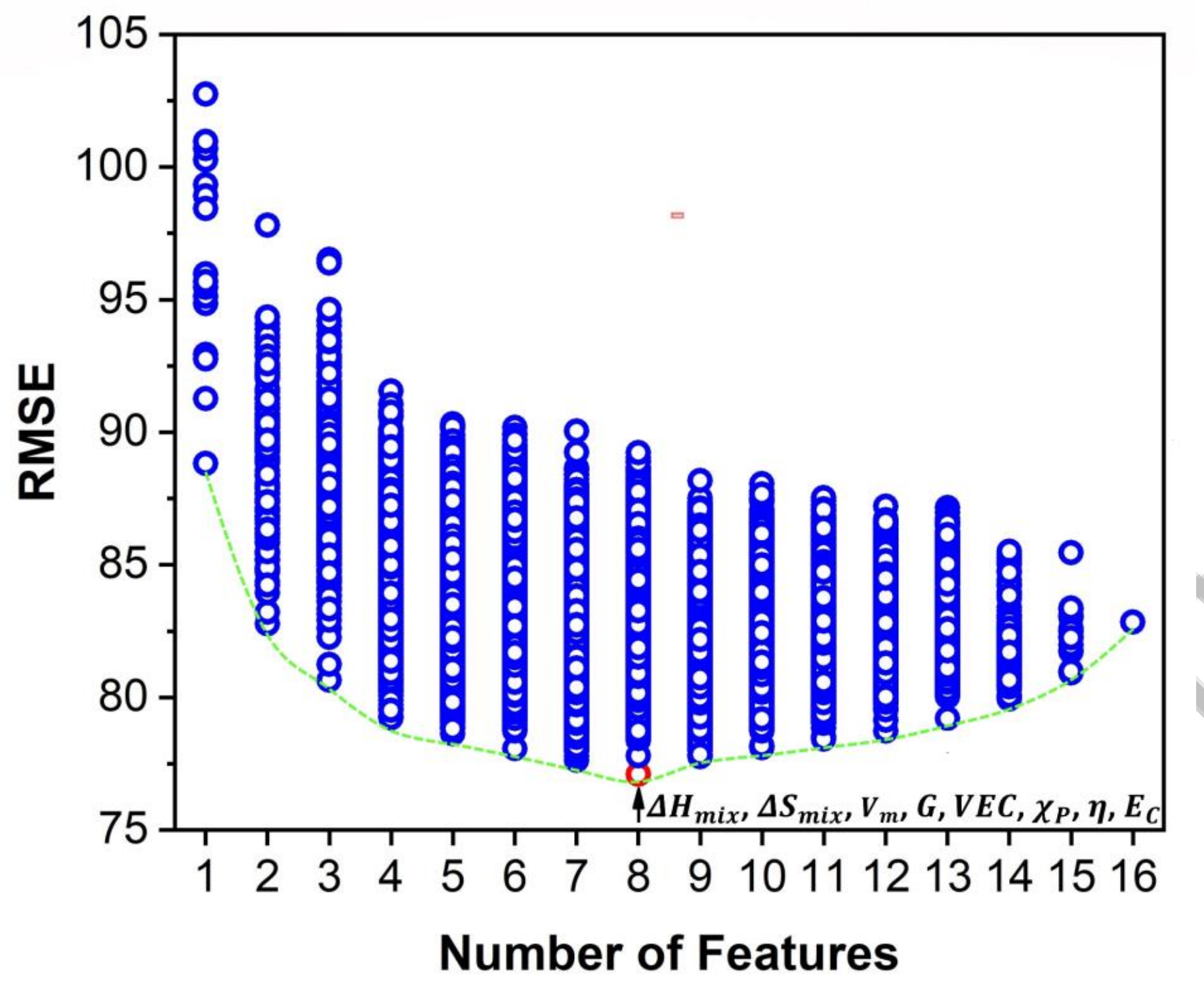

**Figure SF15 –** Best feature subset selection using the exhaustive enumeration with Gradient Boosting regressor and 10-fold cross-validation.

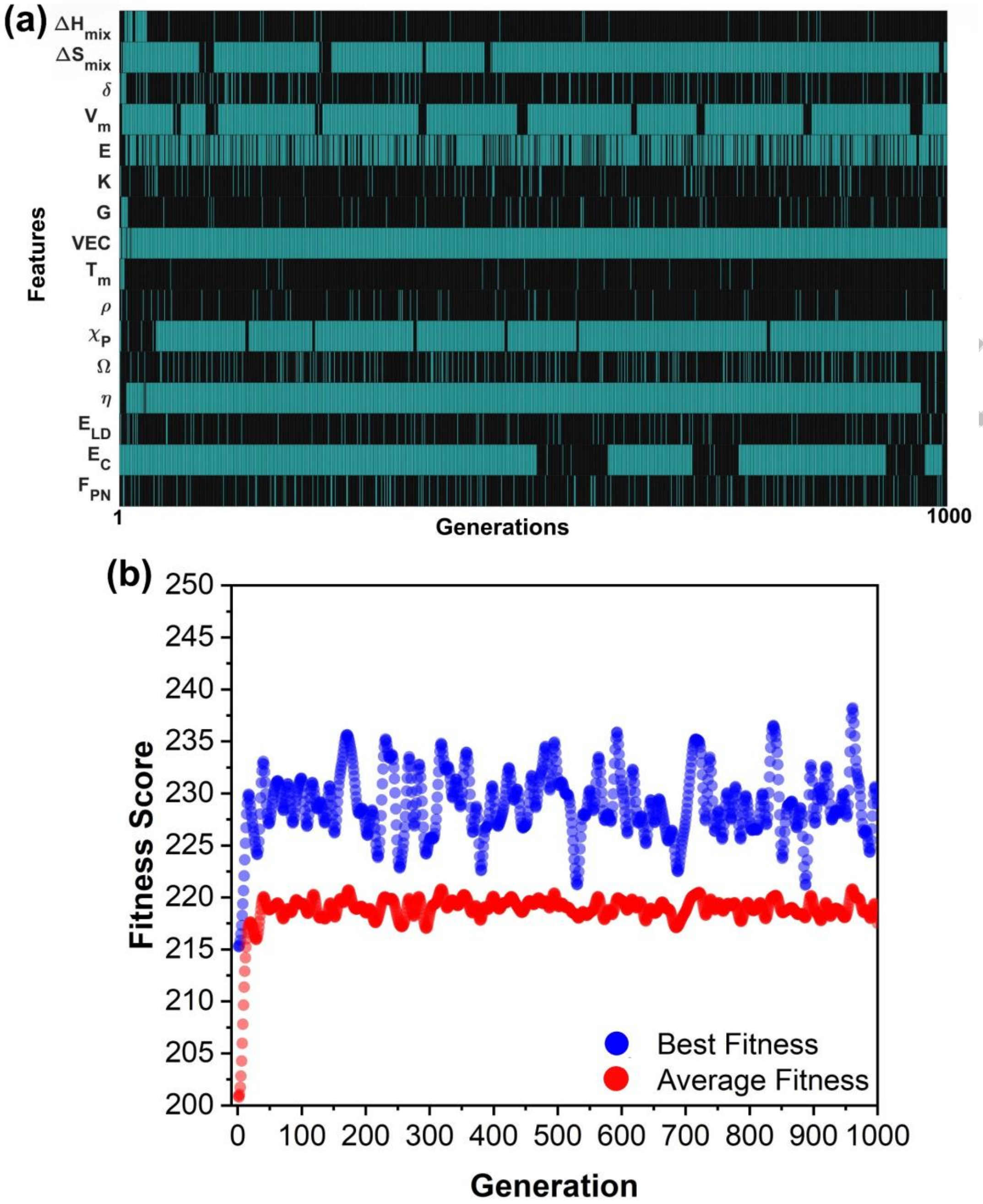

**Figure SF16 –** Genetic Algorithm based feature selection. (a) Occurrence of each feature with the changing number of generations during GA development for the optimal model performance; (b) Variation of fitness score with the number of generations for the average fitness as well as the best fitness architecture of the GA.

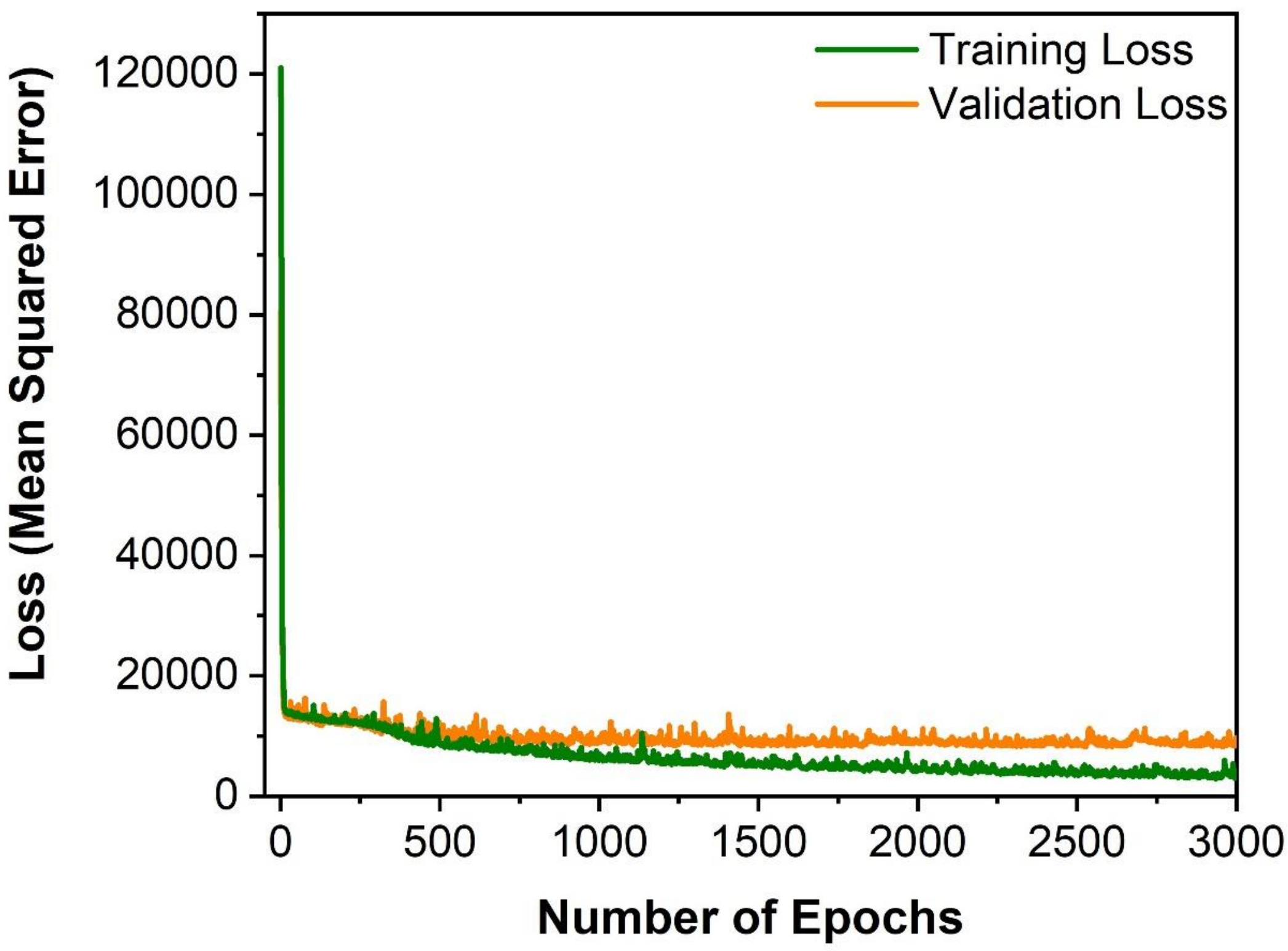

**Figure SF17 –** Variation of loss with the number of epochs for the training and validation datasets for the Neural Network (NN) model trained on DS2. The NN was trained through hyper-parameter optimization using Bayesian optimization and 5-fold cross-validation.

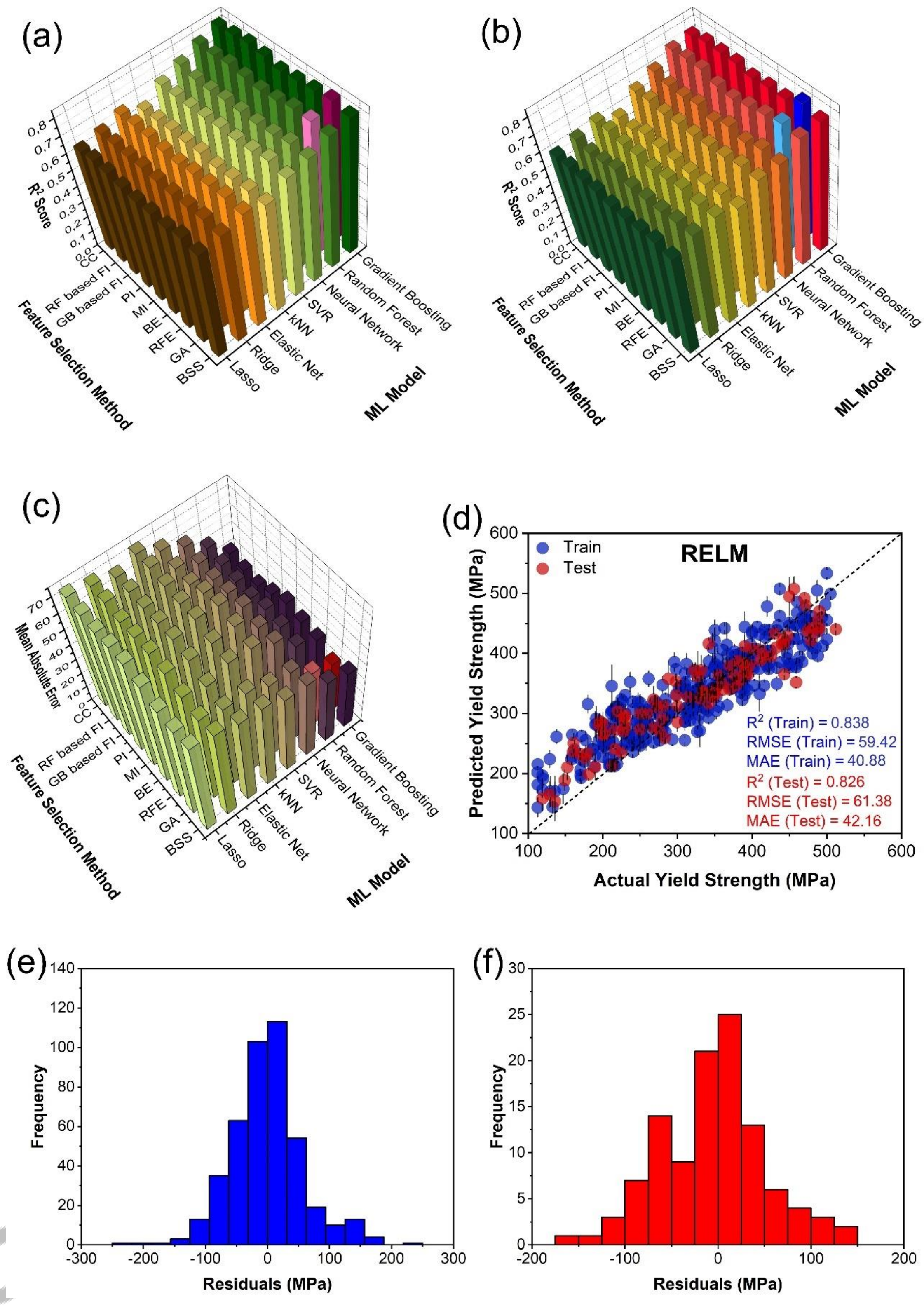

**Figure SF18 –** The performance of various ML models with different feature selection methods. (a) $R^2$ values for the training dataset, (b) $R^2$ values for the test dataset, (c) MAE values for the test dataset; (d) Parity plot for the RELM model trained on the finally filtered elemental properties from DS2; distribution of residuals for (e) training and (f) test datasets for the RELM-DS2 framework.

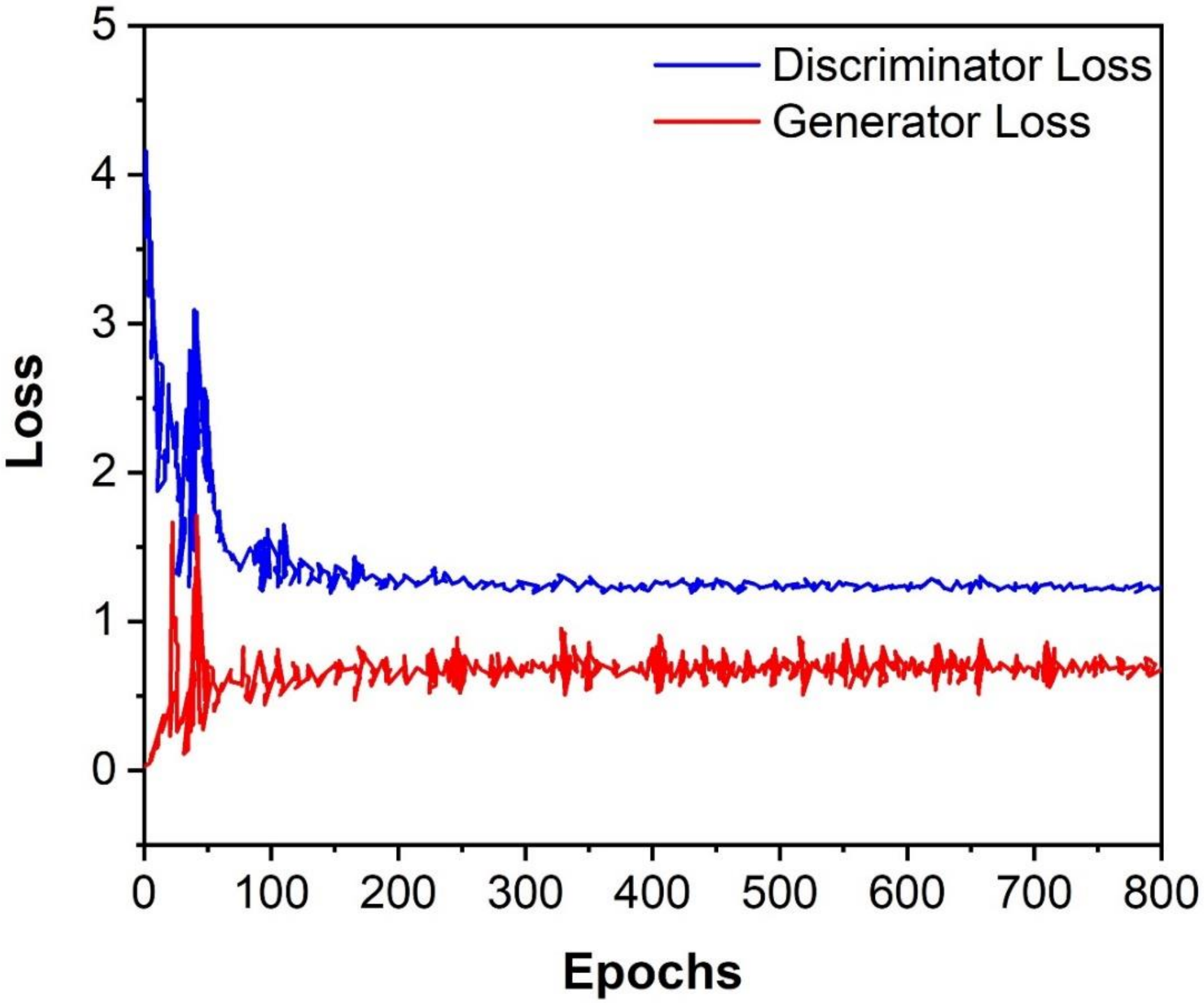

**Figure SF19 –** Variation of discriminator and generator loss (mean squared error) with the number of epochs for the trained GAN.

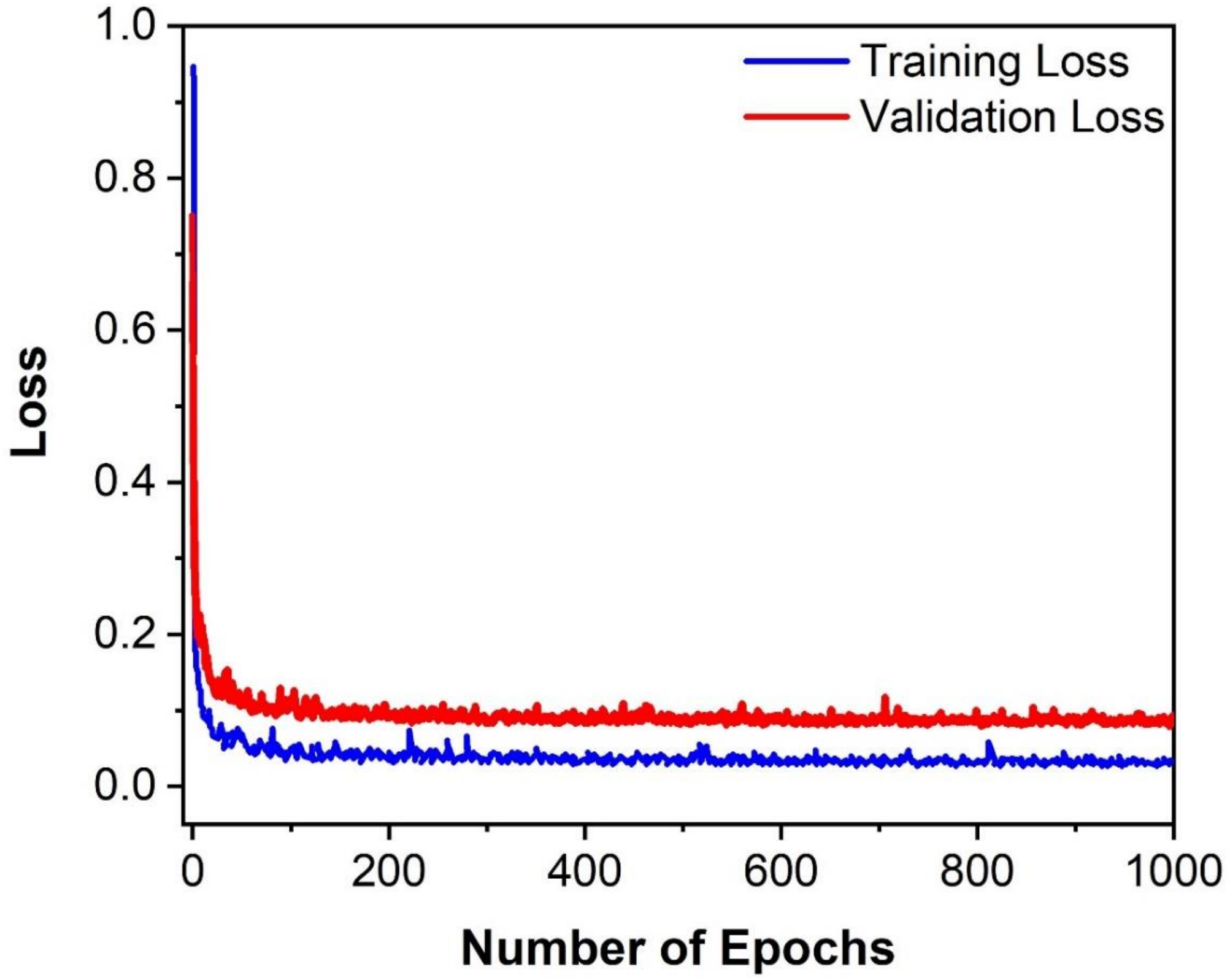

**Figure SF20 –** Variation of loss (mean squared error) with the number of epochs for VAE for the training and validation datasets.

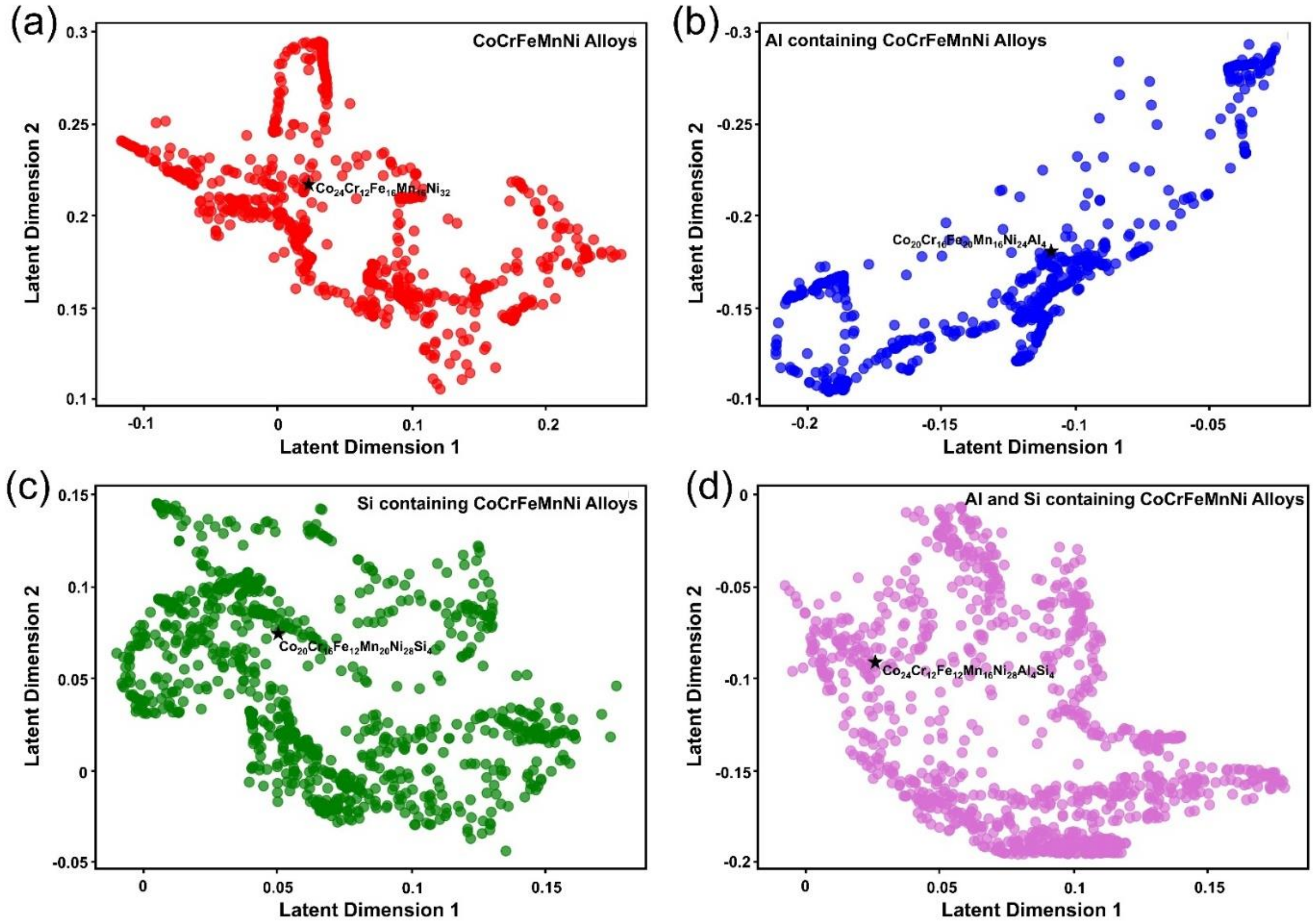

**Figure SF21 –** Latent space description of 5000 sampled compositions using GAN-VAE, each for (a) CoCrFeMnNi, (b) Al containing CoCrFeMnNi, (c) Si containing CoCrFeMnNi and (d) Al and Si containing CoCrFeMnNi alloys. The black stars indicate the sampled alloys which exhibit the highest yield strength based on the predictions of the RELM.

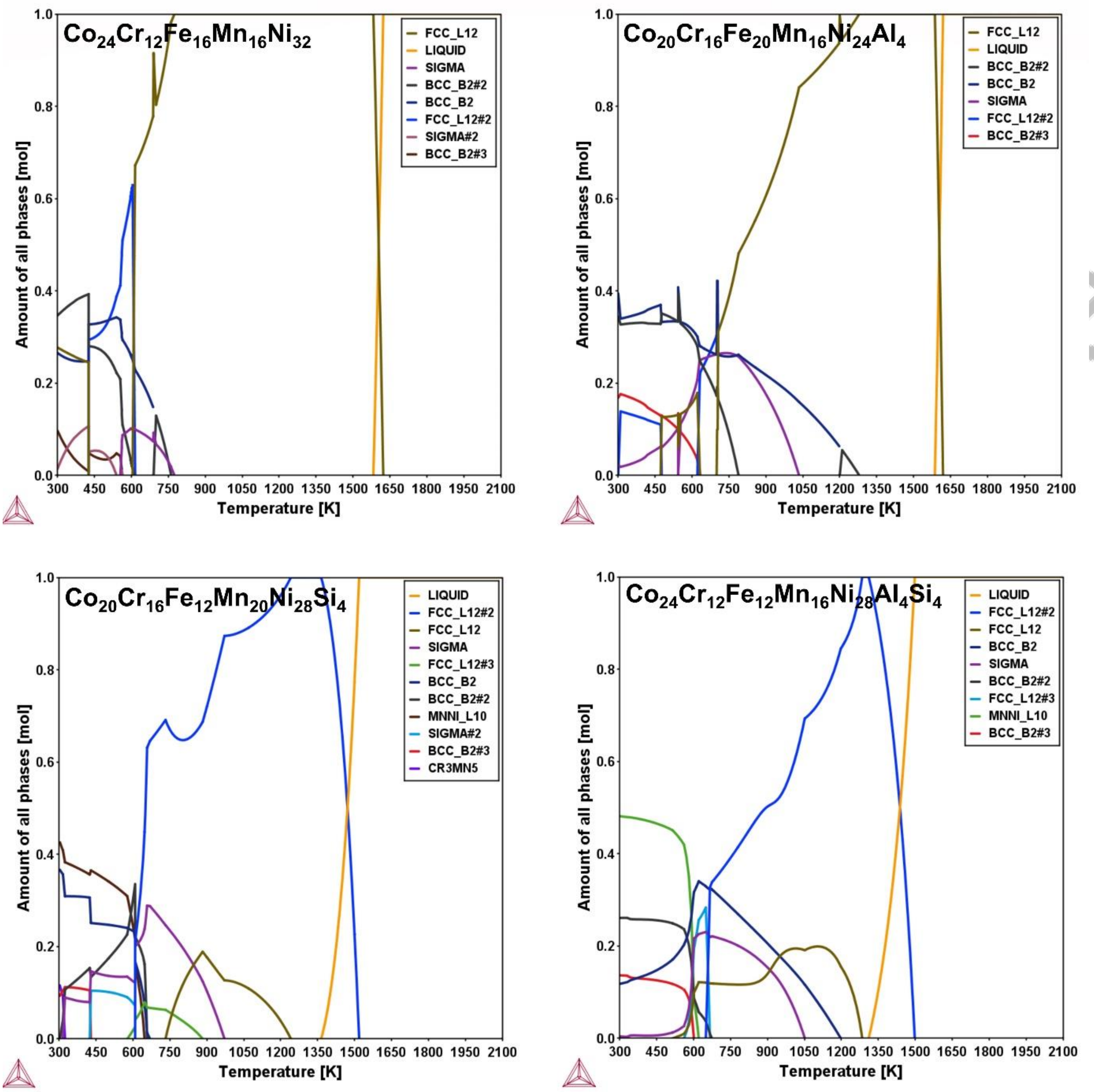

**Figure SF22** – Property plots for $Co_{24}Cr_{12}Fe_{16}Mn_{16}Ni_{32}$, $Co_{20}Cr_{16}Fe_{20}Mn_{16}Ni_{24}Al_4$, $Co_{20}Cr_{16}Fe_{12}Mn_{20}Ni_{28}Si_4$ and $Co_{24}Cr_{12}Fe_{12}Mn_{16}Ni_{28}Al_4Si_4$ alloys, designed using Thermocalc software using *TCHEA5* database.

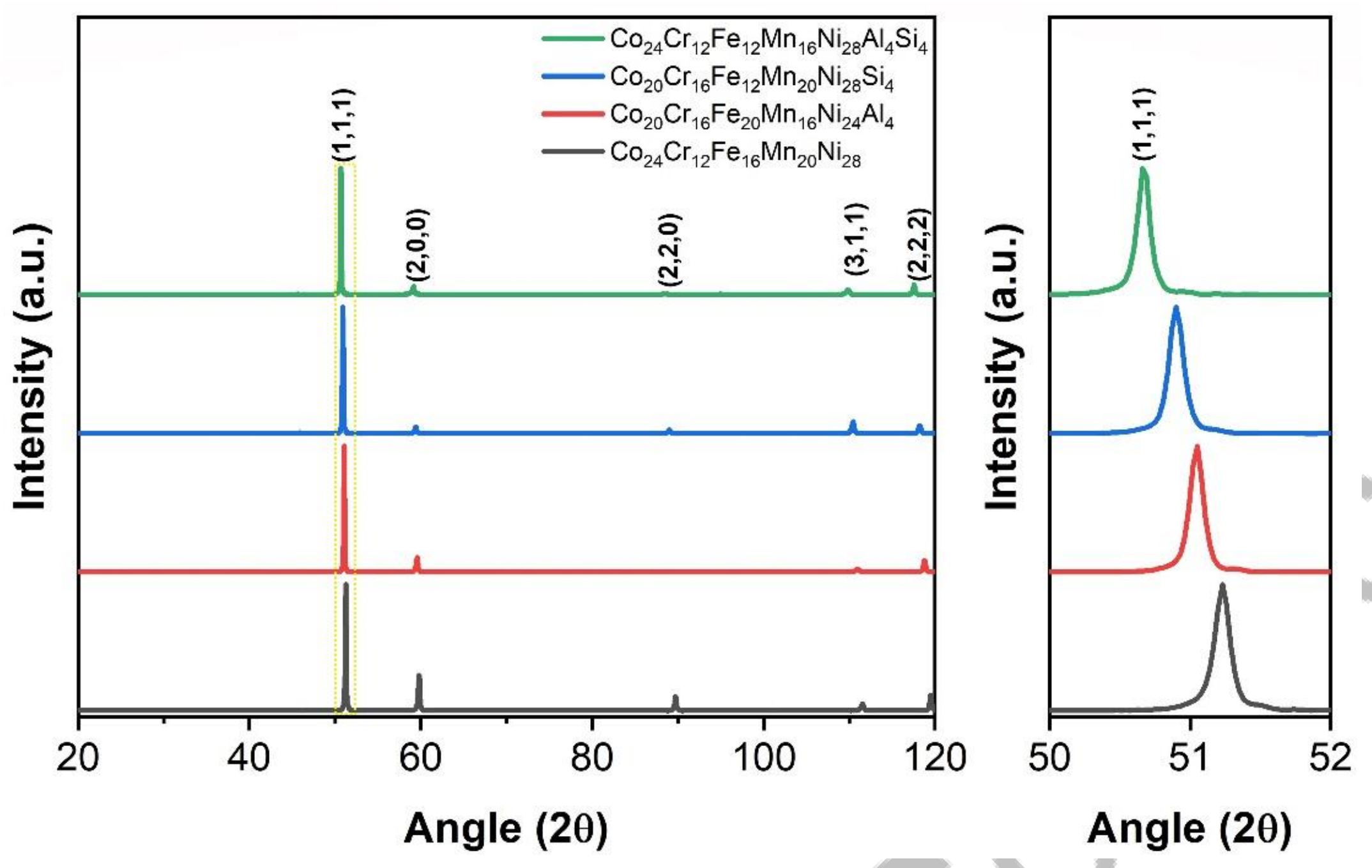

**Figure SF23** – Comparative X-ray diffractogram for as-cast and homogenized $Co_{24}Cr_{12}Fe_{16}Mn_{16}Ni_{32}$, $Co_{20}Cr_{16}Fe_{20}Mn_{16}Ni_{24}Al_4$, $Co_{20}Cr_{16}Fe_{12}Mn_{20}Ni_{28}Si_4$ and $Co_{24}Cr_{12}Fe_{12}Mn_{16}Ni_{28}Al_4Si_4$ alloys. The figure on the right shows the 2θ shift for the first highest intensity peak.

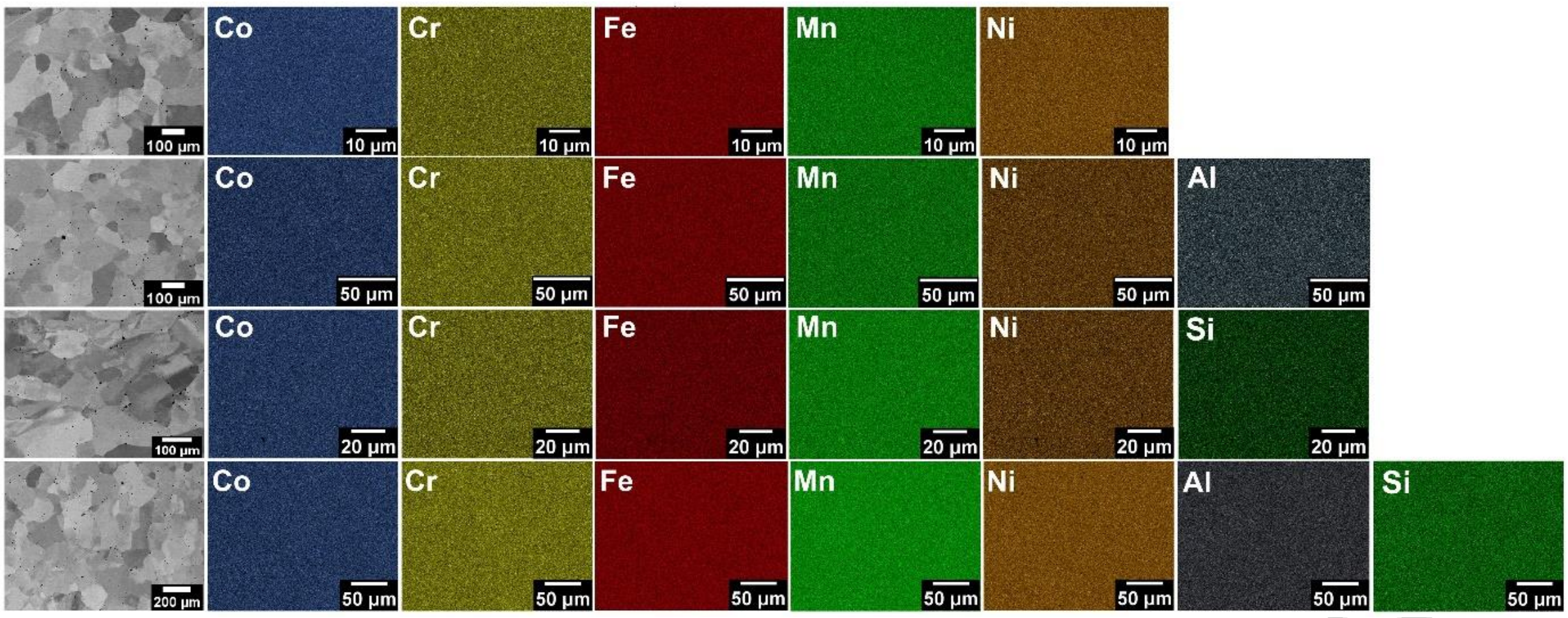

**Figure SF24 –** BSE micrographs for $Co_{24}Cr_{12}Fe_{16}Mn_{16}Ni_{32}$, $Co_{20}Cr_{16}Fe_{20}Mn_{16}Ni_{24}Al_4$, $Co_{20}Cr_{16}Fe_{12}Mn_{20}Ni_{28}Si_4$ and $Co_{24}Cr_{12}Fe_{12}Mn_{16}Ni_{28}Al_4Si_4$ alloys showing the typical grain structure. EDS maps adjoined for various alloy constituents for the four alloys, indicating the chemically homogenous nature of alloys.

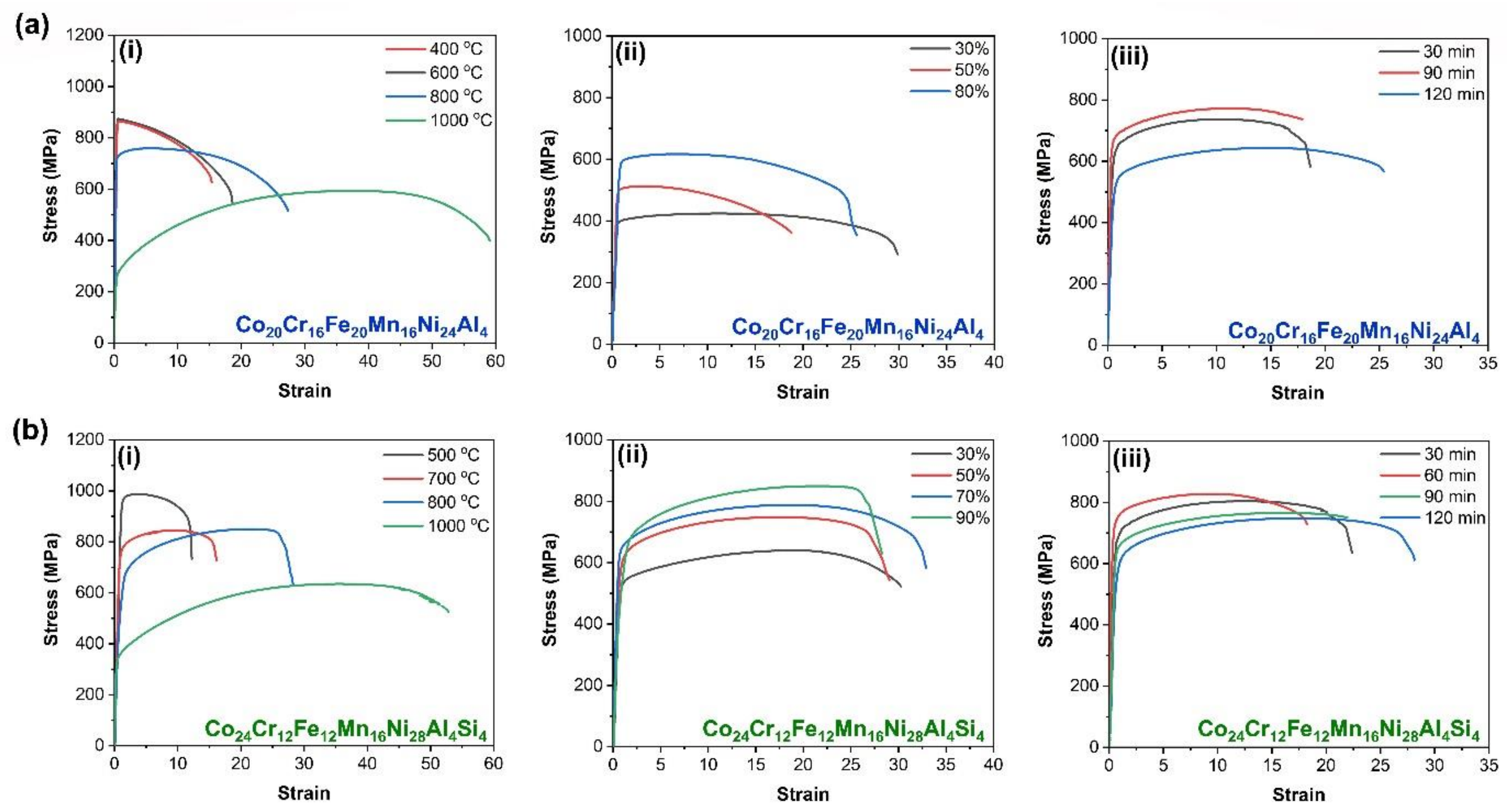

**Figure SF25 –** Stress-strain curves for (a) $Co_{20}Cr_{16}Fe_{20}Mn_{16}Ni_{24}Al_4$ and (b) $Co_{24}Cr_{12}Fe_{12}Mn_{16}Ni_{28}Al_4Si_4$ alloys, for different combinations of varying annealing temperature, % deformation and annealing time values.

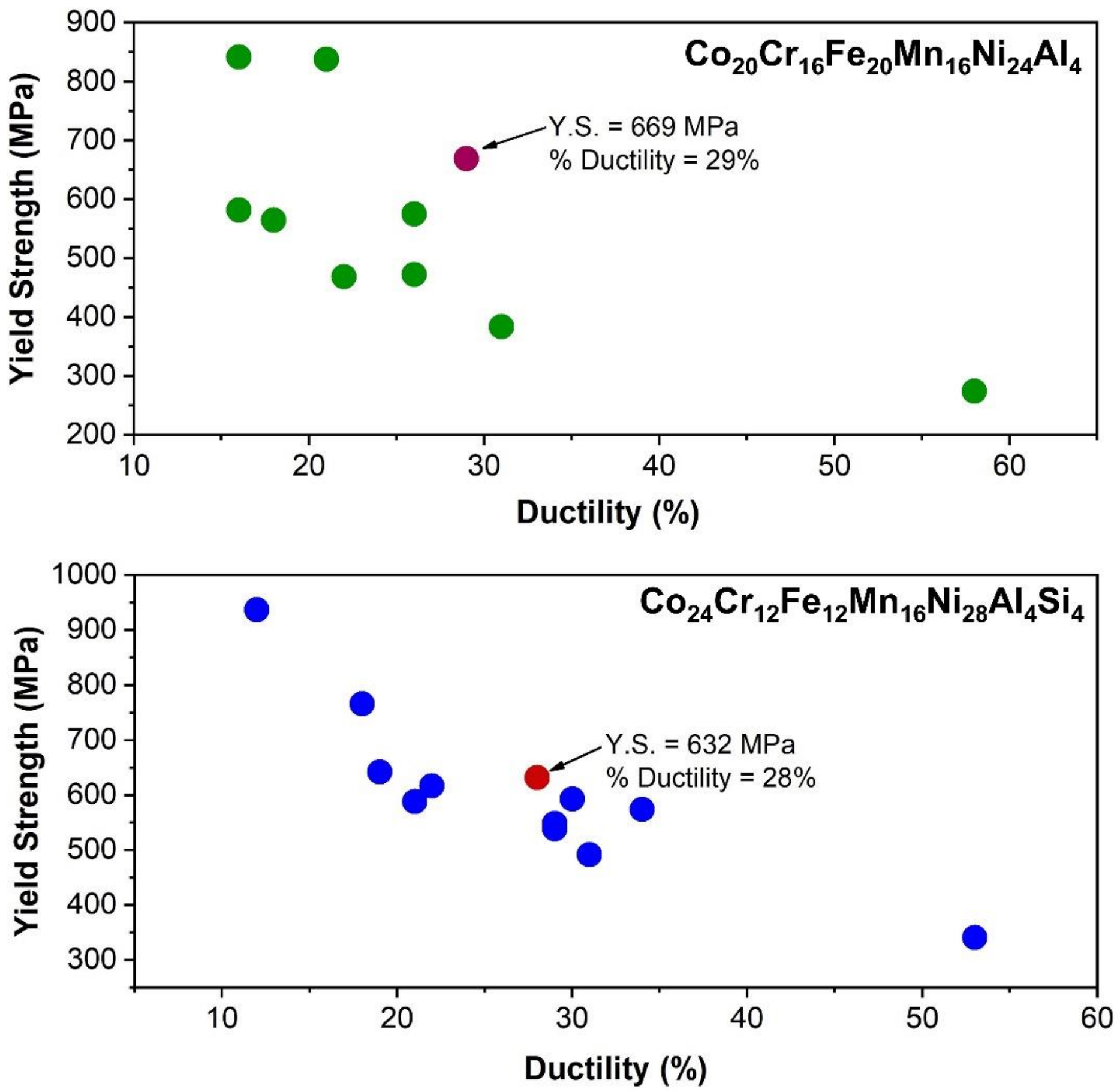

**Figure SF26 –** Simultaneous variation of yield strength and % ductility for $Co_{20}Cr_{16}Fe_{20}Mn_{16}Ni_{24}Al_4$ and $Co_{24}Cr_{12}Fe_{12}Mn_{16}Ni_{28}Al_4Si_4$ alloys. The dissimilar coloured markers indicate the best possible combination of yield strength and ductility for the two alloys.

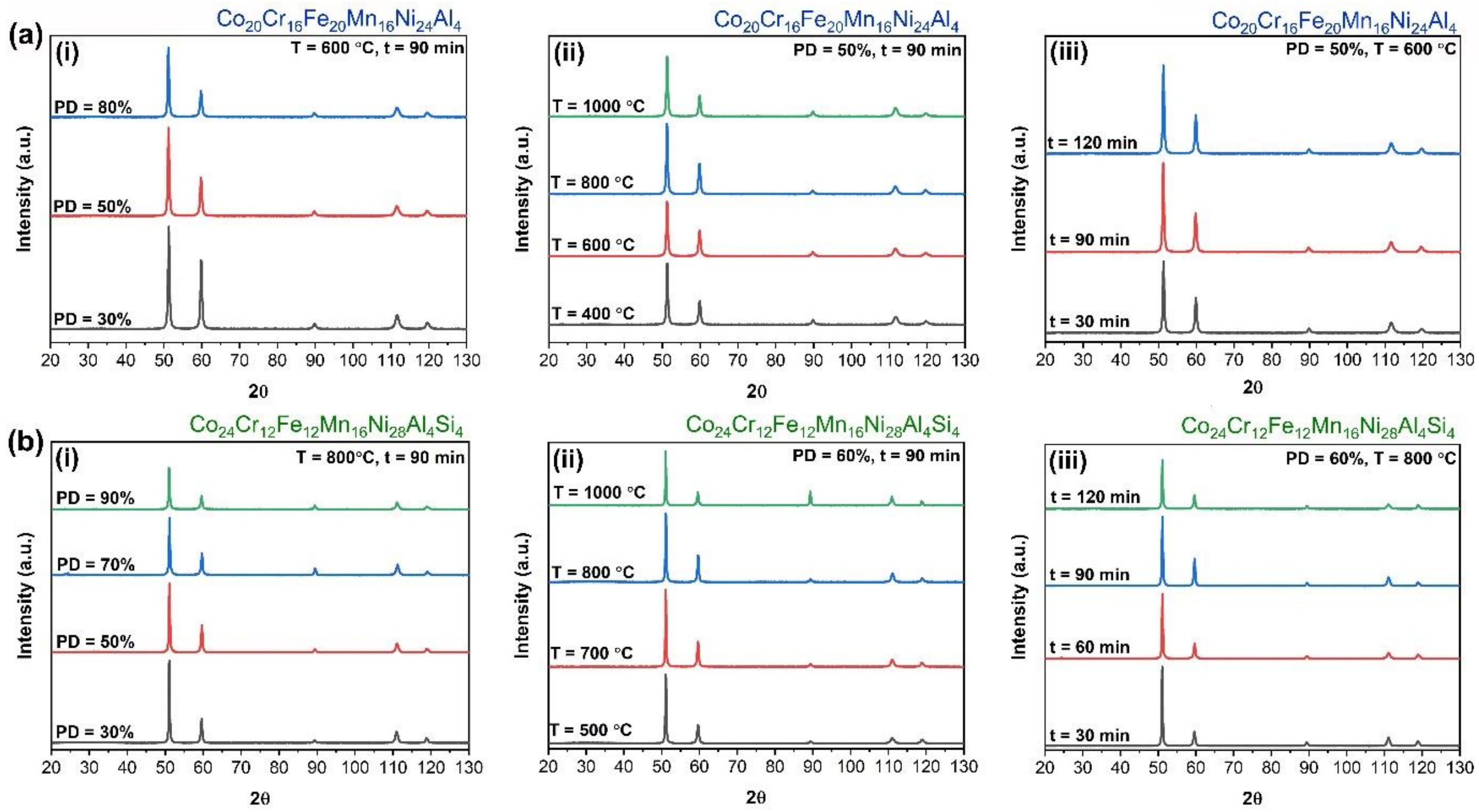

**Figure SF27** – Variation of phase evolution for $Co_{20}Cr_{16}Fe_{20}Mn_{16}Ni_{24}Al_4$ and $Co_{24}Cr_{12}Fe_{12}Mn_{16}Ni_{28}Al_4Si_4$ alloys as a function of varying combinations of annealing temperature, % deformation and annealing time.

# Supplementary Tables

**Table ST1** – Feature candidates for DS1 to predict the yield strength of FCC-HEAs and their distribution.

| | Minimum | Maximum | Mean | Standard Deviation |
|---|---|---|---|---|
| **Al** | 0 | 28.55 | 4.75 | 7.73 |
| **C** | 0 | 4.75 | 0.75 | 0.41 |
| **Co** | 0 | 58.82 | 20.45 | 9.81 |
| **Cr** | 0 | 40 | 16.8 | 9.38 |
| **Cu** | 0 | 50 | 3.72 | 7.92 |
| **Fe** | 0 | 60 | 21.34 | 10.73 |
| **Mn** | 0 | 50 | 8.52 | 11.9 |
| **Mo** | 0 | 20 | 0.58 | 2.64 |
| **Nb** | 0 | 26.8 | 0.34 | 2.53 |
| **Ni** | 0 | 50 | 21.91 | 8.87 |
| **V** | 0 | 14.28 | 0.33 | 1.73 |
| **Ti** | 0 | 25 | 1.16 | 4.56 |
| **Si** | 0 | 6.5 | 0.05 | 0.06 |
| **% deformation** | 5 | 92 | 64.91 | 19.96 |
| **Temperature (K)** | 450 | 1200 | 924.46 | 137.87 |
| **Time (min)** | 0.5 | 360 | 74.09 | 51.4 |
| **Yield Strength (MPa)** | 112 | 1028 | 380.34 | 164.56 |

**Table ST2** – Search space for all relevant hyperparameters and the best-performing hyperparameters for the various ML models (the best hyperparameters are chosen based on the highest validation accuracy across the 10-fold partitioning).

| Model | Hyperparameter | Hyperparameters Grid | Best Hyperparameters |
|---|---|---|---|
| Elastic Net (E-Net) | Alpha | 0.00000001-10.0 | 1.0 |
| | $L_1$ ratio | 0.00001-0.99 | 0.001 |
| | Selection | cyclic, random | Cyclic |
| Support Vector Machine (SVM) | C | 0.00001, 0.0001, 0.001, 0.01, 0.1, 1, 10, 100, 500 | 100 |
| | Kernel | linear, polynomial, radial basis function, sigmoid | radial basis function |
| K-Nearest Neighbours (KNN) | Number of Neighbours | 2, 80 | 8 |
| | Weight | uniform, distance | distance |
| | Metric | euclidean, manhattan | euclidean |
| Random Forest (RF) | Number of estimators | 10-500 | 180 |
| | Criterion | gini, entropy | gini |
| | Maximum depth | 2-30 | 12 |
| | Minimum sample split | 1-30 | 8 |
| | Minimum sample leaf | 1-30 | 6 |
| Lasso Regression (LR) | Alpha | 0.00001-1.0 | 0.01 |
| | Selection | cyclic, random | cyclic |
| Ridge Regression (RR) | Alpha | 0.00001-10 | 0.01 |
| | Solver | auto, svd, cholesky, lsqr, sparse_cg | svd |
| Gradient Boosting (GB) | Number of estimators | 10-500 | 71 |
| | Learning rate | 0.000001-1.0 | 0.01 |
| | Maximum depth | 1-30 | 16 |
| | Minimum sample split | 2-30 | 12 |
| | Maximum features | 0.1-1.0 | 0.43 |
| Neural Network (NN) | Number of hidden layers | 1-4 | 2 |
| | Neurons per layer | 10-100 | 31 |
| | Activation function | relu, sigmoid, tanh | relu |
| | Optimizer | sgd, adam, AdaGrad, RMSprop | adam |
| | Learning Rate | 0.00001-0.1 | 0.001 |
| | Batch Size | 8-64 | 27 |

**Table ST3** – Feature candidates for DS2 and their formulae for calculations ($c_i$ and $c_j$ are the atomic concentration (%) of the $i^{th}$ and $j^{th}$ element, respectively; $r_i$ and $r_j$ are the atomic radius of the $i^{th}$ and $j^{th}$ element, respectively; $w_i$, $\rho_i$, $\chi_i$, $E_{C_i}$ and $G_i$ are the molecular weight, density, Pauling's electronegativity, cohesive energy and shear modulus of the $i^{th}$ element, and $G$ is the average value of the shear modulus of all constituents of an alloy.)

| **Parameter** | **Formula** |
|---|---|
| **Formation Enthalpy ($\Delta H_{mix}$)** (*72*) | $\Delta H_{mix} = 4\left[\sum_{j\neq i}^{N=n}\sum_{i=1}^{N=n}\Delta H_{ij}c_ic_j\right]$ |
| **Mixing Entropy ($\Delta S_{mix}$)** (*73*) | $\Delta S_{mix} = -R\sum_{i=1}^{n}c_i ln c_i$ |
| **Atomic Size Mismatch ($\delta$)** (*74*) | $\delta = \sqrt{\sum_{i=1}^{n}c_i\left(1-\frac{r_i}{\bar{r}}\right)^2} \times 100$ |
| **Molar Volume ($V_m$)** (*75*) | $V_m = \sum_{i=1}^{n}c_iw_i/\rho_i$ |
| **Elastic Modulus ($E$)** (*76*) | $E = \sum_{i=1}^{n}c_iE_iV_{m,i}/V_m$ |
| **Bulk Modulus ($K$)** (*75*) | $K = \sum_{i=1}^{n}c_iK_iV_{m,i}/V_m$ |
| **Shear Modulus ($G$)** (*76*) | $G = \sum_{i=1}^{n}c_iG_iV_{m,i}/V_m$ |
| **Number of itinerant electrons ($e/a$)** (*77*) | $e/a = \sum_{i=1}^{n}c_i(e/a)_i$ |
| **Melting Temperature ($T_m$)** (*75*) | $T_m = \sum_{i=1}^{n}c_i(T_m)_i$ |
| **Density ($\rho$)** (*76*) | $\rho = \sum_{i=1}^{n}c_i\rho_i$ |
| **Pauling's Electronegativity ($\chi_P$)** (*78*) | $\chi_P = \sqrt{\sum_{i=1}^{n}c_i\left(\chi_i-\overline{\chi_P}\right)^2}$ |
| ***Ω* Parameter** (*73*) | $\Omega = \frac{T_m\Delta S_{mix}}{\lvert\Delta H_{mix}\rvert}$ |
| **Modulus Mismatch ($\eta$)** (*79*) | $\eta = \sum_{i=1}^{N}\frac{c_i\times\frac{2(G_i-G)}{(G_i+G)}}{1+0.5\left\lvert c_i\times\frac{2(G_i-G)}{(G_i+G)}\right\rvert}$ |
| **Lattice Distortion Energy ($E_{LD}$)** (*77*) | $E_{LD} = \frac{1}{2}E\times\delta$ |
| **Cohesive Energy ($E_C$)** (*77*) | $E_C = \sum_{i=1}^{n}c_i\,E_{C_i}$ |
| **Peierls-Nabarro factor ($F_{PN}$)** (*77*) | $F_{PN} = \frac{2G}{1-\mu}$ |

**Table ST4** – Selected features from the DS2 candidates for each feature selection method (selected features are marked by green box and rejected features are marked by red box). $\Delta H_{mix}$ - Formation Enthalpy, $\Delta S_{mix}$ - Mixing Entropy, $\delta$ - Atomic Size Mismatch, $V_m$ - Molar Volume, $E$ - Elastic Modulus, $K$ - Bulk Modulus, $G$ - Shear Modulus, *e/a* - Number of itinerant electrons, $T_m$ - Melting Temperature, $\rho$ – Density, $\chi_P$ - Pauling's Electronegativity, $\eta$ - Modulus Mismatch, $E_{LD}$ - Lattice Distortion Energy, $E_C$ - Cohesive Energy and $F_{PN}$ - Peierls-Nabarro factor

| | **Correlation Analysis** | **Random Forest based Feature Importance** | **Gradient Boosting based Feature Importance** | **Permutation Importance** | **Mutual Importance** | **Backward Elimination** | **Recursive Feature Elimination** | **Genetic Algorithm** | **Best Subset Selection** |
|---|---|---|---|---|---|---|---|---|---|
| $\Delta H_{mix}$ | Green | Red | Red | Red | Green | Red | Red | Red | Green |
| $\Delta S_{mix}$ | Green | Green | Green | Red | Red | Green | Green | Green | Green |
| $\delta$ | Green | Green | Red | Red | Green | Green | Green | Red | Red |
| $V_m$ | Green | Green | Green | Green | Green | Green | Green | Green | Green |
| $E$ | Red | Red | Red | Green | Red | Red | Red | Green | Red |
| $K$ | Red | Red | Red | Red | Red | Red | Red | Red | Red |
| $G$ | Green | Red | Red | Red | Green | Red | Green | Red | Green |
| $VEC$ | Green | Red | Green | Green | Green | Green | Green | Green | Green |
| $T_m$ | Green | Green | Green | Green | Green | Green | Green | Red | Red |
| $\rho$ | Red | Red | Red | Green | Green | Red | Green | Red | Red |
| $\chi_P$ | Green | Green | Green | Red | Red | Red | Green | Green | Green |
| $\Omega$ | Red | Red | Red | Red | Green | Red | Red | Red | Red |
| $\eta$ | Green | Green | Green | Red | Green | Green | Green | Green | Green |
| $E_{LD}$ | Red | Red | Red | Red | Green | Red | Red | Red | Red |
| $E_C$ | Green | Green | Green | Green | Green | Green | Green | Green | Green |
| $F_{PN}$ | Red | Red | Red | Red | Red | Red | Red | Red | Red |

**Table ST5** – The performance ($R^2$ score and MAE) of various ML models with different feature selection methods.

| | **Correlation Analysis** | | | **Random Forest based Feature Importance** | | | **Gradient Boosting based Feature Importance** | | | **Permutation Importance** | | | **Mutual Importance** | | | **Backward Elimination** | | | **Recursive Feature Elimination** | | | **Genetic Algorithm** | | | **Best Subset Selection** | | |
|---|---|---|---|---|---|---|---|---|---|---|---|---|---|---|---|---|---|---|---|---|---|---|---|---|---|---|---|
| | **$R^2$ (Train)** | **$R^2$ (Test)** | **MAE (Test)** | **$R^2$ (Train)** | **$R^2$ (Test)** | **MAE (Test)** | **$R^2$ (Train)** | **$R^2$ (Test)** | **MAE (Test)** | **$R^2$ (Train)** | **$R^2$ (Test)** | **MAE (Test)** | **$R^2$ (Train)** | **$R^2$ (Test)** | **MAE (Test)** | **$R^2$ (Train)** | **$R^2$ (Test)** | **MAE (Test)** | **$R^2$ (Train)** | **$R^2$ (Test)** | **MAE (Test)** | **$R^2$ (Train)** | **$R^2$ (Test)** | **MAE (Test)** | **$R^2$ (Train)** | **$R^2$ (Test)** | **MAE (Test)** |
| **Lasso** | 0.64 | 0.60 | 73.65 | 0.66 | 0.61 | 72.33 | 0.66 | 0.61 | 71.69 | 0.65 | 0.62 | 72.83 | 0.65 | 0.61 | 72.96 | 0.66 | 0.62 | 69.60 | 0.66 | 0.63 | 65.20 | 0.69 | 0.65 | 63.81 | 0.68 | 0.64 | 64.85 |
| **Ridge** | 0.66 | 0.62 | 71.38 | 0.67 | 0.61 | 70.95 | 0.67 | 0.63 | 68.66 | 0.68 | 0.66 | 69.30 | 0.68 | 0.63 | 69.47 | 0.68 | 0.64 | 67.75 | 0.69 | 0.65 | 63.70 | 0.70 | 0.67 | 60.29 | 0.69 | 0.68 | 62.28 |
| **Elastic Net** | 0.69 | 0.64 | 67.84 | 0.70 | 0.66 | 68.38 | 0.70 | 0.65 | 68.00 | 0.69 | 0.66 | 68.38 | 0.69 | 0.65 | 68.42 | 0.71 | 0.67 | 65.85 | 0.72 | 0.69 | 60.24 | 0.73 | 0.68 | 56.24 | 0.72 | 0.69 | 58.55 |
| **kNN** | 0.64 | 0.59 | 72.88 | 0.65 | 0.59 | 73.59 | 0.66 | 0.63 | 71.94 | 0.66 | 0.62 | 70.97 | 0.67 | 0.62 | 71.15 | 0.66 | 0.62 | 66.68 | 0.67 | 0.63 | 64.76 | 0.68 | 0.65 | 56.09 | 0.68 | 0.65 | 57.18 |
| **SVR** | 0.71 | 0.67 | 66.80 | 0.71 | 0.68 | 64.95 | 0.71 | 0.67 | 64.42 | 0.72 | 0.68 | 68.22 | 0.73 | 0.68 | 70.03 | 0.73 | 0.69 | 67.25 | 0.74 | 0.70 | 63.83 | 0.76 | 0.72 | 50.57 | 0.74 | 0.72 | 52.16 |
| **Neural Network** | 0.75 | 0.71 | 61.12 | 0.73 | 0.69 | 59.73 | 0.73 | 0.69 | 58.90 | 0.74 | 0.71 | 59.56 | 0.74 | 0.68 | 60.42 | 0.74 | 0.72 | 57.32 | 0.75 | 0.70 | 57.43 | 0.77 | 0.73 | 49.16 | 0.76 | 0.72 | 51.58 |
| **Random Forest** | 0.79 | 0.76 | 52.74 | 0.79 | 0.77 | 50.83 | 0.80 | 0.77 | 49.31 | 0.79 | 0.74 | 48.07 | 0.78 | 0.73 | 50.15 | 0.79 | 0.75 | 48.51 | 0.81 | 0.77 | 45.90 | **0.81** | **0.79** | **38.18** | 0.80 | 0.78 | 42.53 |
| **Gradient Boosting** | 0.82 | 0.77 | 41.60 | 0.83 | 0.79 | 39.38 | 0.84 | 0.79 | 38.23 | 0.83 | 0.79 | 41.77 | 0.81 | 0.77 | 45.20 | 0.82 | 0.78 | 42.50 | 0.83 | 0.78 | 40.43 | **0.84** | **0.82** | **34.22** | 0.82 | 0.79 | 38.29 |

**Table ST6**– Performance of the RELM for the test dataset to evaluate the efficiency of the physics descriptor enhanced RELM with values of true and predicted yield strength as well as the absolute error in prediction.

| Alloy | Reported Yield Strength, MPa ($YS_R$) | Predicted Yield Strength, MPa ($YS_P$) | % Absolute Error $= \left\vert\frac{YS_P - YS_R}{YS_R}\right\vert \times 100$ |
|---|---|---|---|
| $Co_{20}Cr_{20}Fe_{20}Mn_{20}Ni_{20}$ | 280 | 262.52 | 9.36 |
| $Co_{33.34}Mn_{33.34}Ni_{33.34}$ | 261.45 | 255.42 | 3.465 |
| $Co_{20}Cr_{20}Fe_{20}Mn_{20}Ni_{20}$ | 348.41 | 375.44 | 11.64 |
| $Al_{7.4}C_{1.1}Cr_{5.55}Fe_{39.93}Mn_{35.67}Ni_{10.35}$ | 355 | 365.73 | 4.53 |
| $Co_{33.34}Cr_{33.34}Ni_{33.34}$ | 310 | 317.1 | 3.435 |
| $Co_{20}Cr_{20}Fe_{20}Ni_{20}Pd_{20}$ | 389.35 | 399.91 | 4.065 |
| $Co_{24.925}Cr_{24.925}Fe_{24.925}Ni_{24.925}Al_{0.3}$ | 402 | 430.65 | 10.69 |
| $Co_{20}Cr_{20}Fe_{20}Mn_{20}Ni_{20}$ | 366.73 | 387.81 | 8.625 |
| $Co_{33.34}Cr_{33.34}Ni_{33.34}$ | 420 | 394.63 | 9.06 |
| $Co_{10}Cr_{10}Fe_{50}Mn_{30}$ | 230 | 230.57 | 0.375 |
| $Co_{10}Cr_{15}Fe_{40}Ni_{25}V_{10}$ | 463.56 | 443.85 | 6.375 |
| $Co_{33.34}Mn_{33.34}Ni_{33.34}$ | 370 | 386.86 | 6.84 |
| $Fe_{40}Mn_{27}Ni_{26}Co_{5}Cr_{2}$ | 180 | 175.89 | 3.42 |
| $Co_{20}Cr_{20}Fe_{20}Mn_{20}Ni_{20}$ | 275 | 254.07 | 11.415 |
| $Fe_{40}Mn_{27}Ni_{26}Co_{7}$ | 250 | 259.24 | 5.55 |
| $Co_{20}Cr_{20}Fe_{20}Mn_{20}Ni_{20}$ | 250 | 269.95 | 11.97 |
| $Co_{33.34}Cr_{33.34}Ni_{33.34}$ | 281.87 | 282.5 | 0.33 |
| $Cu_{33.34}Cr_{33.34}Ni_{33.34}$ | 351 | 365.25 | 6.09 |
| $Fe_{33.34}Mn_{33.34}Ni_{33.34}$ | 230 | 230.69 | 0.45 |
| $Co_{33.34}Fe_{33.34}Ni_{33.34}$ | 217 | 220.04 | 2.1 |
| $Co_{20}Cu_{20}Fe_{20}Mn_{20}Ni_{20}$ | 343.08 | 334.2 | 3.885 |
| $Cr_{33.34}Fe_{33.34}Ni_{33.34}$ | 230 | 231.57 | 1.02 |
| $Co_{20}Cr_{20}Fe_{20}Mn_{20}Ni_{20}$ | 240 | 265.83 | 16.14 |
| $Cr_{15}Fe_{55}Mn_{20}Ni_{10}$ | 230 | 235.59 | 3.645 |
| $Co_{20}Cr_{20}Fe_{20}Mn_{20}Ni_{20}$ | 347 | 374.69 | 11.97 |
| $Co_{20}Cr_{20}Fe_{20}Mn_{20}Ni_{20}$ | 432.47 | 430.06 | 0.84 |
| $Co_{25}Cr_{25}Fe_{25}Ni_{25}$ | 394.33 | 401.48 | 2.715 |
| $Co_{33.34}Mn_{33.34}Ni_{33.34}$ | 526.26 | 506.68 | 5.58 |
| $Fe_{33.34}Mn_{33.34}Ni_{33.34}$ | 422.42 | 424.36 | 0.69 |
| $Co_{25}Fe_{25}Mn_{25}Ni_{25}$ | 175.97 | 188.59 | 10.755 |
| $Fe_{50}Mn_{50}$ | 459 | 472.24 | 4.32 |
| $Co_{15}Cr_{15}Fe_{40}Mn_{15}Ni_{15}$ | 260 | 255.05 | 2.85 |
| $Co_{25}Fe_{25}Mn_{25}Ni_{25}$ | 309.91 | 297.12 | 6.195 |
| $Co_{10}Cr_{10}Fe_{50}Mn_{30}$ | 250 | 235.15 | 8.91 |
| $Co_{25}Cr_{25}Fe_{25}Ni_{25}$ | 310 | 315.76 | 2.79 |
| $Co_{20}Cr_{20}Fe_{20}Mn_{20}Ni_{20}$ | 390 | 418.75 | 11.055 |
| $Co_{33.34}Fe_{33.34}Ni_{33.34}$ | 392.17 | 398.27 | 2.34 |
| $Co_{33.34}Cr_{33.34}Ni_{33.34}$ | 484 | 441.07 | 13.305 |
| $Co_{20}Cr_{20}Fe_{20}Ni_{20}Pd_{20}$ | 346.5 | 357.39 | 4.71 |
| $Co_{20}Cr_{20}Fe_{20}Mn_{20}Ni_{20}$ | 328.54 | 341.09 | 5.73 |
| $Co_{33.34}Mn_{33.34}Ni_{33.34}$ | 362 | 385.74 | 9.84 |
| $Co_{19}Cr_{19}Fe_{19}Mn_{19}Ni_{19}Al_{5}$ | 321 | 333.34 | 5.76 |
| $Co_{30}Cr_{30}Fe_{25}Ni_{15}$ | 560 | 582.88 | 6.135 |
| $Co_{10}Cr_{15}Fe_{35}Mn_{5}Ni_{25}V_{10}$ | 401 | 392.96 | 3 |
| $Co_{20}Cr_{20}Fe_{20}Mn_{20}Ni_{20}$ | 340.98 | 352.24 | 4.95 |
| $Co_{25}Cr_{25}Fe_{25}Ni_{25}$ | 247.23 | 250.2 | 1.8 |
| $Co_{20}Cr_{20}Fe_{20}Mn_{20}Ni_{20}$ | 190 | 198.1 | 6.39 |
| $Fe_{33.34}Mn_{33.34}Ni_{33.34}$ | 599 | 576.79 | 5.565 |
| $Al_{7.35}C_{0.3}Cr_{5.2}Fe_{44.85}Mn_{32.25}Ni_{10.05}$ | 208 | 219.66 | 8.415 |
| $Co_{24.925}Cr_{24.925}Fe_{24.925}Ni_{24.925}Al_{0.3}$ | 560 | 534.86 | 6.735 |
| $Co_{20}Cr_{20}Fe_{20}Mn_{20}Ni_{20}$ | 211.01 | 222.44 | 8.13 |
| $Co_{10}Cr_{10}Fe_{60}Mn_{10}Ni_{10}$ | 228 | 237.37 | 6.165 |
| $Fe_{40.4}Mn_{34.8}Ni_{11.3}Al_{7.5}Cr_{6}$ | 479 | 464.3 | 4.605 |
| $Co_{20}Cr_{20}Fe_{20}Mn_{20}Ni_{20}$ | 236.5 | 257.81 | 13.515 |